\documentclass[12pt]{article}

\usepackage{amsmath}
\usepackage{amssymb}
\usepackage{graphicx}
\usepackage{ifthen}
\usepackage{verbatim}
\usepackage{psfrag}

\newcommand{\pgrad}[2]
{\frac{\partial #1}{\partial #2}}

\makeatletter
  
  \@addtoreset{equation}{section}
\makeatother

\begin{document}

%
%

\thispagestyle{empty}
\begin{flushright}
\texttt{OU-HET/554}\\
\texttt{hep-th/0601233}\\
January 2006
\end{flushright}
\bigskip
\bigskip
\begin{center}
{\Large \textbf{Amoebas and Instantons}}
\end{center}
\bigskip
\bigskip
\renewcommand{\thefootnote}{\fnsymbol{footnote}}
\begin{center}
Takashi Maeda
\footnote{E-mail: \texttt{maeda@het.phys.sci.osaka-u.ac.jp}}
and 
Toshio Nakatsu
\footnote{E-mail: \texttt{nakatsu@het.phys.sci.osaka-u.ac.jp}}\\
\bigskip
{\small
\textit{Department of Physics, Graduate School of Science,
Osaka University,\\
Toyonaka, Osaka 560-0043, Japan}}
\end{center}
\bigskip
\bigskip
\renewcommand{\thefootnote}{\arabic{footnote}}
\begin{abstract}
We study a statistical model of random plane partitions.
The statistical model has interpretations 
as five-dimensional $\mathcal{N}=1$ supersymmetric 
$SU(N)$ Yang-Mills on $\mathbb{R}^4\times S^1$
and as K\"ahler gravity on local $SU(N)$ geometry.
At the thermodynamic limit
a typical plane partition called the limit shape
dominates in the statistical model.
The limit shape is linked with
a hyperelliptic curve,
which is a five-dimensional version of 
the $SU(N)$ Seiberg-Witten curve.
Amoebas and the Ronkin functions
play intermediary roles between 
the limit shape and the hyperelliptic curve. 
In particular, 
the Ronkin function realizes 
an integration of 
thermodynamical density of the main diagonal partitions,  
along one-dimensional slice of it  
and thereby is interpreted 
as the counting function of gauge instantons. 
The radius of $S^1$ can be identified with 
the inverse temperature of the statistical model.
The large radius limit of the five-dimensional Yang-Mills
is the low temperature limit of the statistical model, 
where the statistical model 
is frozen to a ground state 
that is associated with the local $SU(N)$ geometry.
We also show that the low temperature limit 
corresponds to a certain degeneration of
amoebas and the Ronkin functions 
known as tropical geometry.
\end{abstract}

\setcounter{footnote}{0}
\newpage

\section{Introduction and summary}

It is an old idea that spacetime 
may have more than four dimensions,
with extra coordinates being 
unobservable at available energies.
A first possibility arises in 
the Kaluza-Klein theory.
In the Kaluza-Klein approach,
gravitation and electromagnetism
could be unified in a theory of 
five-dimensional geometry.
The extra dimension
makes $S^1$ that radius
is microscopically small.
The Kaluza-Klein like approach 
has always been one of 
the most intriguing ideas 
in physics.

The idea of the Kaluza-Klein theory is
utilized in superstring theory \cite{superstrings}. 
Superstring theory is a candidate for
a theory of everything.
When superstring theory is completed,
all four-dimensional theories,
such as standard model and general relativity,
can be derived from superstring theory.
In superstring theory,
strings which are one-dimensional objects
play a central role
instead of point particles.
Each elementary particle corresponds to
each vibration mode of string.
While it moves in a spacetime,
a string sweeps out
a two-dimensional surface.
So the motion of a string is
given by a map from
a two-dimensional surface $\Sigma$, 
called the worldsheet,
to a spacetime.
In this sense the spacetime
is often called the target space.
Superstring theory makes sense only
in ten spacetime dimensions
(at least in perturbative treatments).
Is it a pity that the spacetime dimension
is not four?
We don't think so.
We take the ten-dimensional space to be of
the form $M^4\times X$,
where $M^4$ is our four-dimensional 
Minkowski space and
$X$ is a compact six-dimensional space.
From the physical viewpoint,
the most interesting candidate for
$X$ is Calabi-Yau threefolds.
We cannot look at $X$ directly,
however geometrical natures
of $X$ turn up as physical properties
in $M^4$,
such as the gauge symmetry and the matter content.
Studying superstring theory on $M^4\times X$,
one can find that the internal properties of $X$
lead to physical consequences for
the observers living in $M^4$.

Supersymmetric theories \cite{Wess-Bagger}
give laboratories to test these ideas.
Supersymmetric theories are more tractable
than ordinary non-supersymmetric theories,
and many of their observables
can be computed exactly.
Nevertheless, it turns out that these theories
exhibit explicit examples of various phenomena 
in quantum field theories.
Then it has become a priority in particle physics
to understand better the perturbative and
non-perturbative dynamics
of supersymmetric theories.
$\mathcal{N}=2$ supersymmetric gauge theories
have been particularly studied 
at both perturbative and non-perturbative levels.
The extended supersymmetry 
dramatically simplifies 
the dynamics of gauge theories.
$\mathcal{N}=2$ gauge theories also have an interesting
interpretation from the perspective of
superstring theories.
It is called the geometric engineering
\cite{Geometric engineering}.
According to the geometric engineering,
$\mathcal{N}=2$ supersymmetric gauge theory
is realized by the type IIA superstring
on a certain Calabi-Yau threefold.
Some physical properties in the gauge theory
is derived from geometrical natures
of the Calabi-Yau threefold.
The gauge symmetry is closely related with
the ADE singularity
in the Calabi-Yau threefold.
For example,
the $SU(N)$ gauge symmetry
is realized by an $A_{N-1}$-type
singularity.
The geometry which leads to the $SU(N)$
gauge theory is often called
the local $SU(N)$ geometry.
The geometric engineering connects
four-dimensional gauge theories with
geometries of internal spaces.
This gives an example of
the gauge/gravity correspondence.
It goes without saying that
the most important principle in physics is 
gauge theory and general relativity.
The duality between these two theories is
an interesting research area in
particle physics.
AdS/CFT correspondence \cite{AdS/CFT1,AdS/CFT2}
is one of the most
remarkable examples of this duality.

In the last few years,
there has been great progress in
both superstring theories and
supersymmetric gauge theories.
The exact low-energy dynamics of 
$\mathcal{N}=2$ supersymmetric gauge theories
has been revealed by
Seiberg and Witten \cite{Seiberg-Witten}.
They have evaluated exact low-energy effective actions
of $\mathcal{N}=2$ theories
using the holomorphy and a version
of the electromagnetic duality.
The low-energy effective theory
obtained by them led to
numerous achievements in
understanding of the non-perturbative
dynamics of gauge theories.
Their derivation is elegant
but follows in a somewhat indirect way.
Recently,
Nekrasov and Okounkov have calculated
exact low-energy effective actions
in a more direct way
\cite{Nekrasov, Nekrasov-Okounkov}.
They have evaluated the path integrals of
$\mathcal{N}=2$ supersymmetric gauge theories
and give exact formulae for
both perturbative and non-perturbative dynamics.
The Seiberg-Witten solutions emerges
through statistical models of random partitions.
Also, 
Nakajima and Yoshioka derived independently 
the Seiberg-Witten solutions by taking  
the algebraic geometry viewpoint 
\cite{Nakajima-Yoshioka}.

Recent progress in superstring theories is attributed to 
understanding of topological strings.
Topological string is a simplified model of 
superstring theories, 
first proposed by Witten \cite{topological string}, 
which capture topological information 
of the target space. 
In the past ten years or so, 
it has turned out that topological strings 
have an enormous amount of applications.
Their structure is complicated enough to relate them 
to physically interesting theories, 
yet simple enough to be able to obtain exact results. 
There are two different variations in topological strings, 
the A-model and the B-model. 
We are interested in the A-model.
Topological A-model string amplitudes for 
a certain class of geometries 
can be computed using a diagrammatical method, 
called the topological vertex 
\cite{topological vertex 1,topological vertex 2}. 
One of the most interesting applications 
of topological strings 
is the geometric engineering. 
The topological vertex allows us to evaluate 
the topological string amplitude for 
the local $SU(N)$ geometry. 
The result of topological strings 
actually reproduces Nekrasov's formula 
for $\mathcal{N}=2$ $SU(N)$ gauge theory 
\cite{Iqbal,Eguchi}.

The method of the topological vertex
yields an unanticipated
but very exciting connection between
the topological string and a statistical model
of random plane partitions
\cite{Crystal}.
This connection has a surprising
interpretation as quantum gravity.
Quantum theory of gravity is the holy grail of physics.
In quantum gravity spacetime undergoes quantum fluctuations,
which cause wild fluctuations of the geometry and the topology
of spacetime.
These quantum fluctuations make spacetime foamy
at short length scales
\cite{Hawking}.
This idea is very exciting, but we haven't made
a precise understanding of quantum gravity yet.
It is believed that superstring theory 
gives rise to quantum gravity
on the target space.
The topological string can also generate
a quantum gravitational theory.
It is expected that 
the classical part of this target space field theory
is the K\"ahler gravity
\cite{kahler gravity}.
The classical theory seems to receive
quantum deformations caused by
string propagation.
It is conjectured in \cite{quantum foam}
that the statistical model of
random plane partitions
is nothing but the quantum K\"ahler gravity.
The gravitational path integral involving fluctuations
of geometry and topology on the target space
is interpreted as the statistical sum
in the plane partition model.
Namely,
each plane partition corresponds to 
a geometry of the target space.
Then the plane partition model can 
give a precise description of
the target space quantum gravity.

In this article, 
we investigate a certain model of 
random plane partitions 
and its physical applications. 
Our motivation is to understand the relation 
between superstring theories and 
gauge theories, 
and clarify the gauge/gravity correspondence 
between the gauge theories 
and the internal space gravities. 
We think that 
the well-defined statistical model 
gives a useful tool to study these relations. 
Besides, 
our statistical model would be a good laboratory 
for studying quantum gravity.

A plane partition $\pi$ is an array of non-negative integers 
\begin{eqnarray}
\begin{array}{cccc}
\pi_{11} & \pi_{12} & \pi_{13} & \cdots \\
\pi_{21} & \pi_{22} & \pi_{23} & \cdots \\
\pi_{31} & \pi_{32} & \pi_{33} & \cdots \\
\vdots & \vdots & \vdots & ~
\end{array}
\label{pi}
\end{eqnarray}
satisfying 
$\pi_{ij}\geq \pi_{i+1 j}$ and 
$\pi_{ij}\geq \pi_{i j+1}$ for 
all $i,j \geq 1$. 
It is identified with 
the three-dimensional Young diagram  
as depicted in Figure 1-(a). 
The three-dimensional diagram $\pi$ 
is a set of unit cubes such that $\pi_{ij}$ cubes 
are stacked vertically on each $(i,j)$-element of $\pi$. 
The diagram is also regarded 
as a sequence of partitions $\pi(m)$, where $m \in \mathbb{Z}$. 
A partition is identified with the (two-dimensional)
Young diagram.
See Figure 1-(b). 
\begin{figure}[t]
\begin{center}
\includegraphics[scale=0.7]{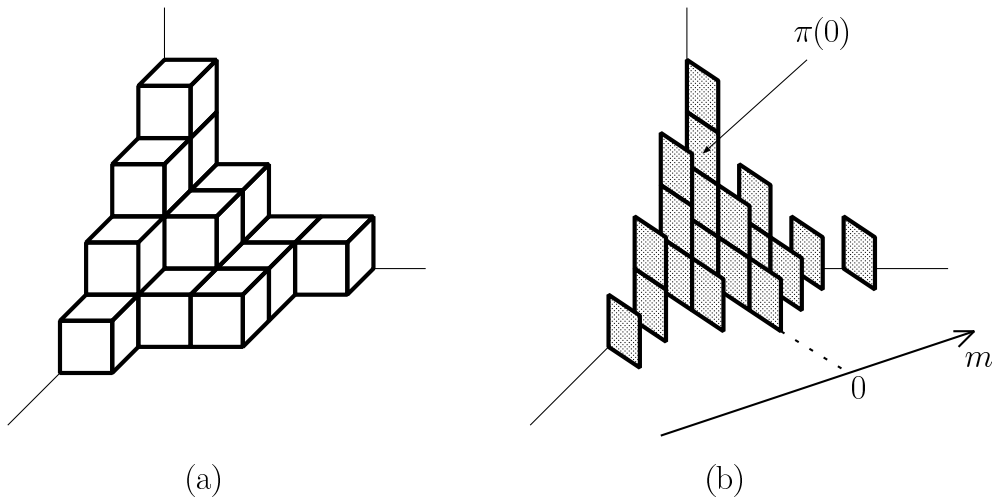}
\caption{\it The three-dimensional Young diagram (a) 
and the corresponding sequence of partitions 
(the two-dimensional Young diagrams) (b).}
\end{center}\label{3d Young diagram}
\end{figure}

Among the series of partitions, 
a partition at the main diagonal $m=0$, 
denoted by $\pi(0)$,
will be called the main diagonal partition of $\pi$
and play an central role in our argument.  
We consider the following model 
of random plane partitions. 
\begin{eqnarray}
Z(q,Q)=\sum_{\pi}\,q^{|\pi|}\,Q^{|\pi(0)|}, 
\label{intro:Z(q,Q)}
\end{eqnarray}
where $q$ and $Q$ are indeterminates. 
$|\pi|$ and $|\pi(0)|$ denote 
respectively the total numbers 
of cubes and boxes of the corresponding diagrams. 
The model with $Q=1$ is well-known \cite{Macdonald}. 
The authors of \cite{Crystal} investigated the $Q=1$ model 
and proposed a connection between the model and topological strings. 
Our plane partition model contains
a new parameter $Q$.
By an identification of $q$ and $Q$ 
with the relevant string theory parameters,  
the partition function 
(\ref{intro:Z(q,Q)})
can be converted into 
topological A-model string amplitude 
on a certain Calabi-Yau threefold. 
We can also retrieve Nekrasov's formulae for 
five-dimensional $\mathcal{N}=1$ supersymmetric Yang-Mills theories 
from the partition function \cite{MNTT1}.

The above model has an interpretation 
as random partitions or the $q$-deformation. 
It can be seen by rewriting the partition function as 
\begin{eqnarray}
Z(q,Q)=\sum_{\lambda}\,Q^{|\lambda|}\, 
\Bigl( \sum_{\pi(0)=\lambda}\, q^{|\pi|} \Bigr), 
\label{intro:Z(q,Q) random partitions}
\end{eqnarray}
where the summation over plane partitions in 
(\ref{intro:Z(q,Q)}) are divided into two branches. 
The partitions $\lambda$ are thought 
as the ensemble of the model by summing first over the 
plane partitions whose main diagonal partitions are $\lambda$.

It is known \cite{Miwa-Jimbo} that 
a partition $\lambda$ has an alternative realization in terms of 
$N$ charged partitions $\{(\lambda^{(r)},p_r)\}_{r=1}^N$. 
The charges $p_r$ are subject to 
the condition $\sum_{r=1}^Np_r=0$. 
Among partitions, 
these coming from the charged empty partitions 
$\{(\emptyset,p_r)\}_{r=1}^N$ turn to play special roles. 
They are called $N$ cores. 
The above realization allows us to read 
the summation over partitions in 
(\ref{intro:Z(q,Q) random partitions}) 
by means of $N$ charged partitions. 
Regarding the model as the $q$-deformed random partitions, 
we will factor the partition function into  
\begin{eqnarray}
Z(q,Q)&=&
\sum_{\{p_r\}}\,
Z^{pert}_{SU(N)}(\{p_r\};q,Q)\,
\sum_{\{\lambda^{(r)}\}} 
Z^{inst}_{SU(N)}(\{\lambda^{(r)}\},\{p_r\};q,Q) 
\nonumber \\
&=&
\sum_{\{p_r\}}\,
Z_{SU(N)}(\{p_r\};q,Q)\,.
\label{intro:factored Z(q,Q)}
\end{eqnarray}
In the first line 
we write the Boltzmann weight for the $N$ core, 
that is read from (\ref{intro:Z(q,Q) random partitions}), 
by $Z^{pert}_{SU(N)}(\{p_r\};q,Q)$. 
In the second line we have included the summation over partitions 
$\lambda^{(r)}$ implicitly in $Z_{SU(N)}(\{p_r\};q,Q)$. 
This factorization turns out useful to find out 
the gauge theoretical interpretations. 
The relevant field theory parameters are 
$a_r,\Lambda$ and $R$, where $a_r$ are the vacuum 
expectation values of the adjoint scalar in the vector 
multiplet and $\Lambda$ is the scale parameter of the 
underlying four-dimensional theory. $R$ is the radius of 
$S^1$ in the fifth dimension. 
We identify these parameters with 
$q,Q$ and $p_r$ in (\ref{intro:factored Z(q,Q)}) as follows.
\begin{eqnarray}
q=e^{-\frac{R}{N}\hbar},~~~~
Q=(R\Lambda)^2,~~~~
p_r=a_r/\hbar\,.
\label{FT parameter introduction}
\end{eqnarray} 
The parameter $\hbar$ is often identified with the string 
coupling constant $g_{st}$.    
The above identification leads \cite{MNTT1} to 
\begin{eqnarray}
Z_{SU(N)}(\{p_r\};q,Q)
=
Z_{\,5d\,\mbox{\scriptsize SYM}}
(\{a_r\};\,\Lambda,R,\hbar), 
\label{intro:Z 5dSYM}
\end{eqnarray}
where the RHS is the exact partition function 
\cite{Nekrasov-Okounkov} for 
five-dimensional $\mathcal{N}=1$ supersymmetric 
$SU(N)$ Yang-Mills with the Chern-Simons term. 
The five-dimensional theory is living on 
$\mathbb{R}^4 \times S^1$. 
The Chern-Simons coupling constant $c_{cs}$ 
is quantized to $N$. 
Actually, 
the identification (\ref{FT parameter introduction}) 
shows that the perturbative part 
and the instanton part of the exact partition function 
for the gauge theory are given by 
$Z_{SU(N)}^{pert}$ and 
$\sum_{\{\lambda^{(r)}\}}Z_{SU(N)}^{inst}$ 
respectively.

\begin{figure}[ht]
\begin{center}
\includegraphics[scale=1]{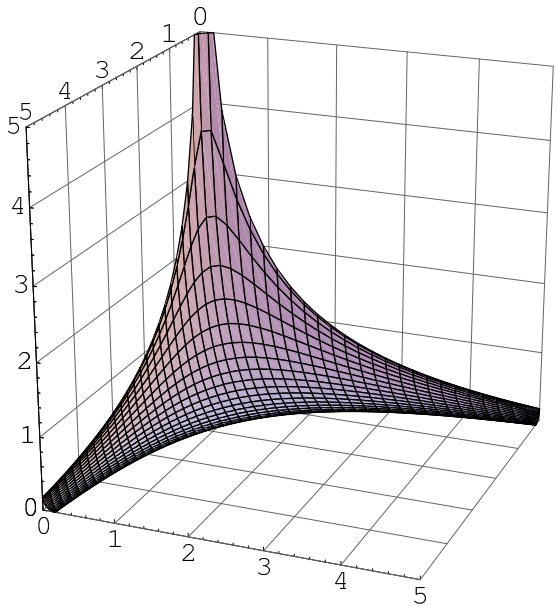}
\end{center}
\caption{\it The limit shape of the simple model
$Z(q,Q=1)=\sum_{\pi}q^{|\pi|}$.}
\label{intro:limit shape}
\end{figure}

The gauge theory prepotential is revealed 
from the exact partition function 
by taking the semiclassical limit, that is, 
the $\hbar \rightarrow 0$ limit.  
This corresponds to the thermodynamic limit of the statistical model. 
At the thermodynamic limit, 
the typical volume of the three-dimensional Young diagrams or 
plane partitions is $\sim \hbar^{-3}$ 
and the variance of the volume is $\sim \hbar^{-4}$. 
Rescaling in all directions by a factor $\hbar$,
the typical three-dimensional Young diagrams approach 
a smooth limit shape 
which is a two-dimensional surface in the octant. 
For the $Q=1$ model, 
such a limit was considered in \cite{Okounkov-Reshetikhin} 
and one can obtain the limit shape 
as in Figure \ref{intro:limit shape}.

In the Seiberg-Witten approach,
hyperelliptic curves known as the Seiberg-Witten curves 
play an important role.
Given a gauge group and matter content,
the potential of the $\mathcal{N}=2$ supersymmetric 
gauge theory has flat directions,
so that the $\mathcal{N}=2$ theory has
a space of physically inequivalent vacua.
This space is often called
the vacuum moduli space.
The vacuum moduli space can be identified with
the moduli space of the Riemann surfaces.
For example, 
each vacua of the $SU(N)$ Yang-Mills theory 
corresponds to a hyperelliptic curve with genus $N-1$
\cite{SU(N) curve}.

The connection between
the plane partition model and 
the $SU(N)$ Yang-Mills suggests a relation between 
the limit shape and the hyperelliptic curve. 
In order to find out the relation,  
we focus our attention on the main diagonal partitions  
and consider the thermodynamic limit of each component 
that appears in the factorization (\ref{intro:factored Z(q,Q)}) 
of the plane partition model.  
We introduce a density $\rho(x|\lambda;p)$
of a charged partition $(\lambda,p)$.
To obtain a finite limit shape
we must scale partitions in a certain manner 
at the thermodynamic limit. 
Thereby 
the density of a partition $\lambda$ 
is scaled to $\rho(u|\lambda)$. 
The asymptotic form of the Boltzmann weight
of the plane partition model is expressed as
an energy functional of the scaled density.
\begin{eqnarray}
\log \bigg(
Q^{|\lambda|}\sum_{\pi(0)=\lambda}q^{|\pi|}
\bigg)
=
-\frac{1}{\hbar^2}
\big\{ E[\rho(\cdot|\lambda)]+O(\hbar)
\big\}
\end{eqnarray}
The minimizer of the energy functional gives 
the typical shape of the main diagonal partitions 
at the thermodynamic limit.  
The minimizer that realizes the thermodynamic limit 
of the component must be found from configurations 
that are expressible in terms of $N$ charged partitions 
with the fixed charges.

The variational problem of the energy functional 
can be solved by using the standard argument \cite{Itzykson}.
The solution $\rho_{\star}^{SU(N)}$ 
turns to be realized using data of the 
hyperelliptic curve.
\begin{eqnarray}
y+y^{-1}
=
\frac{1}{(R\Lambda)^N}
\prod_{r=1}^N(e^{Rz}-\beta_r),
\label{intro:curve}
\end{eqnarray}
where $\beta_r$ are real positive numbers 
determined by the charges.  
This curve is thought to be a five-dimensional 
version of the $SU(N)$ Seiberg-Witten curve 
and 
reduces to the Seiberg-Witten curve 
of four-dimensional $\mathcal{N}=2$ $SU(N)$ Yang-Mills 
at the limit $R\to 0$. 

The more direct link between the limit shape and 
the Seiberg-Witten curve can be revealed.  
To this end, 
we introduce a new object called an amoeba 
\cite{GKZ,Passare-Rullgard}. 
The amoeba of a Laurent polynomial $f(x,y)$ is, 
by definition, 
the image in $\mathbb{R}^2$ of the zero locus of $f(x,y)$ 
under the simple mapping $\mathrm{Log}$ that 
takes each coordinate to the logarithm of its modulus.
We put $(u,v)=\mathrm{Log}(x,y)$ to be coordinates of 
$\mathbb{R}^2$ where amoebas live. 
The most important tool to study amoebas 
is the remarkable Ronkin function $N_f(u,v)$, 
which is a convex function over $\mathbb{R}^2$.
It has been verified in \cite{Okounkov-Reshetikhin}
that the limit shape of 
the statistical model with $Q=1$ 
is expressed by the Ronkin function of an amoeba.
We give pieces of evidence for 
the relation in our statistical model. 
For the $SU(N)$ gauge theory, 
we employ the polynomial. 
\begin{eqnarray}
f_{SU(N)}(x,y)
=
\prod_{r=1}^N(x-\beta_r) 
-(R\Lambda)^N(y+y^{-1})\,.
\end{eqnarray}
The curve $f_{SU(N)}(e^{Rz},y)=0$ is 
the hyperelliptic curve 
(\ref{intro:curve}).
We focus our attention on the Ronkin function over  
the $u$-axis.
We will verify the following relation between
the Ronkin function and the minimizer.
\begin{eqnarray}
N_{f_{SU(N)}}(u,v=0)
=N\int_{-\infty}^u 
d\underline{u}\,\,
\rho_{\star}^{SU(N)}(\underline{u})
\hspace{2mm}+\,\mbox{const}.
\end{eqnarray}

Since the minimizer expresses the gradient of 
the limit shape at the main diagonal,
the above relation shows that
the Ronkin function of $f_{SU(N)}$
is identical with
the limit shape of plane partitions
at least over the main diagonal.
In terms of the internal space geometry,
the limit shape corresponds to 
a low-energy solution of the K\"ahler gravity. 
On the other hand,
the Seiberg-Witten curve describes
the non-perturbative vacuum structure of
the $\mathcal{N}=2$ gauge theory.
Therefore the above connection gives 
the full gauge/gravity correspondence.
The low-energy geometry contains large quantum fluctuations
from the local $SU(N)$ geometry.
These quantum fluctuations would correspond to
the instanton correction in the gauge theory.

One of the most interesting phenomena 
in superstring theory is the mirror symmetry
\cite{Hori-Vafa, Mirror}.
The mirror symmetry is a T-duality, 
in which two very different internal spaces
give rise to equivalent superstring theory.
In terms of topological strings,
the mirror symmetry states that
the topological A-model on a Calabi-Yau threefold
is equivalent to the B-model 
on a mirror Calabi-Yau threefold.
For the local $SU(N)$ geometry,
the mirror Calabi-Yau threefold is specified 
by the hyperelliptic curve $f_{SU(N)}(x,y)=0$
\cite{Toric Mirror}.
Therefore, the above relation
is also a manifestation of the mirror symmetry.

Ronkin's functions appear in the dimer problem  
of bipartite lattice graphs \cite{Kasteleyn,Kenyon}. 
A dimer is a dumb-bell shaped molecule that occupies two adjacent 
lattice sites, and the dimer problem is to determine 
the number of ways of covering the lattice with dimers 
so that all sites are occupied and no two dimers overlap. 
When dimers are magnetically charged, 
the free energy per area is given 
by a Ronkin's function at the thermodynamic limit, 
where the coordinates $(u,v)$ represent 
constant magnetic field $B$ 
\cite{Kenyon-Okounkov-Sheffield}.   
The connection with quiver gauge theories 
has been argued in 
\cite{Hanany-Kennaway}.

The parameter $R$ is identified with 
the radius of $S^1$ in the fifth dimension of the gauge theories. 
In the plane partition model, 
if one keeps $R\Lambda$ fixed, 
this parameter is interpreted as the inverse temperature, 
that is, $R=1/T$, where $T$ denotes the temperature.  
Therefore the large radius limit of the gauge theories 
corresponds to the low temperature limit of the statistical models. 
As the temperature approaches to zero, 
the statistical models get to freeze to the ground states. 
Each ground state is a specific plane partition that is 
determined by the $N$ core at the main diagonal.   
Crystal is the complement of the ground state in the octant.   
It has an interpretation as the gravitational quantum foam  
of the local $SU(N)$ geometry \cite{MNNT}. 
The crystal also has an interpretation in amoebas. 
We will show that 
the low temperature limit $R \rightarrow \infty$ corresponds to 
a certain degeneration of the amoeba and the Ronkin function 
known as tropical geometry \cite{Mikhalkin,RGST,Viro}.  
For instance, 
facet of the crystal is described by 
a piecewise linear function which is the degeneration 
of the Ronkin function $N_{f_{SU(N)}}(u,v)$ at the limit 
$R \rightarrow \infty$.

This article is organized as follows. 
After introducing amoebas briefly in section \ref{section:amoebas},  
we associate amoebas with local Calabi-Yau 
geometries that geometric engineer supersymmetric gauge theories 
having eight supercharges, 
and investigate the amoebas and their Ronkin functions. 
In section \ref{section:plane partition model} 
we study the thermodynamic limit of the plane partition model. 
The relation between the limit shapes and the Ronkin functions 
is shown. 
In section \ref{section:tropical geometry} 
we describe degenerations of the amoebas and the Ronkin functions,  
and relate the degenerations with crystals that are realized 
at the low temperature limit or the large radius limit.

\section{Amoebas from local Calabi-Yau geometries}
\label{section:amoebas}

We associate amoebas with local Calabi-Yau geometries 
that geometric engineer \cite{Geometric engineering} 
supersymmetric gauge theories having eight supercharges. 
The amoebas live in $\mathbb{R}^2$. 
We investigate the Ronkin functions for these amoebas. 
The Ronkin function is a convex function that is strictly 
convex over an amoeba and piecewise linear over the amoeba complement. 
We start this section with an introduction to the notion of 
amoebas \cite{GKZ} and related objects.

\subsection{What is Amoeba ?}
\label{subsection:amoeba review}

\begin{figure}[ht]
\begin{minipage}{0.5\linewidth}
\begin{center}
\includegraphics[width=0.9\linewidth]{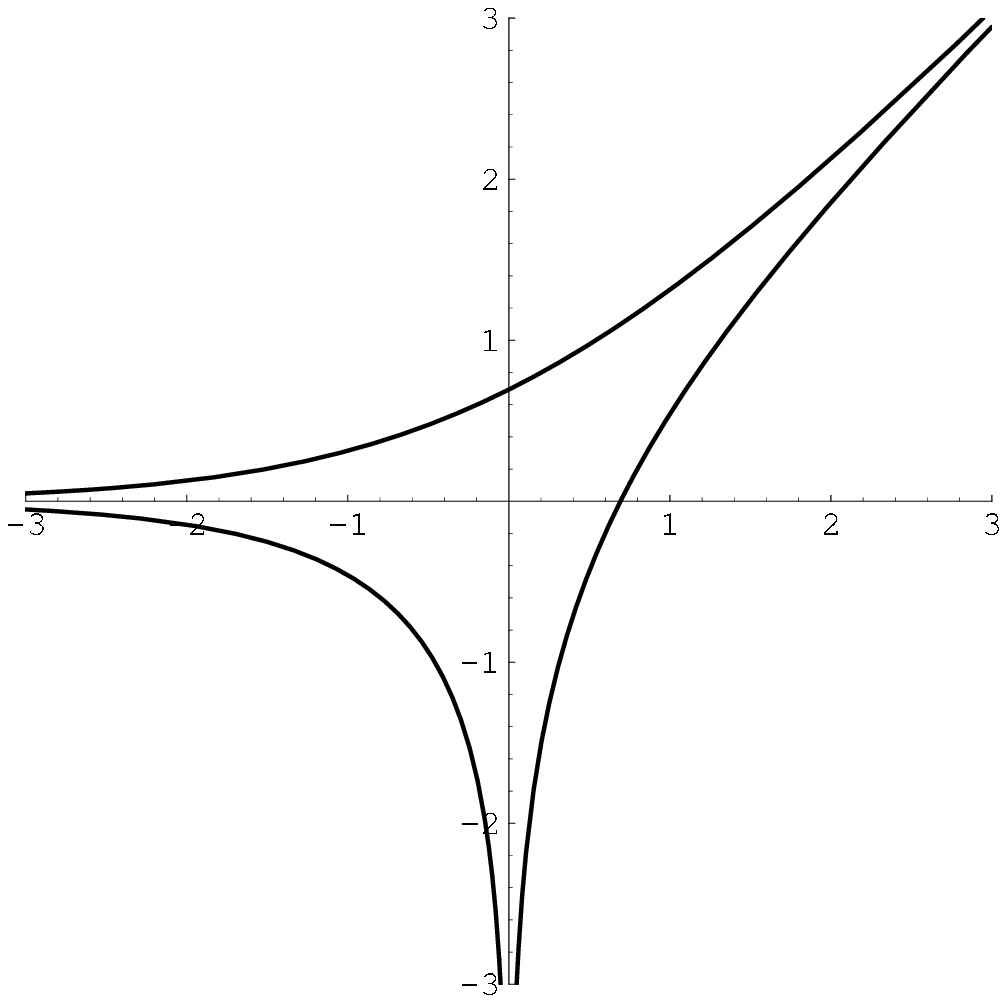}\\
(a)
\end{center}
\end{minipage}
\begin{minipage}{0.5\linewidth}
\begin{center}
\includegraphics[width=0.9\linewidth]{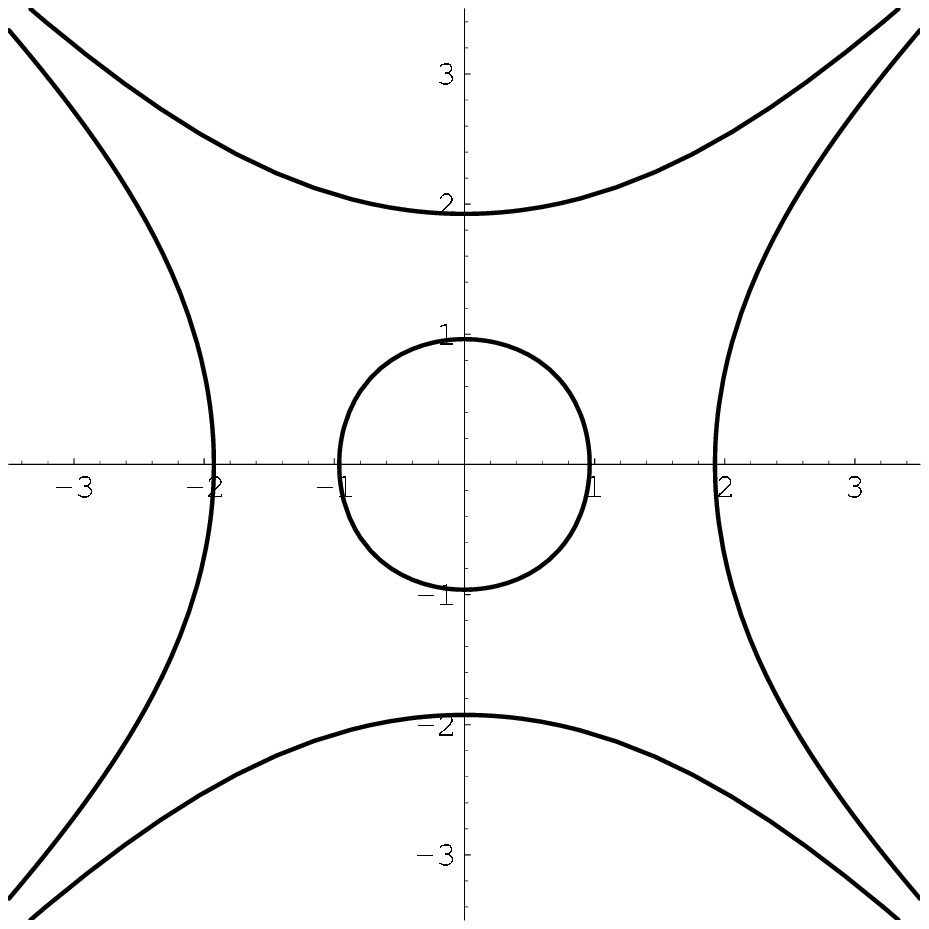}\\
(b)
\end{center}
\end{minipage}
\caption{\it The amoebas of $(a)~1+x+y$ 
and $(b)~5+x+x^{-1}-y-y^{-1}$.}
\label{fig:amoeba}
\end{figure}

Let 
$f(x,y) = \sum_{i,j\in\mathbb{Z}} a_{ij}x^i y^j$ 
be a Laurent polynomial with only a finite number of 
the $a_{ij}$'s being non-zero. 
Let $V_f$ be 
the zero locus of the polynomial in $(\mathbb{C}^*)^2$, 
that is, 
\begin{eqnarray}
V_f=
\Bigl\{ \left. 
(x,y) \in (\mathbb{C}^*)^2 
\,\right|
\,\,f(x,y)=0 
\Bigr\}\,. 
\label{V_f}
\end{eqnarray} 
This $V_f$ is a punctured Riemann surface. 
Define the logarithmic map by 
\begin{eqnarray}
\begin{array}{cclcl}
             \mathrm{Log} & : & (\mathbb{C}^*)^2 
             & \longrightarrow & \mathbb{R}^2 
\\[1mm]
             &   & (x,\,y)            
             & \longmapsto & (\,\log|x|,\, \log|y|\,)\,.
\end{array}
\label{Def of Log map}
\end{eqnarray} 
The amoeba of $f$ is the image of $V_f$. 
\begin{equation}
\mathcal{A}_f = \mathrm{Log}(V_f).
\label{Def of amoeba}
\end{equation}
The amoeba will typically be a subset of $\mathbb{R}^2$
with tentacle-like asymptotes going off to infinity 
and separating the complement 
$^c \mathcal{A}_f = \mathbb{R}^2\backslash \mathcal{A}_f$
into some connected components.
Every connected component of the amoeba complement 
is a convex domain in $\mathbb{R}^2$.
Simple examples of amoebas are shown in 
Figure \ref{fig:amoeba}.

\begin{figure}[ht]
\begin{center}
\includegraphics[scale=0.9]{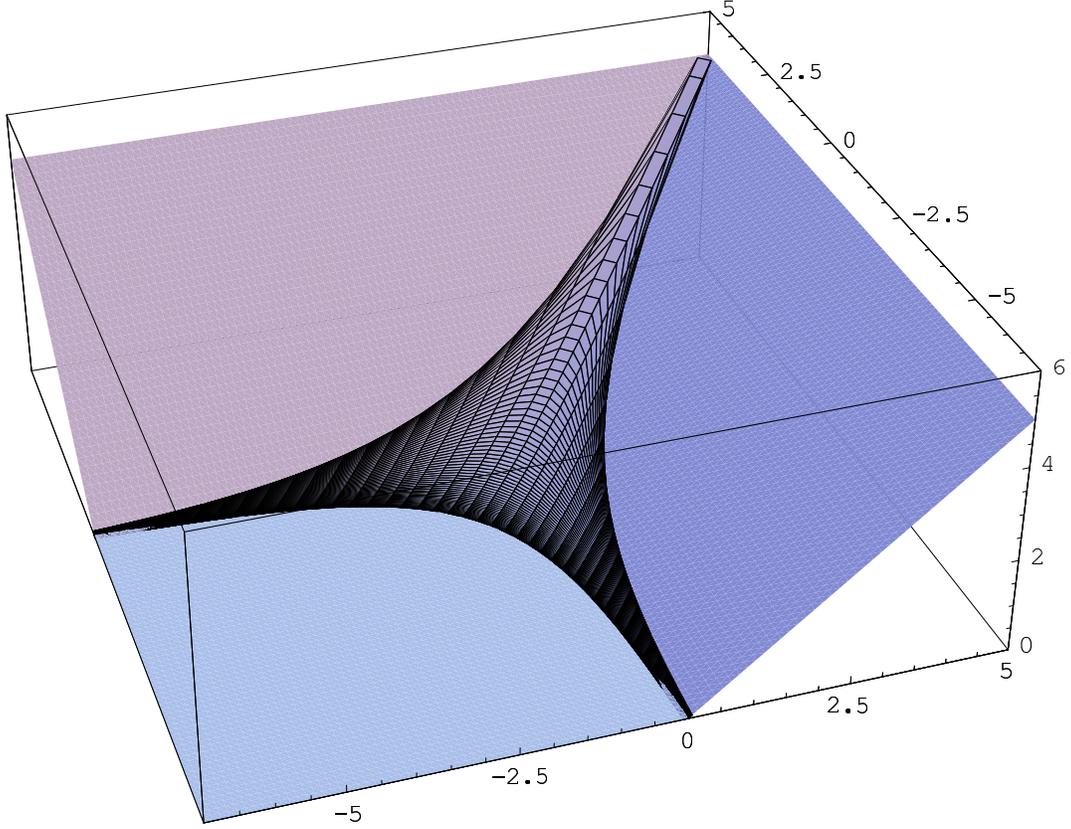}
\end{center}
\caption{\it The Ronkin function of $1+x+y$.}
\label{fig:Ronkin function}
\end{figure}

One of the main tools to study amoebas 
is the Ronkin function \cite{Passare-Rullgard}. 
We now take $(u,v)$ to be Log$(x,y)$,
coordinates on $\mathbb{R}^2$
where amoebas live.
The Ronkin function is defined by
the integral
\begin{equation}
N_f(u,v) =
\frac{1}{(2\pi i)^2}
\int_{\scriptstyle |x| = e^u \atop \scriptstyle|y| = e^v}
\frac{dx}{x} \frac{dx}{y}
\,\log|f(x,y)|\,\,.
\label{Def of Ronkin function}
\end{equation}
The above integrations are over 
the preimage of a point $(u,v)\in \mathbb{R}^2$ 
under the logarithmic map (\ref{Def of Log map}).
It is clear that the integrand is singular 
in the amoeba.
But the Ronkin function  takes real finite values
even over there.
In fact, 
$N_f$ is a convex function which is strictly convex
over $\mathcal{A}_f$
and linear over each connected component of $^c\mathcal{A}_f$.
This particularly means that 
the gradient $\nabla N_f$ has a definite value 
over each connected component of $^c\mathcal{A}_f$. 
As an example, 
the Ronkin function of $1+x+y$ is plotted in 
Figure \ref{fig:Ronkin function}. 

\begin{figure}[ht]
\begin{center}
\includegraphics[scale=0.8]{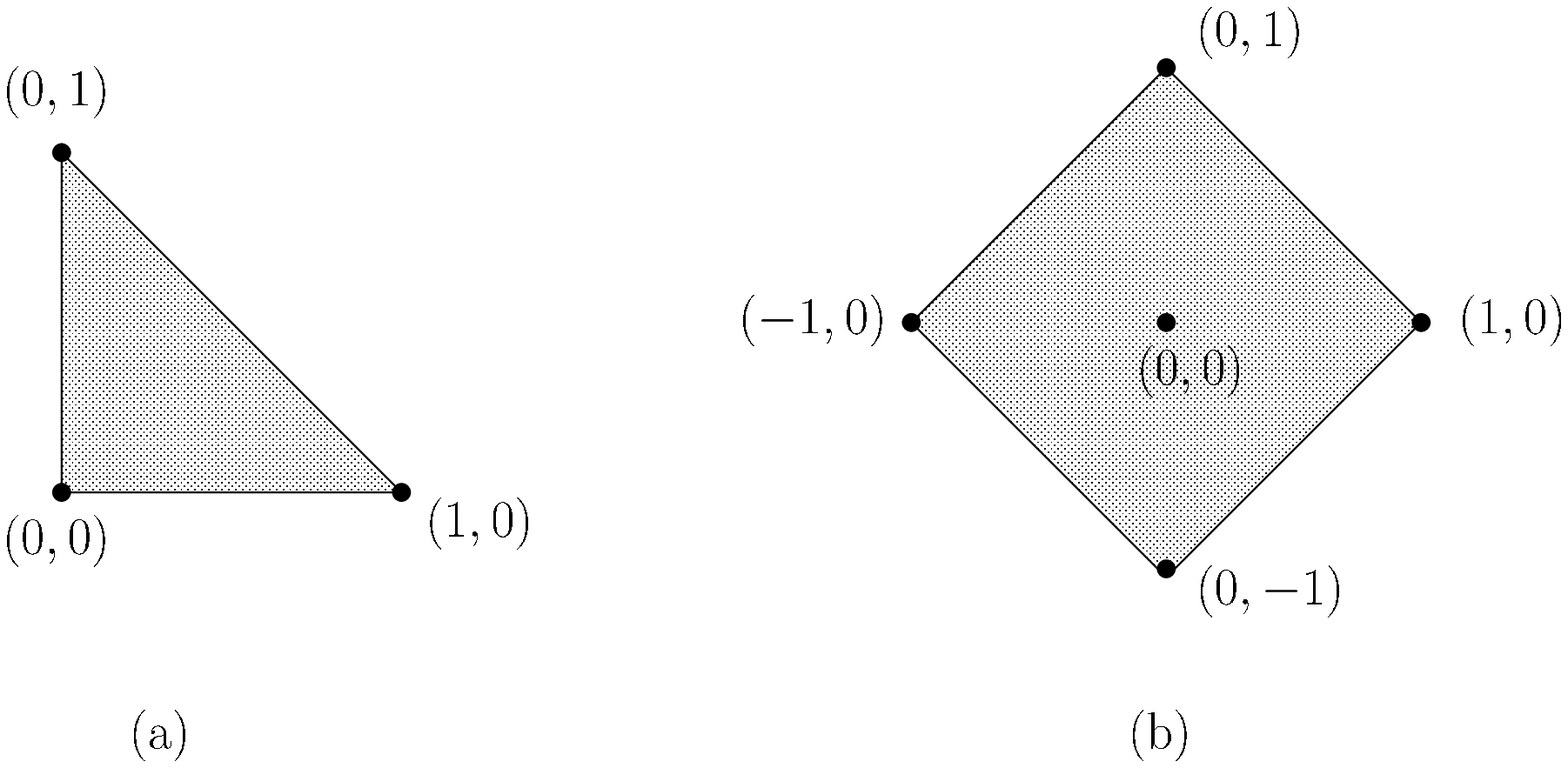}
\end{center}
\caption{\it 
The Newton polygons of $(a)~1+x+y$ and 
$(b)~5+ x +x^{-1}- y-y^{-1}$.}
\label{fig:Newton polygon}
\end{figure}

The amoeba and the Ronkin function reflect 
structure of the Newton polygon of the Laurent polynomial. 
The Newton polygon of $f$ is
the minimal convex polygon which contains
all points $(i,j)\in\mathbb{Z}^2$
corresponding to monomials in the polynomial.
\begin{equation}
\Delta_f = 
\mathrm{convex\ hull\ of}\,\,
\bigl\{(i,j)\in\mathbb{Z}^2 \,|\, 
a_{i,j} \neq 0  \bigr\}.
\label{Newton polygon of f}
\end{equation} 
The Newton polygons of $1+x+y$ and $5+x+x^{-1}-y-y^{-1}$   
can be found in Figure \ref{fig:Newton polygon}. 
\begin{figure}[ht]
\begin{center}
\includegraphics[width=0.9\linewidth]{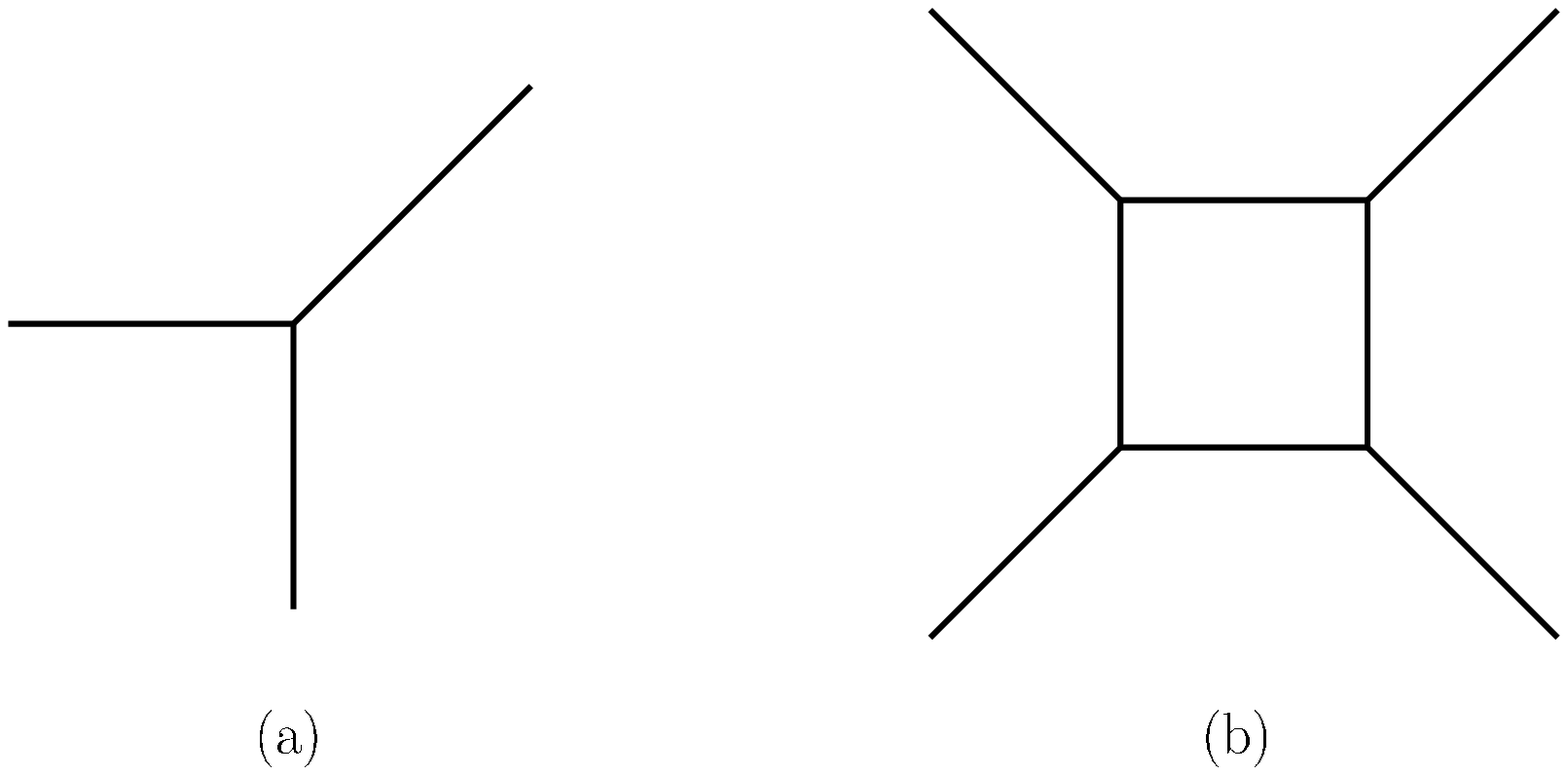}
\end{center}
\caption{\it The spines of the amoebas in Figure \ref{fig:amoeba}}
\label{fig:spine}
\end{figure}

Over each connected component $E$ of 
the amoeba complement $^c\mathcal{A}_f$, 
the gradient $\nabla N_f$ takes value in 
a integer lattice point of the Newton polygon 
\cite{Passare-Rullgard}. 
\begin{equation}
\bigl. \nabla N_f\, \bigr|_{E} \, 
\in \Delta_f \cap \mathbb{Z}^2\,.
\label{Ronkin to Newton}
\end{equation}
This gives an upper bound to the number of 
connected components of $^c\mathcal{A}_f$.
The connected components are always less than 
the lattice points of the Newton polygon.
The symbol $E_{(i,j)}$ will be used to 
denote the connected component 
over which the gradient $\nabla N_f$ equals to $(i,j)$.
For almost all the polynomials considered in this paper, 
the upper bound is attained, that is,
the amoeba complement has the same number of 
the connected components as the lattice points of    
the Newton polygon. 
For such a polynomial $f$, we have 
\begin{eqnarray} 
^c\mathcal{A}_f=
\bigsqcup_{(i,j) \in \Delta_f \cap \mathbb{Z}^2}
E_{(i,j)}\,. 
\label{A_c=sum of E}
\end{eqnarray}

The Ronkin function is piecewise linear over the amoeba complement 
$^c\mathcal{A}_f$. 
For each connected component $E$ of $^c\mathcal{A}_f$, 
let $N_{E}$ be the linear extension of $N_f |_{E}$ to $\mathbb{R}^2$.
When $E=E_{(i,j)}$, 
we have 
\begin{eqnarray}
N_{E_{(i,j)}}(u,v)=
c_{(i,j)}+\langle (u,v), (i,j) \rangle\,, 
\label{N_E_(i,j)}
\end{eqnarray}
where the symbol  $\langle \,,\,\rangle$  
denotes the standard inner product on $\mathbb{R}^2$, 
that is, 
$\langle (u,v), (i,j) \rangle=iu+jv$. 
If the above $(i,j)$ is a vertex of $\Delta_{f}$, 
the constant $c_{(i,j)}$ is computed and 
simply becomes $c_{(i,j)}=\Re (\ln a_{i,j})$. 
According to \cite{Passare-Rullgard}, 
we introduce 
a piecewise linear function $S_f$ over $\mathbb{R}^2$ by 
\begin{eqnarray}
S_f (u,v)
=
\max_{E}
N_{E}(u,v)\,.
\label{S_f}
\end{eqnarray}
Taking account of the convexity of the Ronkin function,  
it follows 
\begin{eqnarray}
N_f(u,v)\geq S_f(u,v)
\label{N geq S}
\end{eqnarray}
for any $(u,v) \in \mathbb{R}^2$. 
The equality holds over the amoeba complement $^c\mathcal{A}_f$. 
So this $S_f$ is a piecewise linear function  
which approximates the Ronkin function 
by putting sharp corners in it.

The piecewise linear function $S_f$ allows us 
to define the spine of the amoeba $\mathcal{A}_f$. 
The spine $\mathcal{S}$ 
is the corner set of $S_f$, that is, 
the set of $(u,v)$ where $S_f(u,v)$ is not smooth. 
The spines of the amoebas in Figure \ref{fig:amoeba} 
are drawn in Figure \ref{fig:spine}.   
Note that $\mathcal{S}\subset\mathcal{A}_f$. 
We can regard the spine $\mathcal{S}$ 
as the 1-skeleton of the convex which is 
dual to a certain triangular subdivision 
of the Newton polygon $\Delta_f$.

\subsection{Amoeba and the Ronkin function for $U(1)$ theory}

Geometric engineering  \cite{Geometric engineering} 
dictates that local geometries realize supersymmetric gauge theories. 
The topological vertex countings 
\cite{topological vertex 1,topological vertex 2} 
of the topological A-model string partition functions 
on local geometries, 
which involve from the worldsheet viewpoint 
sums over holomorphic maps to the target spaces, 
support \cite{Iqbal,Eguchi} the idea. 
The local geometries are noncompact toric Calabi-Yau threefolds. 
If a space is toric, 
many basic and essential characteristics of 
the space are neatly coded and easily 
deciphered from analysis of the corresponding lattices. 
As for the local geometries, 
the data of the space are retrieved from 
a two-dimensional polygon. 
We will regard the polygon as a Newton polygon. 
Thereby we will define the polynomial 
for each local geometry.

\begin{figure}[ht]
\begin{center}
\includegraphics[scale=0.8]{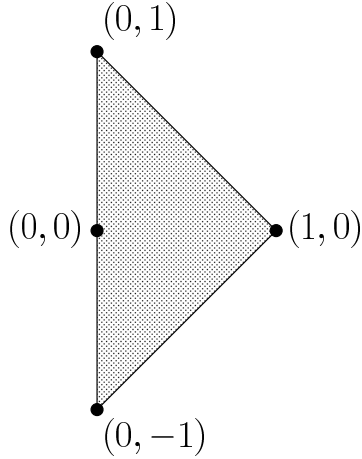}
\end{center}
\caption{\it The Newton polygon for U(1) theory.}
\label{fig:polygon of U(1)}
\end{figure}

We first examine an amoeba for the local geometry 
relevant to the abelian gauge theory. 
In terms of topological strings,
this arises from the topological A-model string 
on $\mathcal{O}\oplus\mathcal{O}(-2)\to\mathbb{P}^1$. 
The toric polygon of this geometry is drawn 
in Figure \ref{fig:polygon of U(1)}. 

We employ the following Laurent polynomial.
\begin{eqnarray}
f_{U(1)}(x,y)
=
x-\beta - R\Lambda \bigl( y + y^{-1} \bigr). 
\label{polynomial of U(1)}
\end{eqnarray}
The parameters $\beta$, $R$ and $\Lambda$ in the above  
are assumed to be real positive numbers. 
In particular, 
we consider the case of $\beta > 2R\Lambda$. 
In the field theory language, 
$R$ is the radius of $S^1$ in the fifth dimension 
and $\Lambda$ denotes the scale parameter of 
the underlying four-dimensional theory.
It is clear that the Newton polygon of this $f_{U(1)}$
is the polygon in Figure \ref{fig:polygon of U(1)}.

We will rescale amoebas by $1/R$. 
This is achieved  
by using the following rescaled coordinates $(u,v)$ 
rather than the original ones. 
\begin{eqnarray}
(u,v) = \frac{1}{R}\,\mathrm{Log}(x,y)=
\bigl(\,\frac{1}{R}\log|x|,\, \frac{1}{R}\log|y|\,\bigr).
\label{rescaled (u,v)}
\end{eqnarray}
To accord with this, 
we also rescale the Ronkin functions by $1/R$. 
In the rest of this paper, 
we describe amoebas by using the above coordinates, 
and call the rescaled Ronkin functions 
simply as the Ronkin functions.

\begin{figure}[t]
\begin{center}
\includegraphics[width=0.5\linewidth]{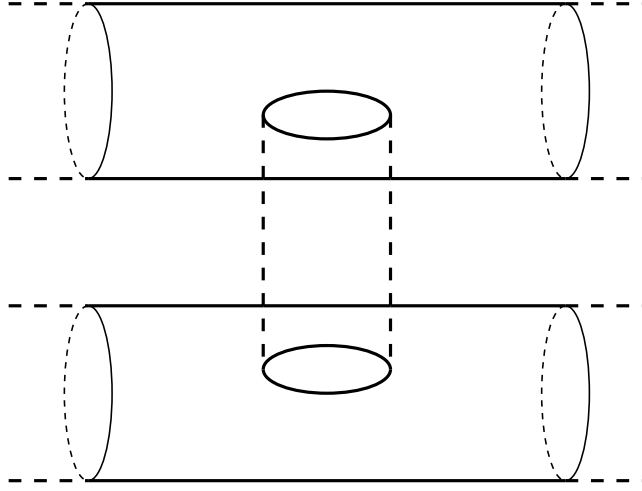}
\end{center}
\caption{\it The zero locus $V_{f_{U(1)}}$.}
\label{fig:V_fU(1)}
\end{figure}
\begin{figure}[t]
\begin{center}
\includegraphics[scale=1]{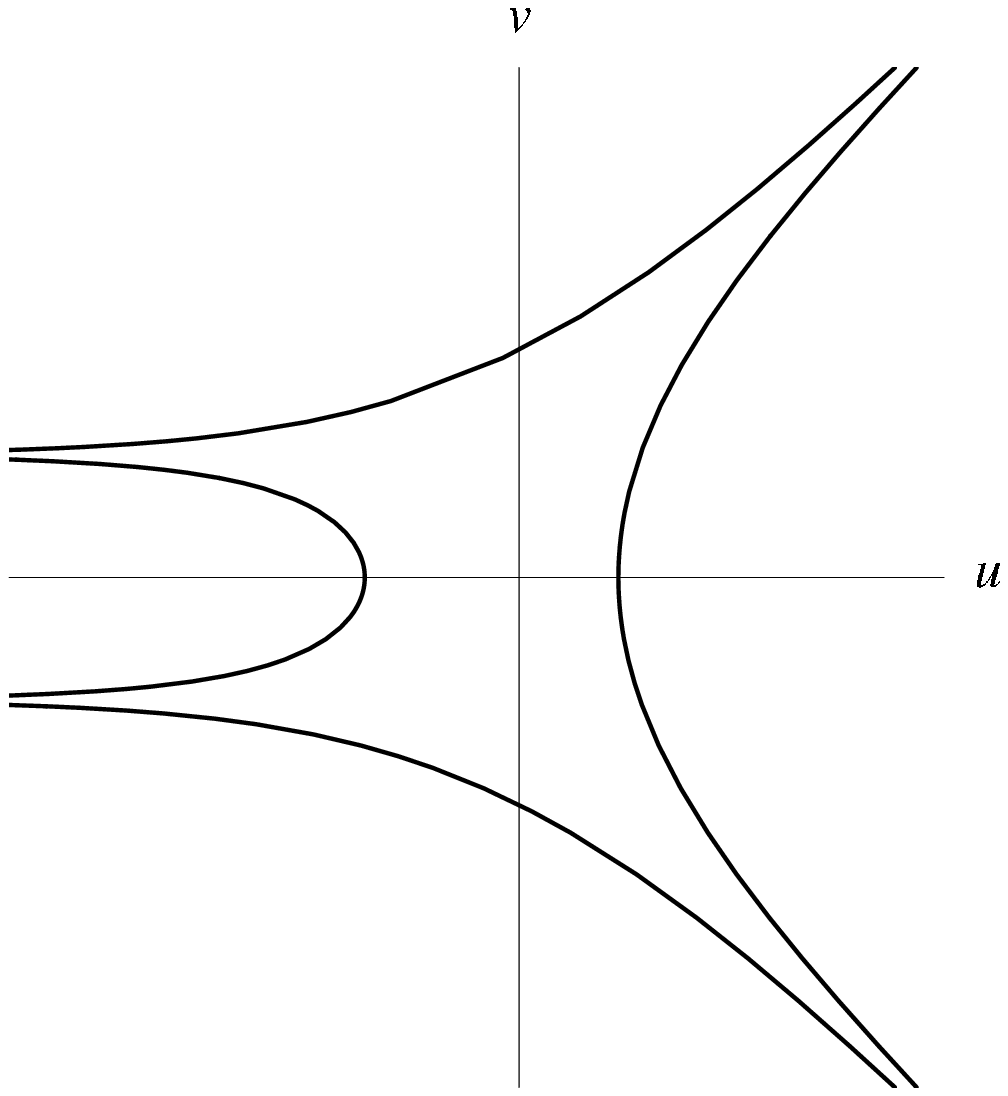}
\end{center}
\caption{\it The amoeba of $f_{U(1)}$.}
\label{fig:amoeba of U(1)}
\end{figure}

Let us now look at the amoeba of $f_{U(1)}$. 
The zero locus $V_{f_{U(1)}}$ in $(\mathbb{C}^*)^2$ 
is a four-punctured sphere. 
If one regards $\mathbb{C}^*$ an infinite cylinder, 
it is a double covering of the cylinder 
(Figure \ref{fig:V_fU(1)}). 
The amoeba $\mathcal{A}_{f_{U(1)}}$ 
is the image $\mathrm{Log}(V_{f_{U(1)}})$.  
Thanks to the condition 
$\beta > 2R\Lambda$, 
this becomes a subset of $\mathbb{R}^2$ 
described by the following inequalities. 
\begin{eqnarray}
| \beta - 2R\Lambda \cosh Rv| 
\le e^{Ru} \le \beta + 2 R\Lambda \cosh Rv\,.
\end{eqnarray} 
The amoeba is depicted in Figure \ref{fig:amoeba of U(1)}. 
The amoeba spreads four tentacles.  
These tentacles asymptote to 
the following straight lines and extend to the infinities. 
\begin{eqnarray}
v=\pm u\,, 
\hspace{10mm}
v=\pm \frac{1}{R}\log 
\left\{\frac{\beta}{2R\Lambda}
+\sqrt{\bigl(\frac{\beta}{2R\Lambda}\bigr)^2-1} 
\right\}\,.
\label{tentackles of A_U(1)}
\end{eqnarray}
They separate the amoeba complement 
$^c\mathcal{A}_{f_{U(1)}}$ 
into four connected components. 
The number of connected components agrees with 
the number of integer lattice points 
of the Newton polygon.

Taking account of the above rescaling, 
the Ronkin function $N_{f_{U(1)}}$ is given by 
\begin{eqnarray}
N_{f_{U(1)}}(u,v)
= 
\frac{1}{R}\,
\frac{1}{(2\pi i)^2}
\int_{\scriptstyle |x| = e^{Ru} \atop \scriptstyle |y| = e^{Rv}}
\frac{dx}{x}\frac{dy}{y}\, 
\log \left| f_{U(1)}(x,y) \right|\,.
\label{Ronkin function of U(1)}
\end{eqnarray} 
Regardless of the simple appearance, 
it is serious to carry out the above integrations. 
However, 
the gradient $\nabla N_{f_{U(1)}}$ 
becomes tractable since we have 
\begin{eqnarray}
\frac{\partial N_{f_{U(1)}}}{\partial u}
&=& 
\frac{1}{(2\pi i)^2}
\int_{\scriptstyle |y| = e^{Rv}}\frac{dy}{y}
\int_{\scriptstyle |x| = e^{Ru}}
\frac{dx}{x-\beta-R\Lambda(y+y^{-1})}\,,  
\label{dN_U(1)/du} 
\\[2mm]
\frac{\partial N_{f_{U(1)}}}{\partial v}
&=& 
\frac{1}{(2\pi i)^2}
\int_{\scriptstyle |x| = e^{Ru}}\frac{dx}{x}
\int_{\scriptstyle |y| = e^{Rv}}
\frac{d(y+y^{-1})}{R\Lambda(y+y^{-1})-x+\beta}\,.  
\label{dN_U(1)/dv}
\end{eqnarray}

In the double integral (\ref{dN_U(1)/du}), 
we can find the following residue integral. 
\begin{eqnarray}
\frac{1}{2\pi i}\int_{|x|=e^{Ru}}
\frac{dx}{x-\beta-R\Lambda(y+y^{-1})}
=\left\{ 
\begin{array}{cc} 
1 & \hspace{5mm}\mbox{if}~~~|\beta+R\Lambda(y+y^{-1})| \leq e^{Ru}\,, 
\\[3mm]
0 & \mbox{otherwise}. 
\end{array}
\right.
\label{U(1) truth function} 
\end{eqnarray}
The integral takes value $1$ or $0$. 
If one puts $|y|=e^{Rv}$, the value depends on 
whether $\theta_y=\arg y$ satisfies a certain condition or not, 
as seen from (\ref{U(1) truth function}).     
This means that the residue integral 
is a truth function of the condition on $\theta_y$.   
The $y$-integration in (\ref{dN_U(1)/du}) gives 
a simple integration over $\theta_y$. 
We thus interpret $\partial_uN_{f_{U(1)}}$ as an integration 
of the truth function over $\theta_y$.   
The similar interpretation is also possible 
for $\partial_vN_{f_{U(1)}}$ by using (\ref{dN_U(1)/dv}).

Over each connected component of the amoeba complement, 
the gradient $\nabla N_{f_{U(1)}}$ takes value in 
an integer lattice point of the Newton polygon. 
It may be helpful to see how this happens by using 
the above interpretation. 
As an example, 
let us consider the connected component $E$ 
that is described by the inequality 
$e^{Ru} > \beta+2R\Lambda \cosh Rv$. 
We have 
$|x|>|\beta +R\Lambda(y+y^{-1})|$ 
for $\forall\, (x,y) \in \mathrm{Log}^{-1}(E)$. 
This means that the residue integral (\ref{U(1) truth function}) 
equals to $1$ and 
we obtain $\partial_uN_{f_{U(1)}}=1$ 
on this connected component.  
Computation of $\partial_vN_{f_{U(1)}}$ goes as well. 
Thereby we obtain $\nabla N_{f_{U(1)}}=(1,0)$. 
The connected component $E$ is $E_{(1,0)}$. 
The other three connected components 
are $E_{(0,0)}$ and $E_{(0,\pm1)}$, 
where the gradient $\nabla N_{f_{U(1)}}$ takes 
$\alpha$ on each $E_{\alpha}$. 
Over the amoeba complement 
the Ronkin function becomes the following piecewise linear function.  
\begin{eqnarray}
\Bigl. 
N_{f_{U(1)}}(u,v)
\Bigr|_{^c\mathcal{A}_{f_{U(1)}}} 
=
\max_{\alpha=(1,0),(0,\pm1),(0,0)}
\Bigl(c_{\alpha}+\langle (u,v), \alpha \rangle \Bigr)\,, 
\label{N_U(1) on complememt} 
\end{eqnarray}
where the constants $c_{\alpha}$ are 
\begin{eqnarray}
&& 
c_{(1,0)}=0\,, 
\hspace{10mm}
c_{(0,\pm 1)}=\frac{1}{R}\log R\Lambda\,, 
\nonumber \\
&&
c_{(0,0)}=\frac{1}{R}\log 
\Bigl\{
R\Lambda 
\Bigl( 
     \frac{\beta}{R\Lambda}
    +\sqrt{\bigl(\frac{\beta}{2R\Lambda}\bigr)^2-1} 
\Bigr)
\Bigr\}\,. 
\label{c_alpha U(1)}
\end{eqnarray}

We examine the Ronkin function over the amoeba. 
In particular, 
we focus on the section of the amoeba along the $u$-axis. 
In the next section we will see that 
the Ronkin functions over the $u$-axis 
relate with instanton countings of gauge theories. 
The section of the amoeba becomes a segment 
$I= [\frac{1}{R}\log\beta^-,\frac{1}{R}\log\beta^+]$, 
where we put $\beta^{\pm}=\beta \pm 2R\Lambda$.
Thanks to the condition $\beta > 2R \Lambda$, 
this segment has a finite length. 
It is convenient to call the segment the band
and the complement of it the gap.
\begin{eqnarray}
\mbox{Band}
\ &:&\ 
I=
\left[
\frac{1}{R} \log\beta^-,\frac{1}{R} \log\beta^+
\right]\,.
\\
\mbox{Gap}
\ &:&\ 
\mathbb{R}\setminus
I\,.
\end{eqnarray}

\begin{figure}[ht]
\begin{center}
\includegraphics[width=0.8\linewidth]{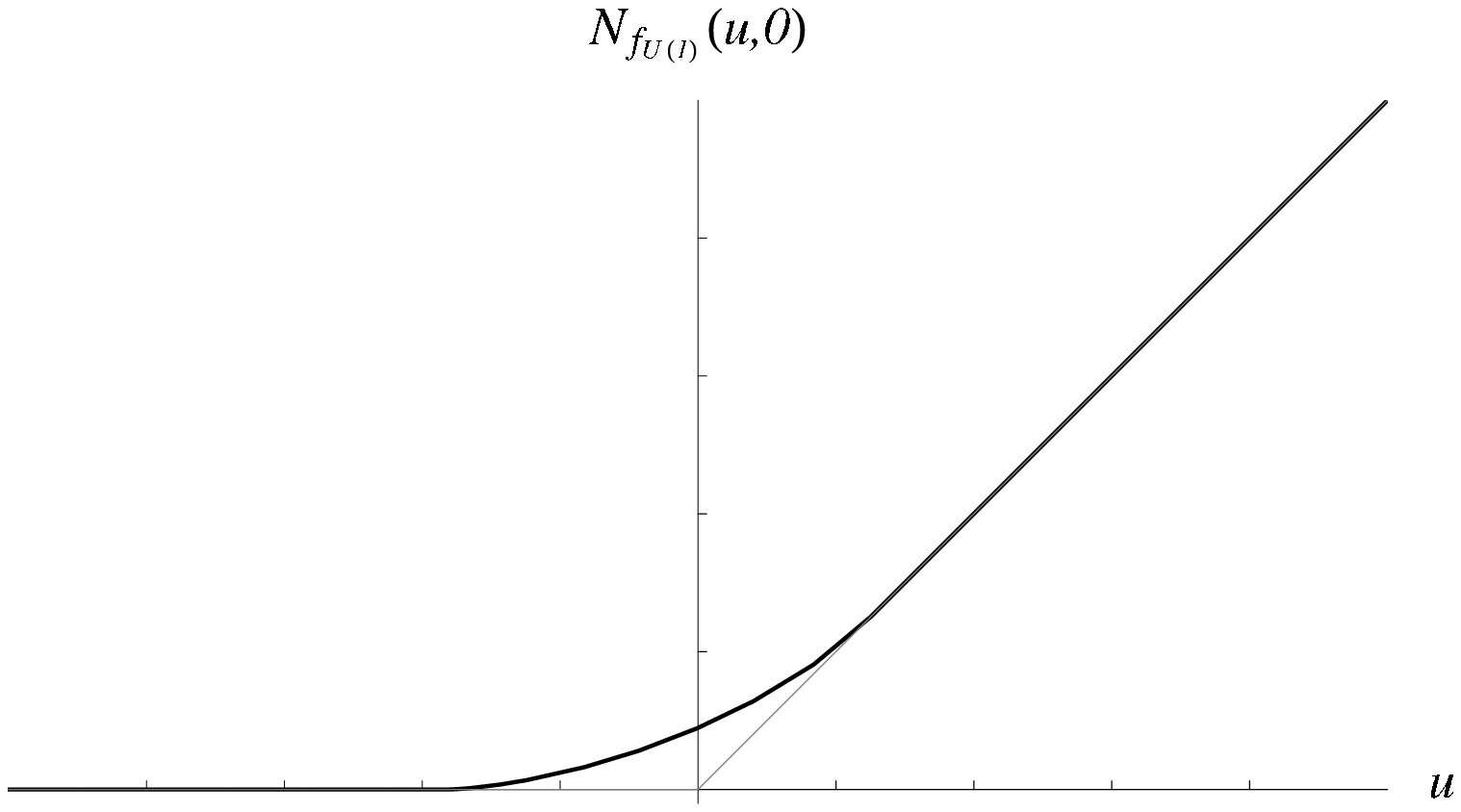}
\end{center}
\caption{\it The Ronkin function of $f_{U(1)}$ over the $u$-axis.}
\label{fig:Ronkin function of U(1) at v=0}
\end{figure}

\begin{figure}[ht]
\begin{center}
\includegraphics[width=0.9\linewidth]{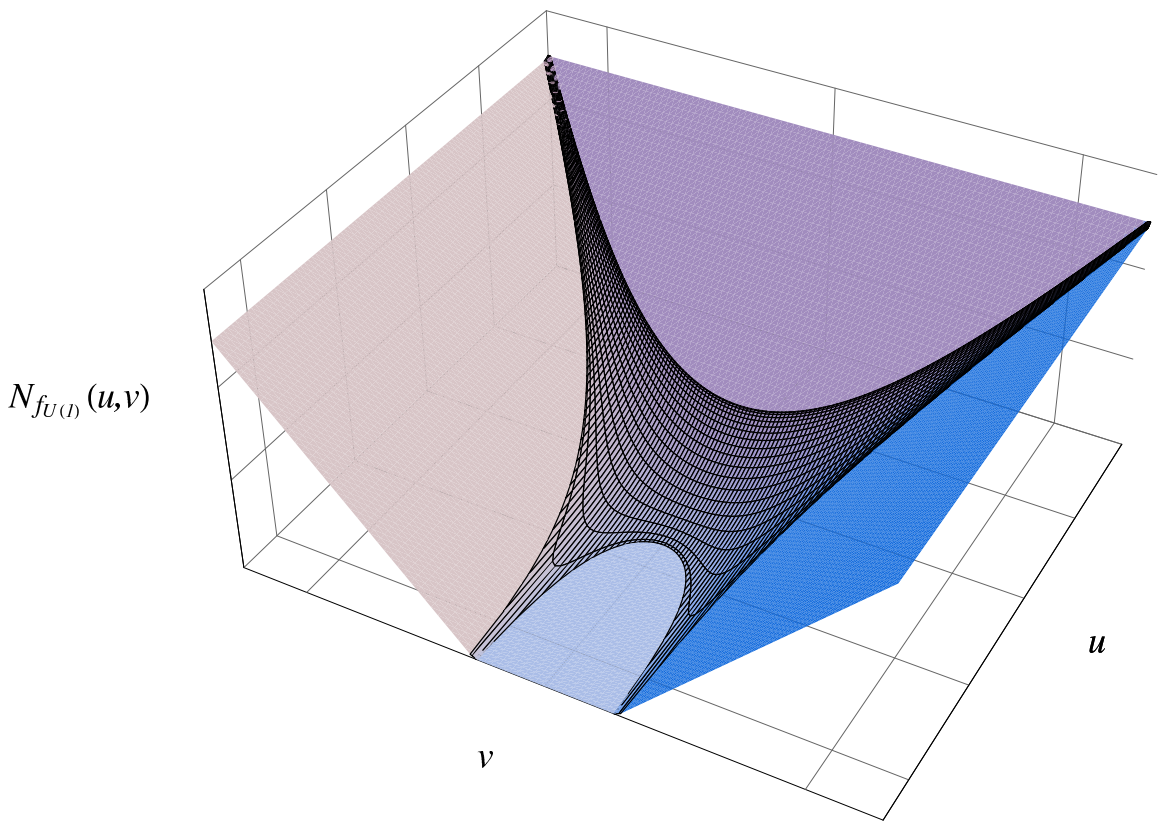}
\end{center}
\caption{\it The Ronkin function of $f_{U(1)}$.}
\label{fig:Ronkin function of U(1)}
\end{figure}

The residue integral (\ref{U(1) truth function}) 
over the gap takes the definite values irrespective of $\theta_y$. 
It always gives $0$ for $u<\frac{1}{R}\log \beta^-$  
and $1$ for $u>\frac{1}{R}\log \beta^+$.  
On the other hand, 
when $u$ is in the band,  
the residue integral takes value $1$ only if 
$\theta_y$ satisfies the inequality. 
\begin{eqnarray}
\cos\theta_y \le \frac{e^{Ru}-\beta}{2R\Lambda}\,.
\label{condition U(1)}
\end{eqnarray} 
Otherwise it gives $0$. 
Therefore we obtain 
\begin{eqnarray}
\pgrad{N_{f_{U(1)}}}{u}\bigg|_{I}
&=&
\frac{1}{2\pi} \int_{\cos \theta_y 
\le \frac{e^{Ru}-\beta}{2R\Lambda}} 
d\theta_y \cdot 1
\notag
\\
&=&
1 + \
\frac{1}{\pi}
\arccos\left(\frac{e^{Ru}-\beta}{2R\Lambda}\right)\,,   
\label{gradient on amoeba in U(1)}
\end{eqnarray} 
where the branch of the arccosine is fixed by choosing 
$\arccos(0) = - \pi / 2$. 
To summarize, 
the gradient of the Ronkin function along the $u$-axis 
becomes 
\begin{eqnarray}
\pgrad{N_{f_{U(1)}}}{u}\bigg|_{v=0}
=
\left\{
\begin{array}{lc}
0\,,
&
~~u < \frac{1}{R}\log\beta^-
\\[2mm]
1 + 
\frac{1}{\pi}
\arccos\left(\frac{e^{Ru}-\beta}{2R\Lambda}\right)\,,
&
~~u \in I
\\[2mm]
1\,,
&
~~\frac{1}{R}\log\beta^+ < u\,.
\end{array}
\right. 
\label{gradient on u-axis in U(1)}
\end{eqnarray} 
By using this, we plot 
the Ronkin function over the $u$-axis 
in Figure \ref{fig:Ronkin function of U(1) at v=0}.
For this particular case, 
our computation is not limited on the $u$-axis 
and can be done over $\mathbb{R}^2$.  
We plot the Ronkin function over $\mathbb{R}^2$ 
in Figure \ref{fig:Ronkin function of U(1)}. 

The expression (\ref{gradient on u-axis in U(1)}) 
is arranged by using the terminology of complex geometry. 
The zero locus $V_{f_{U(1)}}$ 
is a complex curve in $(\mathbb{C}^*)^2$. 
It can be written as 
\begin{eqnarray}
y+y^{-1}=\frac{1}{R\Lambda}
\bigl(e^{Rz}-\beta \bigr)\,, 
\label{f_U(1)=0}
\end{eqnarray}
where $z=u+i\theta/R$ 
is a cylindrical coordinate of $\mathbb{C}^*$. 
The curve is a double cover of the $\mathbb{C}^*$. 
The branch points locate 
at two ends of the band $I$. 
The holomorphic function $y$ has a cut along the band 
on the Riemann sheet and takes values at the unit circle there. 
In particular, it is $\pm 1$ at the branch points. 
Using $y$, we can rewrite (\ref{gradient on u-axis in U(1)})
as follows. 
\begin{eqnarray}
\pgrad{N_{f_{U(1)}}}{u}\bigg|_{v=0}
=
\Re \Bigl(
1 + \frac{1}{\pi i}\log y(u-i0)
\Bigr),
\label{d_uN_fU(1)}
\end{eqnarray}
where we choose $\arg y(u-i0)$ so that 
it increases along the $u$-axis from $-\pi$ to $0$.

\subsection{Amoeba and the Ronkin function for $SU(N)$ theory}

\begin{figure}[ht]
\begin{center}
\includegraphics[scale=0.8]{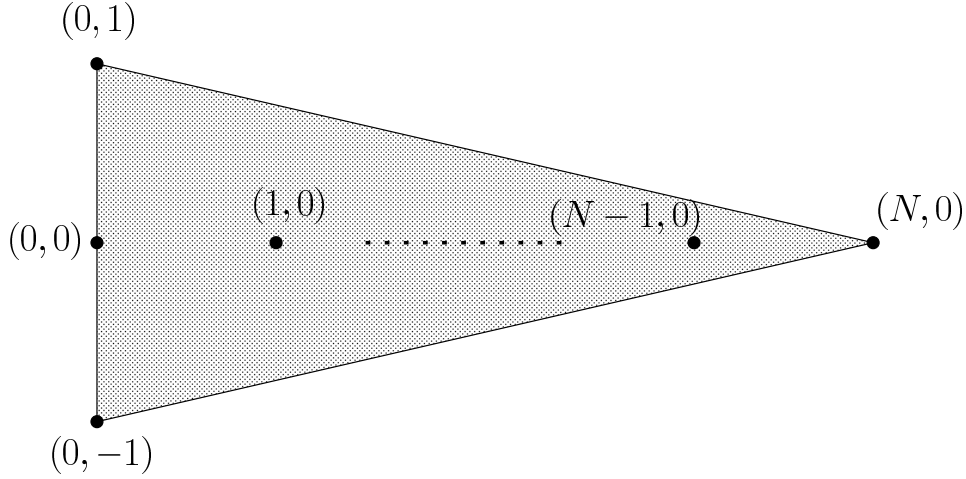}
\end{center}
\caption{\it The Newton polygon for $SU(N)$ theory.}
\label{fig:polygon of SU(N)}
\end{figure}

Five-dimensional $\mathcal{N}=1$ supersymmetric $SU(N)$ Yang-Mills theory 
is realized by a local $SU(N)$ geometry. 
The geometry is an ALE space with $A_{N-1}$ singularity 
nontrivially fibered over $\mathbb{P}^1$. 
Fibration of the space reflects 
the Chern-Simons term of this five-dimensional theory. 
The polygon of this geometry is set out
in Figure \ref{fig:polygon of SU(N)}.

By understanding the above polygon as the Newton polygon, 
we employ the following Laurent polynomial. 
\begin{eqnarray}
f_{SU(N)}(x,y)
=
Q_N(x) 
- (R\Lambda)^N \bigl( y + y^{-1} \bigr)\,, 
\label{f_SU(N)}
\end{eqnarray}
where $Q_N(x)$ denotes a monic polynomial of degree $N$
with real coefficients. 
Let $\beta_1,\ldots, \beta_N$ be roots of $Q_N(x)$. 
\begin{eqnarray}
Q_N(x) &=& x^N + b_1 x^{N-1} + \cdots + b_N \\*
&=& \prod_{r=1}^{N}( x - \beta_r)\,. 
\label{Q_N(x)}
\end{eqnarray}
All the roots are assumed to be real positive numbers. 
They are arranged in numerical order 
$0 < \beta_1 < \cdots < \beta_N$.  
We also make $\Lambda$ very small so that 
$Q_N(x)\pm 2(R\Lambda)^N$ have 
$N$ distinct real positive roots.

\begin{figure}[ht]
\begin{center}
\includegraphics[width=0.8\linewidth]{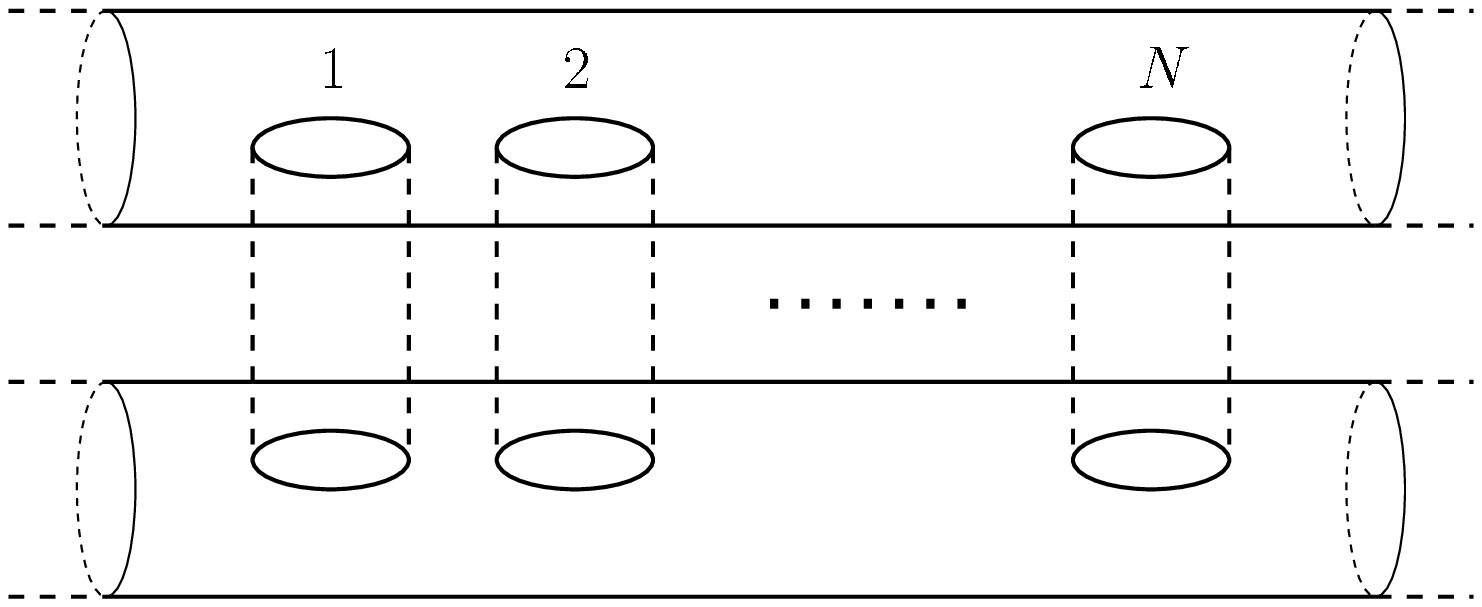}
\end{center}
\caption{\it The zero locus $V_{f_{SU(N)}}$.}
\label{fig:V_fSU(N)}
\end{figure}
\begin{figure}[ht]
\begin{center}
\includegraphics[width=0.4\linewidth]{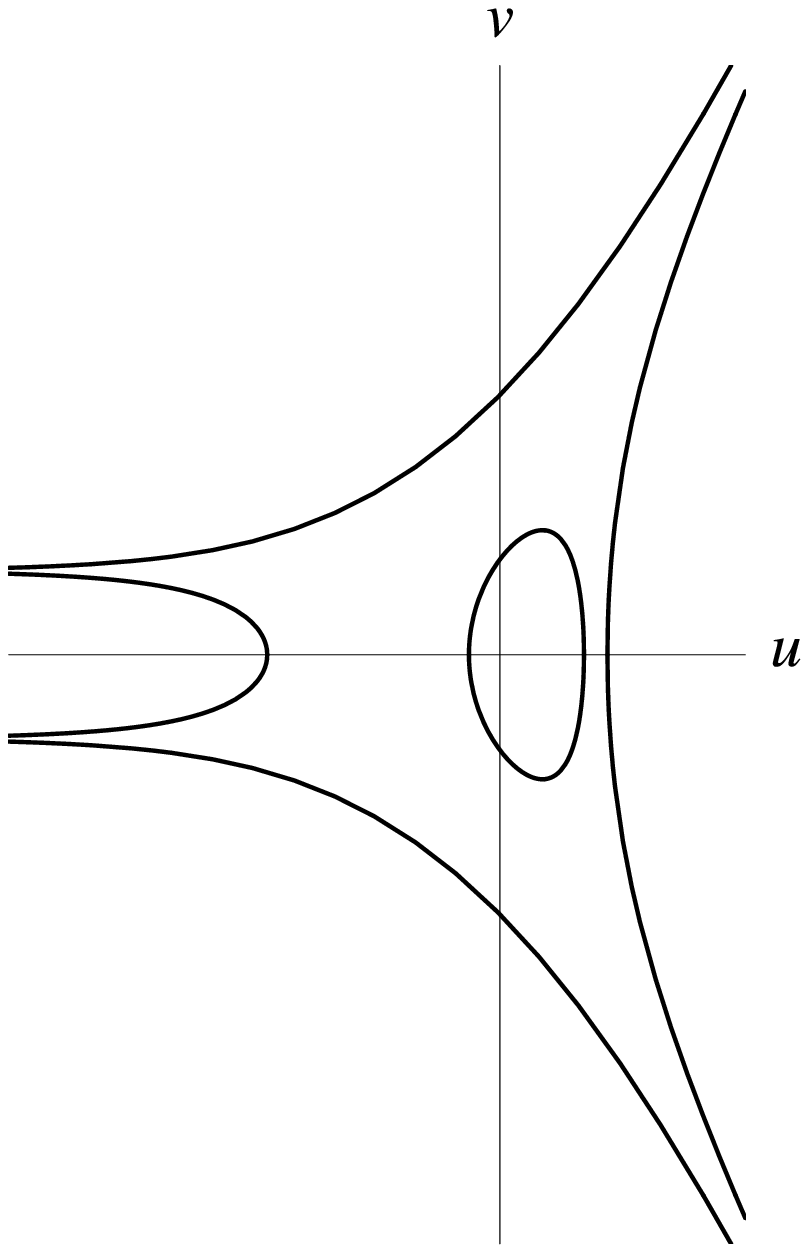}
\end{center}
\caption{\it The amoeba of $f_{SU(2)}$.}
\label{fig:amoeba for SU(N)}
\end{figure}

We observe the amoeba of $f_{SU(N)}$. 
The zero locus $V_{f_{SU(N)}}$ in $(\mathbb{C}^*)^2$ 
is a four-punctured surface. 
It is a hyperelliptic curve of genus $N-1$ 
(Figure \ref{fig:V_fSU(N)}).  
The amoeba $\mathcal{A}_{f_{SU(N)}}$ is 
the image $\mathrm{Log}\bigl(V_{f_{SU(N)}}\bigr)$.  
The hyperelliptic involution $y \leftrightarrow y^{-1}$ 
induces the involution $v \leftrightarrow -v$ on the amoeba. 
Shape of the amoeba depends sharply 
on the parameters $\Lambda$ and $\beta_r$. 
When the parameters are set as above, 
the amoeba spreads its four tentacles and 
these asymptotes are 
\begin{eqnarray}
v=\pm Nu\,, 
\hspace{10mm}
v=\pm \delta_N\,, 
\label{SU(N) tentackles}
\end{eqnarray}
where $\delta_N$ is a positive constant given by 
\begin{eqnarray}
\delta_N 
=\frac{1}{R}
\ln 
\left\{ 
\frac{|b_N|}{2(R\Lambda)^N}
+\sqrt{ \Bigl(\frac{|b_N|}{2(R\Lambda)^N}\Bigr)^2-1} 
\right\}\,. 
\label{delta_N}
\end{eqnarray}
The amoeba complement $^c\mathcal{A}_{f_{SU(N)}}$ 
has $N+3$ connected components 
that correspond exactly 
to integer lattice points of  
the Newton polygon $\Delta_{f_{SU(N)}}$. 
Four of them are unbounded components 
lying between the tentacles. 
All the other components are bounded components 
enclosed by the amoeba. 
That is to say, 
there are $N-1$ holes in the amoeba. 
An example of the amoeba is drawn in 
Figure \ref{fig:amoeba for SU(N)}. 

Consider the section of the amoeba along the $u$-axis. 
Let us write 
\begin{eqnarray}
Q_N(x)-2(R\Lambda)^N\cos \theta_y
=\prod_{r=1}^N 
\bigl(x-\beta_r(\theta_y) \bigr)\,. 
\label{beta_r(theta_y)}
\end{eqnarray} 
All the $\beta_r(\theta_y)$'s become real positive numbers 
since $\Lambda$ is made so small. 
They are distinct and 
are arranged in numerical order 
$0< \beta_1(\theta_y) < \cdots < \beta_N(\theta_y)$. 
Bands of the amoeba,  
which are the longitudinal section of the amoeba along the $u$-axis, 
consist of $N$ segments with finite length. 
\begin{eqnarray}
&&
\mbox{Bands : }
\bigsqcup_{r=1}^N I_r\,,~~~~
I_r
=
\left\{
\frac{1}{R} \log\beta_r(\theta_y) \in \mathbb{R}\ 
:\ 0 \leq \theta_y < 2\pi
\right\}\,. 
\label{bands for SU(N) amoeba}
\end{eqnarray}
Similarly to the $U(1)$ case,
we call the complement of the bands the gap. 
Two ends of the $r$-th band $I_r$ are 
$\frac{1}{R}\log\beta_r^+$ and $\frac{1}{R}\log\beta_r^-$, 
where we put 
$\beta^+_r=\beta_r(0)$ and $\beta_r^-=\beta_r(\pi)$. 
The $r$-th band is a segment between 
$\frac{1}{R}\log\beta_r^-$ and $\frac{1}{R}\log\beta_r^+$ 
on the $u$-axis. 
Whether $\beta_r^- < \beta_r^+$ or $\beta_r^+ < \beta_r^-$ 
depends on $N$ and $r$. 
\begin{eqnarray}
\begin{array}{ll}
\beta_r^- < \beta_r^+ & ~~~\mbox{if $Nr=$even}\,, 
\\[2mm]
\beta_r^+ < \beta_r^- & ~~~\mbox{if $Nr=$odd}\,.
\end{array}
\end{eqnarray}

The Ronkin function $N_{f_{SU(N)}}$ is given by 
\begin{eqnarray}
N_{f_{SU(N)}}(u,v)
= 
\frac{1}{R}\,
\frac{1}{(2\pi i)^2}
\int_{\scriptstyle |x| = e^{Ru} \atop \scriptstyle |y| = e^{Rv}}
\frac{dx}{x}\frac{dy}{y}\, 
\log \left| f_{SU(N)}(x,y) \right|\,.
\label{Ronkin function of SU(N)}
\end{eqnarray} 
We will concentrate our attention on the Ronkin function 
over the $u$-axis. 
The gradient along the $u$-axis can be written as 
\begin{eqnarray}
\pgrad{N_{f_{SU(N)}}}{u}\bigg|_{v=0}
=
\sum_{r=1}^{N}
\frac{1}{(2\pi i)^2}
\int_{|y|=1} \frac{dy}{y} 
\int_{|x| = e^{Ru}}
\frac{dx}{x - \beta_r(\theta_y)}\,.
\label{nabla N_f_SU(N) integral}
\end{eqnarray}
There are $N$ poles in the above integrand. 
Each band holds exactly one pole and 
every pole depends on $\theta_y$. 
It is obvious that
every summand in (\ref{nabla N_f_SU(N) integral}) 
takes the same form as in the previous $U(1)$ case.

Consider the $r$-th summand in (\ref{nabla N_f_SU(N) integral}). 
The residue integral vanishes unless $u$ satisfies 
the inequality $e^{Ru} \geq \beta_r(\theta_y)$. 
When $u$ is off the $r$-th band $I_r$, 
this condition is irrespective of $\theta_y$.  
We have 
\begin{eqnarray}
\frac{1}{(2\pi i)^2}
\int_{|y|=1}\frac{dy}{y} 
\int_{|x| = e^{Ru}} \frac{dx}{x - \beta_r(\theta_y)}
=
\left\{
\begin{array}{ll}
0 & 
\hspace{3mm}\mbox{if}~~
u < \min 
\bigl(\,\frac{1}{R}\log\beta_r^{+}, \frac{1}{R}\log\beta_r^{-}\,\bigr)\,, 
\\[3mm]
1 & 
\hspace{3mm}\mbox{if}~~
u > \max 
\bigl(\,\frac{1}{R}\log\beta_r^{+}, \frac{1}{R}\log\beta_r^{-}\,\bigr) 
\,.
\end{array}
\right.
\end{eqnarray} 
On the other hand, 
when $u$ is on the $r$-th band $I_r$, 
we have to read 
the inequality $e^{Ru} \geq \beta_r(\theta_y)$ as  
a condition on $\theta_y$.  
It should be noted that this reading  
depends on whether
$\beta_r^- < \beta_r^+$ or $\beta_r^+ < \beta_r^-$. 
We first consider the case of $\beta_r^- < \beta_r^+$.  
The function $Q_N(x)$ increases over $[\beta_r^-,\beta_r^+]$. 
Thus the inequality can be translated to 
$Q_N(e^{Ru}) \ge Q_N(\beta_r(\theta_y))= 2(R\Lambda)^N \cos\theta_y$ 
over the band $I_r$. 
So we obtain the following condition for
$\theta_y$.
\begin{eqnarray}
\cos\theta_y 
\leq
\frac{Q_N(e^{Ru})}{2(R\Lambda)^N}\,.
\label{condition 1 in SU(N)}
\end{eqnarray} 
Therefore the $r$-th summand of this case becomes 
\begin{eqnarray}
\frac{1}{(2\pi i)^2}
\int_{|y|=1}\frac{dy}{y} 
\int_{|x| = e^{Ru}} 
\frac{dx}{x - \beta_r(\theta_y)}
&=&
\frac{1}{2\pi}
\int_{\cos\theta_y 
         \leq
        \frac{Q_N(e^{Ru})}{2(R\Lambda)^N}}
d\theta_y \cdot 1 
\nonumber \\
&=&
N - r + 1 +
\frac{1}{\pi}
\arccos\left\{
\frac{Q_N(e^{Ru})}{2(R\Lambda)^N}
\right\}\,, 
\label{r-th summand SU(N)}
\end{eqnarray}
where the branch of the arccosine is fixed by choosing 
$\arccos (0) = \left( r-N-\frac{1}{2} \right)\pi$. 
We turn to consider the case of $\beta_r^+ < \beta_r^-$. 
In this case, $Q_N(x)$ decreases over $[\beta_r^+ , \beta_r^-]$.  
So the inequality means 
$Q_N(e^{Ru}) 
\le Q_N(\beta_r(\theta_y))
= 2(R\Lambda)^N \cos\theta_y$
over the band $I_r$. 
The condition for $\theta_y$ becomes
\begin{eqnarray}
\cos\theta_y 
\geq
\frac{Q_N(e^{Ru})}{2(R\Lambda)^N}\,. 
\label{condition 2 in SU(N)}
\end{eqnarray} 
The $r$-th summand of this case leads to the same expression 
as (\ref{r-th summand SU(N)}).

\begin{figure}[h]
\begin{center}
\includegraphics[width=0.9\linewidth]{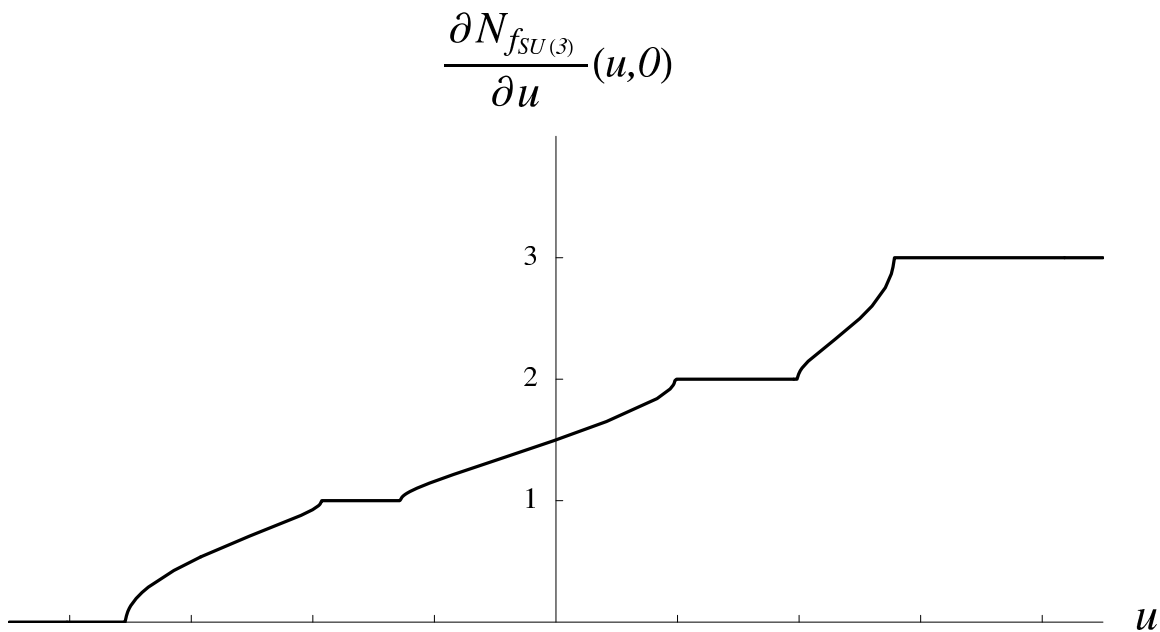}
\end{center}
\caption{\it The gradient of the Ronkin function of $f_{SU(3)}$.}
\label{fig:gradient of Ronkin function}
\end{figure}

\begin{figure}[h]
\begin{center}
\includegraphics[width=0.9\linewidth]{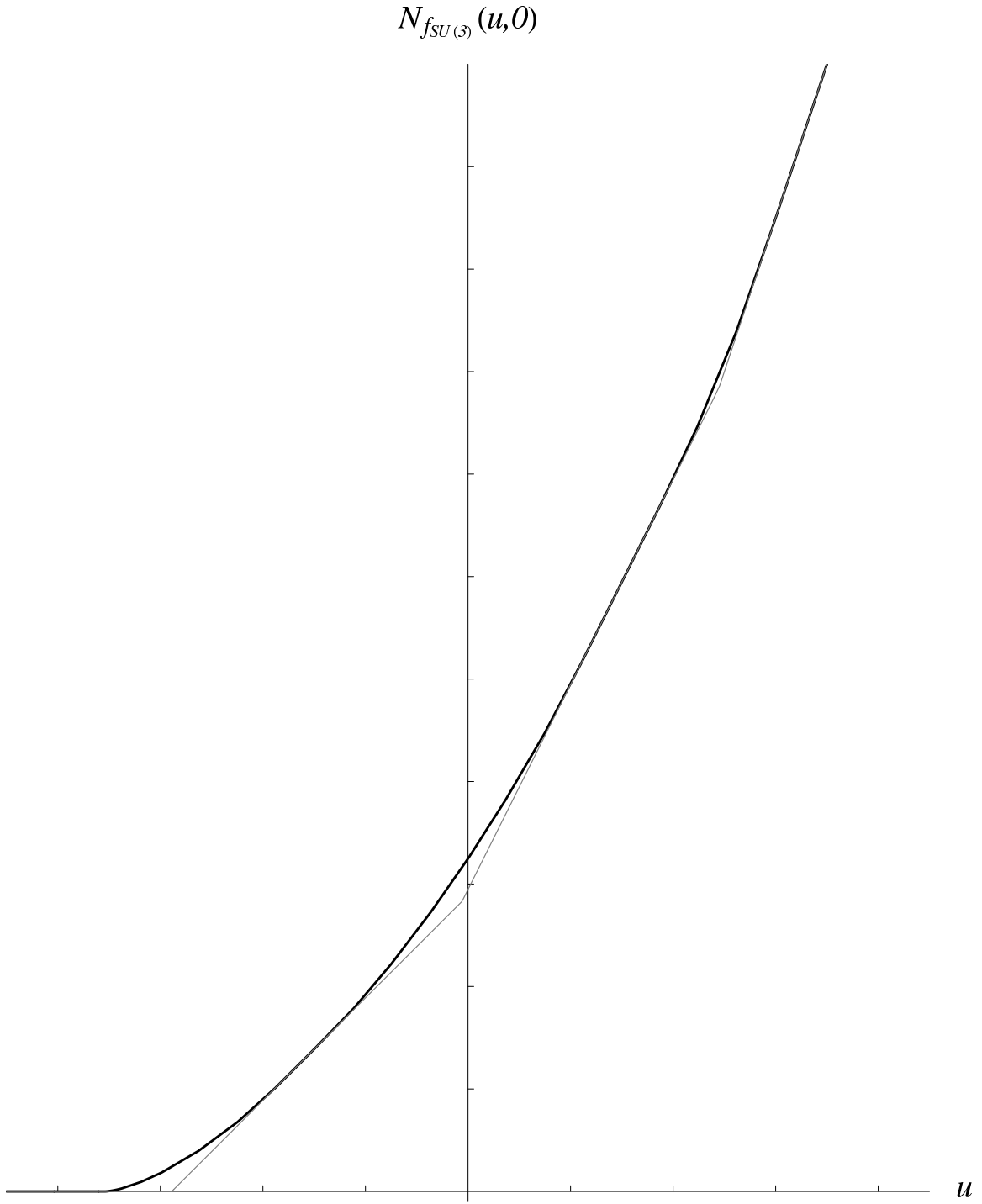}
\end{center}
\caption{\it The Ronkin function of $f_{SU(3)}$ over the $u$-axis.}
\label{fig:Ronkin function of SU(N)}
\end{figure}

By summing up the above ingredients, 
we obtain the following expression for  
the gradient of the Ronkin function along 
the $u$-axis. 
\begin{eqnarray}
\pgrad{N_{f_{SU(N)}}}{u}\bigg|_{v=0}
=
\left\{
\begin{array}{ll}
r
&
\begin{array}{l}
   \mbox{if}~~
   \max \bigl(\frac{1}{R}\log\beta_r^{\pm}\bigr)
   < u < 
   \min \bigl(\frac{1}{R}\log\beta_{r+1}^{\pm}\bigr) 
\\[-0.5mm]
   \hspace{5mm}(r=0,1,\cdots,N)\,,
\end{array}
\\[7mm]
N+ \frac{1}{\pi}\arccos
    \left\{\frac{Q_N(e^{Ru})}{2(R\Lambda)^N}\right\}
&
\begin{array}{l}
    \mbox{if}~~u\in I_s 
\\[-0.5mm]
    \hspace{5mm}(s=1,2,\cdots,N)\,.
\end{array}
\end{array}
\right.
\label{nabla N_f_SU(N) over u-axis}
\end{eqnarray}
We put 
$\beta_0^{\pm} = -\infty$ and $\beta_{N+1}^{\pm} = \infty$ 
in the above. 
The graph of $\partial_u N_{f_{SU(N)}}\big|_{v=0}$ 
is depicted in Figure 
\ref{fig:gradient of Ronkin function}.
Integrating 
(\ref{nabla N_f_SU(N) over u-axis}), 
we acquire the Ronkin function over the $u$-axis. 
The Ronkin function is constant over the connected component 
$E_{(0,0)}$ of the amoeba complement 
and its height there is 
$c_{(0,0)} = \delta_N+\frac{1}{R}\log(R\Lambda)^N$. 
This gives the boundary condition for the integration. 
See Figure \ref{fig:Ronkin function of SU(N)}.

We also arrange the expression (\ref{nabla N_f_SU(N) over u-axis}) 
in terms of complex geometry. 
The zero locus $V_{f_{SU(N)}}$ in $(\mathbb{C}^*)^2$ is written as 
\begin{eqnarray}
y + y^{-1}=\frac{Q_N(e^{Rz})}{(R\Lambda)^N}\,,
\label{f_SU(N)=0}
\end{eqnarray}
where $z$ denotes the cylindrical coordinate of $\mathbb{C}^*$. 
This hyperelliptic curve 
is a double cover of the $\mathbb{C}^*$ 
with $2N$ branch points (Figure \ref{fig:V_fSU(N)}). 
The branch points locate at the ends of the bands 
(\ref{bands for SU(N) amoeba}). 
The holomorphic function $y$ has $N$ cuts along 
the bands on the Riemann sheet 
and takes values at the unit circle over the bands. 
In particular, it is $\pm 1$ at the branch points. 
Using $y$, we can rewrite (\ref{nabla N_f_SU(N) over u-axis}) 
as follows. 
\begin{eqnarray}
\pgrad{N_{f_{SU(N)}}}{u}\bigg|_{v=0}
=
\Re \Bigl(
N + \frac{1}{\pi i}\log y(u-i0)
\Bigr),
\label{d_uN_fSU(N)}
\end{eqnarray}
where we choose $\arg y(u-i0)$ so that 
it increases along the $u$-axis from $-N\pi$ to $0$.

{\it Note added :}~ 
It was shown in \cite{Kenyon-Okounkov} 
(see also \cite{Kenyon-Okounkov-Sheffield}) that 
expression of the gradient like (\ref{d_uN_fSU(N)}) 
holds commonly for the Ronkin functions of Harnack curves, 
where a curve is Harnack if and only if the map from the curve 
to its amoeba is two-to-one over the amoeba interior. 
This was pointed out by A. Okounkov after the original 
version of this paper was submitted to e-print archives. 
The authors are grateful to him.

\clearpage

\section{Ronkin's functions and $5d$ SUSY gauge theories}
\label{section:plane partition model}

By identifying the parameters of the Laurent polynomials  
with the gauge theory order parameters in suitable manners, 
their Ronkin functions relate with instanton countings of 
five-dimensional supersymmetric gauge theories. 
The so-called Nekrasov formulae for 
the exact gauge theory prepotentials 
have a description in terms of random plane partitions. 
The Ronkin functions over the $u$-axis 
turn to be identified with  
integrations of scaled densities of the main diagonal partitions 
in the statistical models at the thermodynamic limit. 
Thus the Ronkin functions are the counting functions.

\subsection{A model of random plane partitions}
\label{subsection:rpp}

\begin{figure}[ht]
\begin{center}
\includegraphics[width=0.8\linewidth]{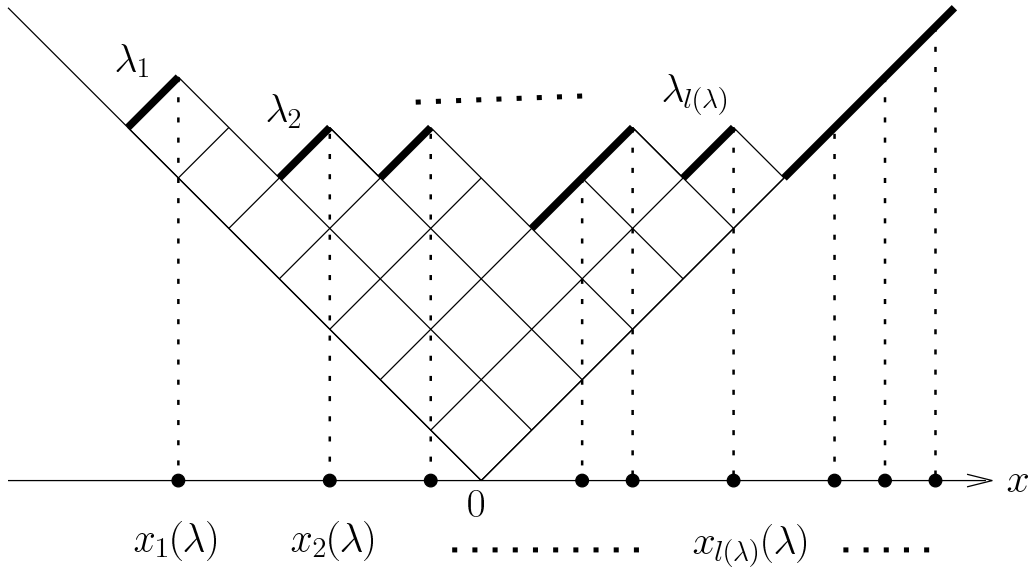}
\end{center}
\caption{\it The Young diagram and the Maya diagram}
\label{fig:maya}
\end{figure}

\begin{figure}[ht]
\begin{center}
\includegraphics[width=0.8\linewidth]{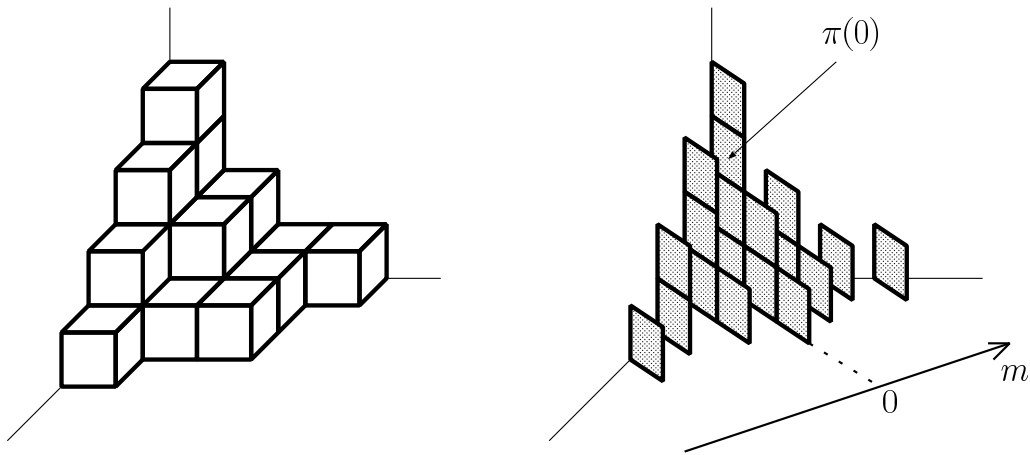}
\end{center}
\caption{\it The $3d$ Young diagram and the diagonal slices.}
\label{fig:plane partition}
\end{figure}

A partition $\lambda=(\lambda_1,\lambda_2,\cdots)$ 
is a sequence of non-negative integers 
satisfying $\lambda_{i} \geq \lambda_{i+1}$ for all $i\geq 1$.
Partitions are often identified with the Young diagrams 
as seen in Figure \ref{fig:maya}. 
The size is defined by $|\lambda|=\sum_{i \geq 1}\lambda_i$, 
which is the total number of boxes of the diagram.
A plane partition $\pi$ is an array of 
non-negative integers satisfying 
$\pi_{ij}\geq \pi_{i+1 j}$ and $\pi_{ij}\geq \pi_{i j+1}$ 
for all $i,j \geq 1$. 
Plane partitions are identified 
with the three-dimensional 
Young diagrams. The three-dimensional diagram $\pi$ 
is a set of unit cubes such that $\pi_{ij}$ cubes 
are stacked vertically on each $(i,j)$-element of $\pi$. 
See Figure \ref{fig:plane partition}. 
The size is defined by 
$|\pi|=\sum_{i,j \geq 1}\pi_{ij}$,  
which is the total number of cubes of the diagram. 
Diagonal slices of $\pi$ become partitions.    
Let $\pi(m)$ denote the partition 
along the $m$-th diagonal slice. 
In particular, 
$\pi(0)$ is the main diagonal partition.  
This series of partitions satisfies the condition
\begin{eqnarray}
\cdots \prec \pi(-2) \prec \pi(-1) \prec 
\pi(0) \succ \pi(1) \succ \pi(2) \succ \cdots,
\label{time evolution}
\end{eqnarray}
where $\mu \succ \nu$ means the interlace relation 
between two partitions $\mu$ and $\nu$.  
\begin{eqnarray}
\mu \succ \nu ~~~
\Longleftrightarrow ~~~
\mu_1 \geq \nu_1 \geq \mu_2 \geq \nu_2 
\geq \mu_3 \geq \cdots.
\end{eqnarray}

We first consider a statistical model of 
plane partitions defined by the following 
partition function.
\begin{eqnarray}
Z(q,Q)
&\equiv& 
\sum_{\pi}\, q^{|\pi|}\,Q^{|\pi(0)|}\,, 
\label{Z(q,Q)}
\end{eqnarray}
where $q$ and $Q$ are indeterminate.   
The Boltzmann weight consists of two parts. 
The first contribution comes from the energy 
of plane partitions, and 
the second contribution is a chemical potential 
for the main diagonal partitions.  
The condition (\ref{time evolution}) suggests that 
plane partitions are certain evolutions of partitions 
by the discretized time $m$. 
This leads to a hamiltonian formulation for the model.  
In particular, the transfer matrix approach 
developed in \cite{Okounkov-Reshetikhin} 
express the partition function (\ref{Z(q,Q)}) 
in terms of two-dimensional conformal field theory.

We can interpret the random plane partitions  
as a $q$-deformation of random partitions. 
It may be seen by rewriting (\ref{Z(q,Q)}) as 
\begin{eqnarray}
Z(q,Q)=\sum_{\lambda}\,Q^{|\lambda|}\, 
\Bigl( \sum_{\pi(0)=\lambda}\, q^{|\pi|} \Bigr)\,. 
\label{Z(q,Q) random partitions}
\end{eqnarray}
Partitions $\lambda$ in the above are thought of 
as the ensemble of the model after summing over  
plane partitions that main diagonal partitions are $\lambda$. 
By using the transfer matrix approach \cite{Okounkov-Reshetikhin}, 
the above factorization yields 
\begin{eqnarray}
Z(q,Q)=
\sum_{\lambda}Q^{|\lambda|}
s_{\lambda}(q^{\frac{1}{2}},q^{\frac{3}{2}},\cdots)^2, 
\label{Z(q,Q) schur}
\end{eqnarray}
where $s_{\mu}(q^{\frac{1}{2}},q^{\frac{3}{2}},\cdots)$ 
is the Schur function 
$s_{\mu}(x_1,x_2,\cdots)$ of infinitely many variables 
specialized at 
$x_i=q^{i-\frac{1}{2}}$ \cite{Macdonald}.

The statistical model gives rise to 
a description of five-dimensional ${\cal N}=1$ 
supersymmetric Yang-Mills theories \cite{MNTT1}. 
To contact with the $SU(N)$ gauge theory,  
we need to identify the indeterminates $q$ and $Q$ 
with the following field theory variables. 
\begin{eqnarray}
q=e^{-\frac{R}{N}\hbar},~~~
Q=(R\Lambda)^2, 
\label{q and Q}
\end{eqnarray}
where $R$ is the radius of $S^1$ in the fifth dimension, 
and $\Lambda$ is the lambda parameter of the underlying 
four-dimensional field theory.

\subsection{Thermodynamic limit and variational problem}

The thermodynamic limit is achieved  
by letting $\hbar \rightarrow 0$. 
By summing up partitions $\lambda$ 
in (\ref{Z(q,Q) schur}), we obtain 
\begin{eqnarray}
Z(q,Q)=\prod_{n=1}^{+\infty}(1-Qq^n)^{-n}.
\label{MacMahon}
\end{eqnarray}
The average numbers of cubes and boxes 
of plane partitions and  the main diagonal partitions 
can be computed from (\ref{MacMahon}). 
By using the identification (\ref{q and Q}), 
the mean values near the thermodynamic limit 
become respectively of orders $\hbar^{-3}$ and $\hbar^{-2}$. 
In particular, we have  
\begin{eqnarray}
\lim_{\hbar \rightarrow 0}
\Bigl(\frac{\hbar}{N}\Bigr)^3 
\langle |\pi| \rangle
=\frac{2}{R^3}Li_3(Q)\,,
\hspace{10mm} 
\lim_{\hbar \rightarrow 0}
\Bigl(\frac{\hbar}{N}\Bigr)^2 
\langle |\pi(0)| \rangle
=\frac{1}{R^2}Li_2(Q)\,.  
\end{eqnarray}
This implies that, as $\hbar$ goes to zero, 
a plane partition $\pi$ that dominates 
is a plane partition of order $\hbar^{-3}$. 
Similarly, 
its main diagonal partition $\pi(0)$ or $\lambda$
becomes a partition of order $\hbar^{-2}$. 
To realize the thermodynamic limit, 
it becomes necessary 
to rescale plane partitions and partitions 
relevantly.

\subsubsection{Scaling partitions}

We provide a description of 
the scaling of partitions at the thermodynamic limit. 
Use of the Maya diagram becomes helpful. 
For a partition $\lambda$, 
the Maya diagram $K(\lambda) \subset \mathbb{Z}$ 
is a set of the integers 
$x_i(\lambda)\equiv -\lambda_i+i$,  
where $i \in \mathbb{Z}_{\geq 1}$. 
An example of the Maya diagram can be found in 
Figure \ref{fig:maya}. 
For the later convenience, 
we associate  partitions with charges. 
We denote such a charged partition 
by $(\lambda,p)$, where $p \in \mathbb{Z}$ is the charge. 
The Maya diagram $K(\lambda,p)$ is a set of the integers 
$x_i(\lambda)+p$, 
that is the parallel transport of $K(\lambda)$ by $p$ 
along the $x$-axis.

For a charged partition $(\lambda,p)$ 
or the Maya diagram $K(\lambda,p)$, 
we introduce the density as 
\begin{eqnarray}
\rho_{bare}(x|\lambda; p)\equiv 
\sum_{i=1}^{+\infty}\delta(x-x_{i}(\lambda)-p)\,. 
\label{bare density}
\end{eqnarray}
The above is not sensitive to the charge 
since $p$ can be absorbed into the shift of $x$. 
We can modify (\ref{bare density}) as 
\begin{eqnarray}
\rho_{reg}(x| \lambda ; p)=
\sum_{i=1}^{+\infty}\delta(x-x_{i}(\lambda)-p)
-\sum_{i=1}^{+\infty}\delta(x-i)\,, 
\label{reg density}
\end{eqnarray}
where the subtraction is prescribed so that it satisfies 
the normal-ordering condition. 
\begin{eqnarray}
\int_{-\infty}^{+\infty}dx 
\rho_{reg}(x| \lambda ; p)=-p\,. 
\label{normal ordering condition}
\end{eqnarray}

Let $\lambda$ be a partition of order $\hbar^{-2}$ and 
$p$ be of order $\hbar^{-1}$.    
We may think of $x_i(\lambda)$ as quantities of order $\hbar^{-1}$. 
We regard that elements of such a partition could be suffixed 
by $s \in \mathbb{R}_{\geq 0}$ rather than $i \in \mathbb{Z}_{\geq 1}$, 
and that these two kinds of indices relate with each other 
by $i=s/\hbar$ when $\hbar$ is nearly zero.  
As $\hbar \rightarrow 0$, 
$x_i(\lambda)$ are rescaled to a certain function $u(s|\lambda)$, 
which takes value in $\mathbb{R}$. 
\begin{eqnarray}
x_i(\lambda)=\frac{N}{\hbar}u(s|\lambda)+O(\hbar^0)\,. 
\label{u(s|lambda)} 
\end{eqnarray}
As for the charge we rescale $p$ to $a \in \mathbb{R}$ by 
\begin{eqnarray}
p=\frac{N}{\hbar}a\,. 
\label{a}
\end{eqnarray}

It should be noticed that $x_1(\lambda),x_2(\lambda),\cdots$  
is a strictly increasing series which satisfies 
the conditions that 
$x_i(\lambda) \leq i$ for $\forall i$ 
and that 
$x_i(\lambda)=i$ for $i \geq l(\lambda)$, 
where $l(\lambda)$ is the length of $\lambda$. 
The function $u(s|\lambda)$  
becomes a strictly increasing function,  
and satisfies the conditions that 
$u(s|\lambda) \leq s/N$ for $\forall s$ 
and that 
$u(s|\lambda)=s/N$ for $s \geq \zeta$, 
where $\zeta$ is a finite constant obtained from $l(\lambda)$. 
The inverse of $u(s|\lambda)$ exists and is denoted by 
$s(u|\lambda)$. 
It becomes a nondecreasing function over $\mathbb{R}$, 
and satisfies the conditions that 
$s(u|\lambda)=Nu$ for $u \geq \xi$ and 
$s(u|\lambda)=0$ for $u \leq -\xi$, where 
$\xi$ is a certain positive constant 
which one may take $\xi=\zeta/N$.

As $\hbar \rightarrow 0$,  
we also scale the density (\ref{bare density}) 
by using (\ref{u(s|lambda)}) and (\ref{a}). 
In particular, taking account of (\ref{u(s|lambda)}),  
we rescale $x$ to $u \in \mathbb{R}$ by $x=Nu/\hbar$. 
Then the scaling limit is read as 
\begin{eqnarray}
\lim_{\hbar \rightarrow 0} 
\rho_{bare}(x=Nu/\hbar\,|\lambda;\,p=Na/\hbar) 
=
\rho(u-a|\lambda)\,, 
\label{rescaled bare density}
\end{eqnarray}
where 
\begin{eqnarray}
\rho(u|\lambda)\equiv
\frac{1}{N}
\frac{ds(u|\lambda)}{du}\,. 
\label{density}
\end{eqnarray}

The behavior of the function $s(u|\lambda)$ implies that  
the scaled density $\rho(u|\lambda)$ takes values in $[0,1]$.
It also implies that 
$d\rho(u|\lambda)/du$ has a compact support in $\mathbb{R}$ 
and satisfies 
\begin{eqnarray}
\int_{-\infty}^{+\infty}du 
\frac{d\rho(u|\lambda)}{du}=1\,. 
\label{normalization}
\end{eqnarray} 
Similarly to (\ref{rescaled bare density}), 
the regularized density  
$\rho_{reg}(x|\lambda;p)$ is scaled to 
$\rho(u-a|\lambda)-\rho(u|\emptyset)$. 
The scaled density of the empty partition is 
$\rho(u|\emptyset)=\theta(u)$, 
where $\theta(u)$ denotes the step function, that is, 
$\theta(u)=1$ for $u \geq 0$ and $0$ for $u <0$. 
Applying a partial integration 
and combining with (\ref{normalization}), 
we can translate the condition (\ref{normal ordering condition}) 
as follows. 
\begin{eqnarray}
\int_{-\infty}^{+\infty}du 
u \frac{d\rho(u|\lambda)}{du}=0\,. 
\label{vanishing U(1) charge condition}
\end{eqnarray}

The scaled density $\rho(u|\lambda)$ 
relates with the shape of the Young diagram. 
A piecewise linear curve   
that traces the upper side of the Young diagram  
positioned as in Figure \ref{fig:maya} 
and extends over $\mathbb{R}$ by $|x|$, 
is called profile of $\lambda$. 
The scaling limit of the profile becomes the graph of 
a certain function $P(u|\lambda)$.  
This function relates with the scaled density by 
\begin{eqnarray}
\frac{dP(u|\lambda)}{du}
=-1+2\rho(u|\lambda)\,. 
\label{f(u|lambda)}
\end{eqnarray}

\subsubsection{The variational problem}

In the description (\ref{Z(q,Q) schur}), 
each partition $\lambda$ has the Boltzmann weight 
$Q^{|\lambda|}
s_{\lambda}(q^{\frac{1}{2}},q^{\frac{3}{2}},\cdots)^2$, 
where the parameters are identified with 
$q=\exp(-R\hbar/N)$ and $Q=(R\Lambda)^2$. 
When partitions are scaled according to (\ref{u(s|lambda)}), 
the asymptotic form of their weight is computed in \cite{MNTT2}  
by using the product formula of 
the  (specialized) Schur function \cite{Macdonald},  
and is expressed as a functional of the scaled density. 
It can be read  
\begin{eqnarray}
\log Q^{|\lambda|}s_{\lambda}
(q^{\frac{1}{2}},q^{\frac{3}{2}},\cdots)^2=
-\frac{1}{\hbar^2}
\Bigl\{ E[\rho(\cdot|\lambda)] 
+ O(\hbar) \Bigr\}\,,
\label{asymptotic BW}
\end{eqnarray}
where 
\begin{eqnarray}
E[\rho(\cdot|\lambda)]
&=&
N^2
\int_{-\infty <u < \underline{u}<+\infty} 
du d\underline{u}\,\,
\rho(u|\lambda)
\bigl(1-\rho(\underline{u}|\lambda)\bigr) 
\log \Bigl\{
\frac{\sinh \frac{R}{2}(\underline{u}-u)}
     {\frac{R}{2}\Lambda} \Bigr\}^2 
\nonumber \\
&&     
+N^2 
\int_{-\infty}^{+\infty}
du \frac{Ru^3}{6}
\frac{d\rho(u|\lambda)}{du}\,.
\label{energy functional}
\end{eqnarray}

Near the thermodynamic limit, 
partitions of order $\hbar^{-2}$ dominate 
in the statistical model.   
Their Boltzmann weights are measured 
by using the energy functional $E$ as $e^{-E/\hbar^2}$. 
Since the thermodynamic limit is achieved by letting 
$\hbar \rightarrow 0$, 
this means that it is realized 
by a classical configuration that minimizes 
the energy functional.
We are thus led to consider a variational problem 
of the energy functional and to find the minimizer 
as the stationary configuration.

The minimizer of the energy functional 
(\ref{energy functional}) 
should be found from among certain admissible configurations.  
The admissible configurations are thought of functions $\rho(u)$ 
that could be obtained as the scaled densities of partitions. 
They need to take values in $[0,1]$. 
Furthermore, taking account of 
(\ref{normalization}) and (\ref{vanishing U(1) charge condition}), 
we require that $d\rho(u)/du$ has a compact support in $\mathbb{R}$ 
and satisfies the conditions 
\begin{eqnarray}
\int_{-\infty}^{+\infty}du\,\,\frac{d\rho(u)}{du}&=&1\,, 
\label{admissible condition 1} \\
\int_{-\infty}^{+\infty} du\,\,u\,\frac{d\rho(u)}{du}&=&0 \,.
\label{admissible condition 2}
\end{eqnarray}

To argue the variational problem, 
it is convenient to rewrite the energy functional 
in the following form by partial integrations. 
\begin{eqnarray}
E[\rho(\cdot)]
&=&
\frac{N^2}{2}
\int_{u \neq \underline{u}} 
du d\underline{u}\,\,
\frac{d\rho(u)}{du}
\frac{d\rho(\underline{u})}{d\underline{u}}
\gamma(|\underline{u}-u|;\Lambda;R) 
\nonumber \\
&&
+N^2 
\int_{-\infty}^{+\infty}
du \frac{Ru^3}{6}
\frac{d\rho(u)}{du}\,,
\label{energy functional 2}
\end{eqnarray}
where the function $\gamma(u;\Lambda;R)$ has been 
introduced by the conditions 
\begin{eqnarray}
&&
\frac{\partial^2\gamma}{\partial u^2}(u;\Lambda;R)\,=\, 
\log \Bigl(
\frac{\sinh \frac{R}{2}u}{\frac{R}{2}\Lambda}
\Bigr)\,,
\label{condition 1 for gamma}
\\
&&
\gamma(0;\Lambda;R)\,=\, 
\frac{\partial \gamma}{\partial u}(0;\Lambda;R)
\,=0\,.
\label{condition 2 for gamma}
\end{eqnarray}

It is clear from the expression (\ref{energy functional 2}) 
that variations $\rho \rightarrow \rho+\delta \rho$ 
of the energy functional lead to the stationary equation 
\begin{eqnarray}
\mbox{PP}
\int_{-\infty}^{+\infty}
d\underline{u}\,\,
\frac{\partial \gamma}{\partial u}
(|u-\underline{u}|;\Lambda;R) 
\frac{d\rho(\underline{u})}{d\underline{u}}
=-\frac{Ru^2}{2}
\hspace{7mm}
\mbox{on}
\hspace{3mm}
\mbox{supp}\,\bigl(\frac{d\rho}{du}\bigr)\,. 
\label{stationary equation 1}
\end{eqnarray}
The integration symbol in the above means principle part. 
By differentiating Eq.(\ref{stationary equation 1}) twice 
with respect to $u$, we get 
\begin{eqnarray}
\mbox{PP}
\int_{-\infty}^{+\infty}
d\underline{u}\,\,
\coth 
\frac{R}{2}(u-\underline{u}) 
\frac{d\rho(\underline{u})}{d\underline{u}}=-1
\hspace{7mm}
\mbox{on}
\hspace{3mm}
\mbox{supp}\,\bigl(\frac{d\rho}{du}\bigr)\,. 
\label{stationary equation 2}
\end{eqnarray} 
The variational problem is now stated to find out 
a solution $\rho_{\star}(u)$ 
of the integral equation (\ref{stationary equation 2}) 
from among the admissible configurations.

The standard argument \cite{Itzykson} 
allows us to reformulate this as a problem 
to find out a certain analytic function $\Phi(z)$.  
It may be helpful to consider an analytic function   
that is realized as the integral transform of 
a continuous function $\chi(u)$ over the real axis. 
\begin{eqnarray}
\Phi_{\chi}(z)=
\int_{-\infty}^{+\infty}du\,\,
\coth \frac{R}{2}(z-u) \,\chi(u)\,. 
\end{eqnarray} 
The above function is regarded 
as an analytic function on $\mathbb{C}^*$ 
since it is periodic with the period $2\pi i/R$. 
On the real axis, 
when one approaches from the upper or the lower half plane, 
it takes   
\begin{eqnarray}
\Phi_{\chi}(u\pm i0)=
\mbox{PP}
\int_{-\infty}^{+\infty}
d\underline{u}\,\,
\coth 
\frac{R}{2}(u-\underline{u})\, 
\chi(\underline{u})\,\,
\mp \frac{2\pi i}{R}\chi(u)\,. 
\end{eqnarray}

\begin{figure}[t]
\begin{center}
\includegraphics[width=0.8\linewidth]{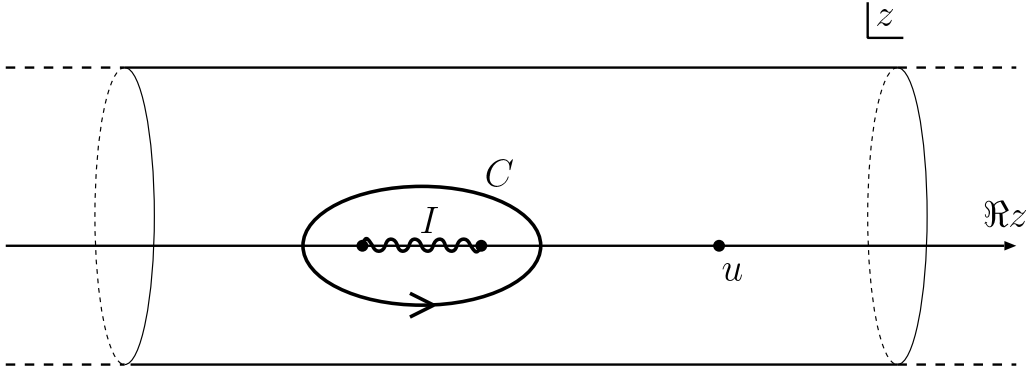}
\end{center}
\caption{\it The cut $I$ and the contour $C$ on the Riemann sheet.}
\label{fig:Riemann sheet for U(1)}
\end{figure}

Observing the above formula by putting $\chi=d\rho/du$, 
we deduce the variational problem as follows:  
Let $\Phi(z)$ be an analytic function  
that is periodic with the period $2\pi i/R$   
and behave at the infinities as
\begin{eqnarray}
\Phi(z) \longrightarrow \pm 1 
\hspace{5mm}\mbox{as}\hspace{2mm}
\Re z \rightarrow \pm \infty\,.
\label{condition 1 for Phi}
\end{eqnarray}
We further suppose that $\Phi(z)$ has a cut 
on a bounded region $I$ along the real axis   
and satisfies 
\begin{eqnarray}
&&
\begin{array}{ll}
\Re \hspace{1mm}\Phi(u)\,=\,-1 & 
\hspace{6mm}\mbox{on}\hspace{2mm}I\,, \\ 
\Im \hspace{1mm}\Phi(u)\,=\,0  &
\hspace{6mm}\mbox{on}\hspace{2mm}\mathbb{R}\setminus I\,, 
\end{array}
\label{condition 2 for Phi}
\end{eqnarray}
and
\begin{eqnarray}
\frac{R}{4\pi i}\oint_C dz\,\,z\Phi(z)\,=\,0\,,
\label{condition 3 for Phi}
\end{eqnarray}
where $C$ denotes a contour which encircles the above 
$I$ anticlockwise (Figure \ref{fig:Riemann sheet for U(1)}). 
Then the solution $\rho_{\star}$ of the variational problem 
is given by the imaginary part of $\Phi$ along the real axis   
as follows. 
\begin{eqnarray}
\frac{d\rho_{\star}(u)}{du}=
\mp \frac{R}{2\pi} 
\Im \,\Phi(u\pm i0)\,.
\label{minimizer} 
\end{eqnarray}
In particular, 
we have 
$\mbox{supp}\,(d\rho_{\star}/du)=I$.

We note that 
the conditions (\ref{condition 1 for Phi}), 
(\ref{condition 2 for Phi}) and (\ref{condition 3 for Phi}) 
are translations of 
(\ref{normalization}), (\ref{stationary equation 2}) 
and (\ref{vanishing U(1) charge condition}) 
respectively.

\subsection{$U(1)$ gauge theory and the Ronkin function}
\label{subsection:U(1) theory and the Ronkin function}

We consider the case of the $U(1)$ gauge theory. 
We put $N=1$. 
The partition function (\ref{Z(q,Q)}) 
can be identified with the partition function 
for four-dimensional $\mathcal{N}=2$ supersymmetric 
$U(1)$ gauge theory on noncommutative $\mathbb{R}^4$.

To find out a solution of the variational problem 
posed at the end of the previous subsection, 
we test an analytic function of the following form.
\begin{eqnarray}
\Phi(z)= 
\frac{d}{dz}
\Bigl(-z+\frac{2}{R}\log y \Bigr)\,, 
\label{Phi U(1)}
\end{eqnarray}
where 
\begin{eqnarray} 
y+y^{-1}= 
\frac{1}{R\Lambda}
\bigl(e^{Rz}-\beta \bigr)\,. 
\label{U(1) curve}
\end{eqnarray}
Here $\beta$ is a real parameter 
and is assumed $\beta > 2R\Lambda$.

It is easy to see that the above function 
satisfies the condition (\ref{condition 1 for Phi}). 
Along the real axis, it has a single cut on 
$[\,\frac{1}{R}\log(\beta-2R\Lambda),\,\frac{1}{R}\log(\beta+2R\Lambda)\,]$. 
We can also see that it satisfies the condition 
(\ref{condition 2 for Phi}). 
We examine the condition (\ref{condition 3 for Phi}). 
Let $C$ be the contour that encircles  
the above cut anticlockwise on the $z$-plane. We will evaluate 
the contour integral in (\ref{condition 3 for Phi}) as follows. 
\begin{eqnarray}
\frac{R}{4\pi i}\oint_C dz\,\,z\Phi(z)
&=&
\frac{1}{2\pi i}\oint_C zd \log y 
\nonumber \\
&=& 
\frac{1}{2\pi iR} 
\int_{|y|=1} 
\frac{dy}{y} 
\log \left\{ 
R\Lambda(y+y^{-1})+\beta 
\right\}. 
\end{eqnarray}
The last integration can be computed by applying the classical 
Jensen formula in complex analysis and gives rise to 
\begin{eqnarray}
\frac{R}{4\pi i}\oint_C dz\,\,z\Phi(z)
=
-\frac{1}{R}
\log 
\frac{\beta-\sqrt{\beta^2-4(R\Lambda)^2}}{2(R\Lambda)^2}\,. 
\end{eqnarray}
The vanishing of the contour integration fixes $\beta$ as  
\begin{eqnarray}
\beta 
=1+(R\Lambda)^2\,. 
\end{eqnarray}

We thus obtain the solution
$\rho^{U(1)}_{\star}(u)$ 
by letting $\beta=1+(R\Lambda)^2$ in (\ref{U(1) curve}).  
Let us summarize this as follows. 
\begin{eqnarray}
\frac{d\rho_{\star}^{U(1)}}{du}(u)
=\mp \frac{1}{\pi}\,\Im \,\frac{d\log y}{dz}(u\pm i0)\,, 
\label{U(1)drho}
\end{eqnarray}
where 
\begin{eqnarray}
y+y^{-1}=
\frac{1}{R\Lambda}
\bigl(e^{Rz}-1-(R\Lambda)^2 \bigr)\,. 
\label{U(1)SWcurve}
\end{eqnarray}
We note that the above solution is precisely 
that obtained in \cite{MNTT2} 
by the WKB analysis of the fermion wave function.

We specialize the Laurent polynomial $f_{U(1)}$, 
which is given in (\ref{polynomial of U(1)}),  
by choosing $\beta=1+(R\Lambda)^2$. 
The above solution can be expressed by using 
the Ronkin function of this $f_{U(1)}$. 
By comparing (\ref{U(1)drho}) with (\ref{d_uN_fU(1)}), 
we find  
\begin{eqnarray}
\rho_{\star}^{U(1)}(u)=
\frac{\partial N_{f_{U(1)}}(u,0)}{\partial u}\,. 
\label{dN_fU(1)=rho_U(1)}
\end{eqnarray}
This means that the Ronkin function over the $u$-axis 
is an integration of the scaled density of the main diagonal 
partitions that realizes the thermodynamic limit. 
\begin{eqnarray}
N_{f_{U(1)}}(u,0)=
\int_{-\infty}^u 
d\underline{u}\,\, 
\rho_{\star}^{U(1)}(\underline{u})
+ \mbox{const.}
\label{N_fU(1)=integrated rho_U(1)}
\end{eqnarray}
Let $P_{\star}^{U(1)}(u)$ be the limit shape 
of the Young diagrams obtained from $\rho_{\star}^{U(1)}(u)$ 
by using the relation (\ref{f(u|lambda)}). 
We can write (\ref{N_fU(1)=integrated rho_U(1)}) as follows. 
\begin{eqnarray}
N_{f_{U(1)}}(u,0)=
\frac{1}{2}
\Bigl(P_{\star}^{U(1)}(u)+u\Bigr)
+ \mbox{const.}
\label{N_fU(1)=P_U(1)}
\end{eqnarray}

\subsection{Random plane partitions and $SU(N)$ gauge theory}
\label{subsetion:rpp and SU(N) Yang-Mills}

\begin{figure}[ht]
\begin{center}
\includegraphics[width=0.8\linewidth]{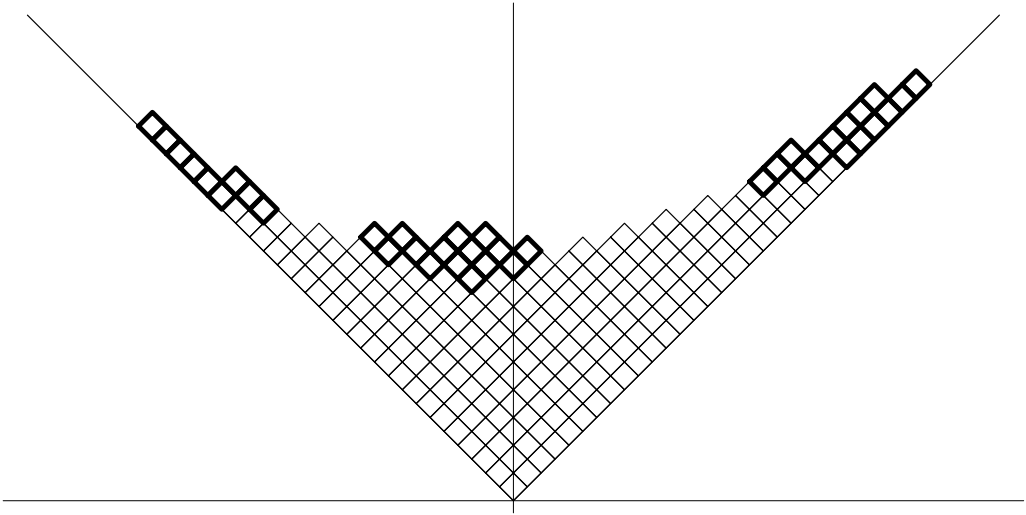}
\end{center}
\caption{\it Partition corresponding to three charged partitions 
$(\lambda_1,p_1) = ((2,1),-7)$, 
$(\lambda_2,p_2) = ((3,2),-1)$ and 
$(\lambda_3,p_3) = ((3,2),8)$
by (\ref{1:N by Maya diagrams}).
Three clusters of bold boxes correspond respectively to
the three partitions.} 
\label{fig:1:N by Young diagrams}
\end{figure}

There is a bijective correspondence 
between charged partitions $(\mu,p)$ and 
$N$ charged partitions $\{(\lambda^{(r)},p_r)\}^N_{r=1}$. 
This arises from a division algorithm for the Young diagrams 
analogous to that for integers. 
The correspondence is neatly described by using the Maya diagrams 
as follows. 
\begin{eqnarray}
K(\mu,p)=
\bigcup_{r=1}^N\, \,
\iota_r\bigl(K(\lambda^{(r)},p_r)\bigr)\, ,
\label{1:N by Maya diagrams}
\end{eqnarray}
where $\iota_r$ $\mathbb{Z}\hookrightarrow \mathbb{Z}$ 
denotes an injective map given by $\iota_r(m)=Nm+r-N$, and 
we have $\iota_r\bigl(K(\lambda^{(r)},p_r)\bigr)=
N\,K(\lambda^{(r)},p_r)+r-N$. It can be seen that 
the charges are conserved in (\ref{1:N by Maya diagrams}). 
\begin{eqnarray}
p=\sum_{r=1}^Np_r\,.
\label{charge conservation}
\end{eqnarray} 
In terms of the Young diagrams 
the correspondence (\ref{1:N by Maya diagrams}) 
is seen as Figure \ref{fig:1:N by Young diagrams}. 

By using the above correspondence, 
the factorization (\ref{Z(q,Q) random partitions}) 
or (\ref{Z(q,Q) schur}) is also expressed 
in terms of $N$ charged partitions 
$\{(\lambda^{(r)},p_r)\}^N_{r=1}$. 
Since the factorization was done by using neutral partitions,  
the charge conservation says that 
the charges $p_r$ satisfy the $SU(N)$ condition 
\begin{eqnarray}
\sum_{r=1}^Np_r=0\,.
\label{SU(N) condition}
\end{eqnarray}
After including the summation over partitions $\lambda^{(r)}$ 
implicitly in each component of the factorization, 
let us factor the partition function into 
\begin{eqnarray}
Z(q,Q)= 
\sum_{\{p_r\}}
Z_{SU(N)}(\{p_r\};q,Q)\,. 
\label{N-factored RPP}
\end{eqnarray}

The above factorization becomes 
a bridge between the random plane partitions 
and five-dimensional $\mathcal{N}=1$ supersymmetric $SU(N)$ Yang-Mills. 
Let $a_r$ be the vacuum expectation values of the adjoint scalar 
in the vector multiplet. 
We identify the parameters $q,Q$ and $p_r$ 
with the gauge theory parameters as follows. 
\begin{eqnarray}
q=e^{-\frac{R}{N}\hbar},
\hspace{4mm}
Q=(R\Lambda)^{2}, 
\hspace{4mm}
p_r=a_r/\hbar\,. 
\label{mapping RPP to gauge theory}
\end{eqnarray}
It is shown \cite{MNTT1} that the above identification 
leads to 
\begin{eqnarray}
Z_{SU(N)}(\{p_r\};q,Q)=
Z_{5d\, \mbox{\scriptsize{SYM}}}(\{a_r\};\Lambda,R,\hbar)\,, 
\label{exact partition function for SU(N)SYM}
\end{eqnarray}
where the RHS is the exact partition function 
\cite{Nekrasov-Okounkov} 
for five-dimensional $\mathcal{N}=1$ supersymmetric $SU(N)$ 
Yang-Mills plus the Chern-Simons term 
having the coupling constant equal to $N$. 
The five-dimensional theory is living on $\mathbb{R}^4 \times S^1$, 
where the radius of $S^1$ in the fifth dimension is $R$. 
The scale parameter of the underlying four-dimensional theory is $\Lambda$.  
In the cases that the Chern-Simons coupling constant takes other values, 
the partition functions can be also retrieved from the statistical model 
by adding another potential term for the main diagonal partitions 
in (\ref{Z(q,Q)}) \cite{MNNT}.

The realization of partitions in terms of multiple charged partitions 
also plays important roles in multi-instanton calculus on ALE spaces 
\cite{instanton countings on ALE 1},\cite{quiver gauge theory}.
Combinatorial aspect that becomes key to the calculus 
is further elucidated in \cite{instanton countings on ALE 2}.

\subsubsection{Scaling multiple charged partitions}

Let us consider $N$ charged partitions 
$\{(\lambda^{(r)},p_r)\}^N_{r=1}$,   
where each partition $\lambda^{(r)}$ is of order $\hbar^{-2}$ 
and each charge $p_r$ is of order $\hbar^{-1}$. 
Each charged partition is treated by the following scalings. 
\begin{eqnarray}
x_{i_r}(\lambda^{(r)}) &=&
u(s_r|\lambda^{(r)})/\hbar+O(\hbar^0)\,,  
\label{u(s|lambda) for r-th partition} \\
p_r &=& a_r/\hbar\,, 
\label{a_r}
\end{eqnarray}
where $s_r \in \mathbb{R}_{\geq 0}$ has been introduced by 
$i_r=s_r/\hbar$.

Together with the above scalings, 
the correspondence leads to 
a scaling of the charged partition $(\mu,p)$ 
in the forms (\ref{u(s|lambda)}) and (\ref{a}).  
The density of $(\mu,p)$ is expressed 
in terms of those of $(\lambda^{(r)},p_r)$ as follows. 
\begin{eqnarray}
\rho_{bare}(x|\mu;p)=
\frac{1}{N}\sum_{r=1}^N 
\rho_{bare}\bigl(\frac{x-r+N}{N}|\lambda^{(r)};p_r \bigr)\,. 
\label{1:N by bare density}
\end{eqnarray}
We will obtain the scaling limit of this relation.   
We rescale $x$ to $u$ by $x=Nu/\hbar$. 
For each charged partition, 
the inverse of $u(s_r|\lambda^{(r)})$ exists, 
as follows from (\ref{u(s|lambda) for r-th partition}), 
and is denoted by $s_r(u|\lambda^{(r)})$.  
It is a nondecreasing function over $\mathbb{R}$,  
and satisfies the conditions 
that $s_r(u|\lambda)=u$ for $u \geq \xi_r$ 
and 
$s_r(u|\lambda^{(r)})=0$ for $u \leq -\xi_r$, 
where $\xi_r$ is a certain positive constant. 
By using (\ref{u(s|lambda) for r-th partition}) and (\ref{a_r}),  
the scaling limit of each density becomes 
\begin{eqnarray}
\lim_{\hbar \rightarrow 0} 
\rho_{bare}(x=u/\hbar\,|\lambda^{(r)};\,p_r=a_r/\hbar) 
=
\rho_r(u-a_r|\lambda^{(r)})\,, 
\label{rescaled r-th bare density}
\end{eqnarray}
where 
\begin{eqnarray}
\rho_r(u|\lambda^{(r)})\equiv
\frac{ds_r(u|\lambda^{(r)})}{du}\,. 
\label{r-th density}
\end{eqnarray} 
As for the charged partition $(\mu,p)$, 
by taking account of 
(\ref{rescaled bare density}) and (\ref{density}), 
the scaling limit of the density is $\rho(u-a|\mu)$. 
Therefore the relation (\ref{1:N by bare density}) 
leads to 
\begin{eqnarray}
\rho(u-a|\mu)=
\frac{1}{N}\sum_{r=1}^N
\rho_r(u-a_r|\lambda^{(r)})\,, 
\label{1:N by rescaled density}
\end{eqnarray}
where $Na=\sum_{r=1}^Na_r$.

Each scaled density $\rho_r(u|\lambda^{(r)})$ 
takes values in $[0,1]$. 
We also see that 
$d\rho_r(u|\lambda^{(r)})/du$ has a compact support 
in $\mathbb{R}$ and satisfies 
\begin{eqnarray}
\int_{-\infty}^{+\infty}du \,
\frac{d\rho_r(u|\lambda^{(r)})}{du}
&=&1\,,
\label{r-th normalization} 
\\
\int_{-\infty}^{+\infty}du\, 
u \frac{d\rho_r(u|\lambda^{(r)})}{du}
&=&0\,. 
\label{r-th U(1) charge condition}
\end{eqnarray}

At this stage it is convenient to 
impose the $SU(N)$ condition (\ref{SU(N) condition}) on the charges.  
This means that $a_r$ are taken so that $\sum_{r=1}^Na_r=0$. 
It follows from (\ref{1:N by rescaled density}) that 
$d\rho(u|\mu)/du$ is supported on $\cup_{r=1}^N I_r$, 
where $I_r$ denotes the support of $d\rho_r(u-a_r|\lambda^{(r)})/du$ 
and is thought to be a certain segment centered around $a_r$. 
These bands become disjointed 
when $a_r$ are sufficiently separated from one another. 
In such a circumstance, by using (\ref{1:N by rescaled density}), 
it is possible to rewrite 
(\ref{r-th normalization}) and 
(\ref{r-th U(1) charge condition})
as follows. 
\begin{eqnarray}
\int_{I_r}du\,\frac{d\rho(u|\mu)}{du}
&=&
1/N\,, 
\label{r-th normalization by mu}
\\
\int_{I_r}du\,u\frac{d\rho(u|\mu)}{du}
&=&
a_r/N\,.
\label{r-th U(1) charge condition by mu}
\end{eqnarray}

\begin{figure}[t]
\begin{center}
\includegraphics[width=0.9\linewidth]{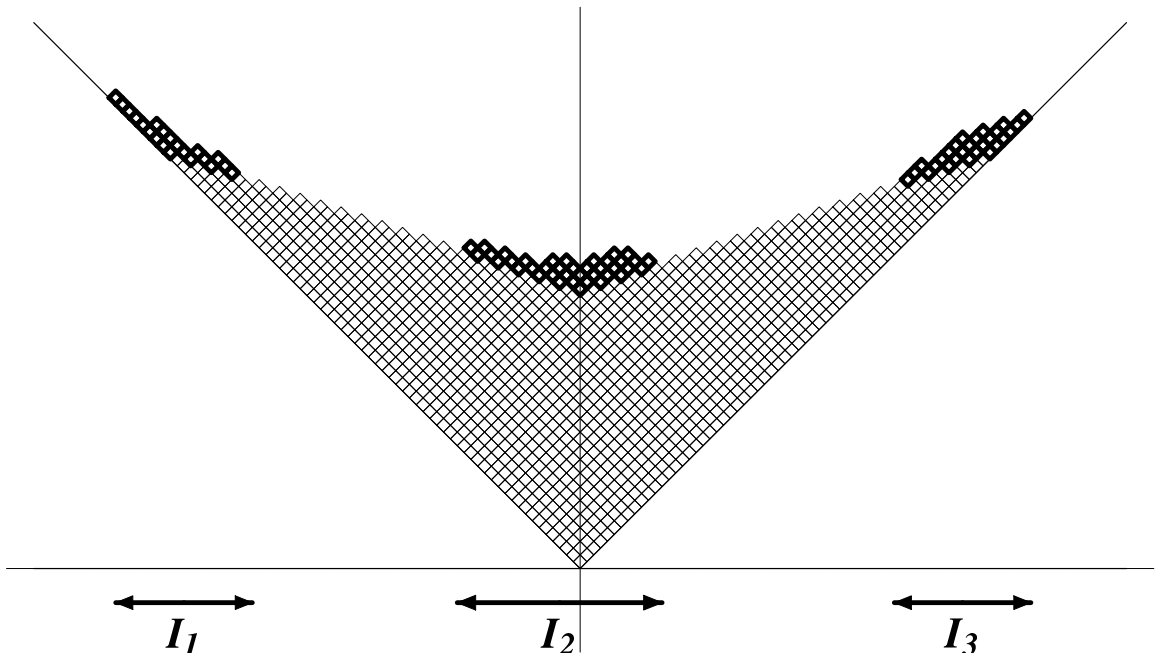}
\end{center}
\caption{\it A typical behavior of the density at small $\hbar$.} 
\label{fig:density at small h}
\end{figure}

\subsubsection{The variational problem} 
\label{subsubsection:SU(N) variation problem}

We would like to consider the thermodynamic limit 
of each component that appears in the factorization 
(\ref{N-factored RPP}). 
In what follows, 
we impose the $SU(N)$ condition on the charges. 
We also fix $a_r$ so that they are sufficiently 
separated from one another. 
Without any loss it is enough to 
consider the case of $a_1<a_2<\,\cdots\,<a_N$. 
By taking account of 
the previous study of the $U(1)$ gauge theory, 
the thermodynamic limit we are investigating now 
is realized by a configuration that minimizes 
the energy functional with keeping $a_r$ fixed. 
Therefore, 
the relevant variational problem of the energy functional 
should be restricted within configurations of 
the scaled densities 
that are expressible in terms of 
$N$ charged partitions with the fixed charges. 
Such configurations may be obtained by restricting 
the admissible configurations relevantly.  
By observing (\ref{r-th normalization by mu}) 
and (\ref{r-th U(1) charge condition by mu}), 
we require that $d\rho(u)/du$ is supported on 
$\cup_{r=1}^NI_r$, where $I_r$ are 
certain disjointed segments, and satisfies there 
\begin{eqnarray}
\int_{I_r}du\,\frac{d\rho(u)}{du}
&=&
1/N\,, 
\label{SU(N) admissible condition 1}
\\
\int_{I_r}du\,u\frac{d\rho(u)}{du}
&=&
a_r/N\,.
\label{SU(N) admissible condition 2}
\end{eqnarray}
The variational problem is formulated to find out a solution 
$\rho_{\star}^{SU(N)}$ of the integral equation 
(\ref{stationary equation 2}) from among the admissible 
configurations of the above type.

\begin{figure}[ht]
\begin{center}
\includegraphics[width=0.8\linewidth]{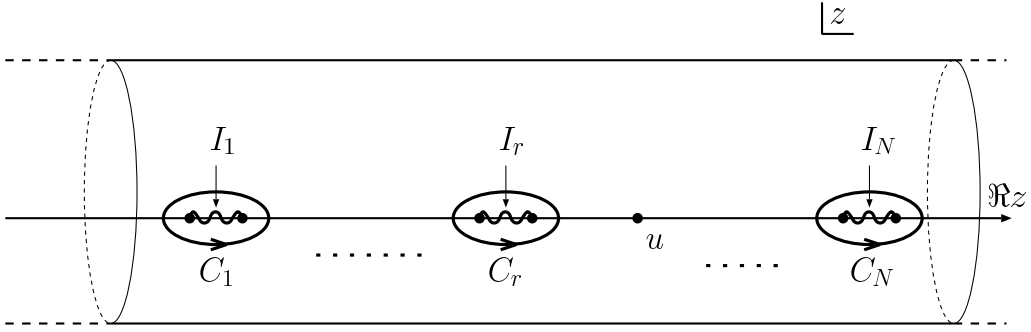}
\end{center}
\caption{\it The cuts $I_r$ and the contours $C_r$ on the Riemann sheet.}
\label{fig:Riemann sheet for SU(N)}
\end{figure}

We can also reformulate this as a problem to find out 
a certain analytic function: 
Let $\Phi(z)$ be an analytic function 
that is periodic with the period $2\pi i/R$,  
and behave at the infinities as
$\Phi(z) \longrightarrow \pm 1$ as 
$\Re\,z \rightarrow \pm \infty$.
We suppose that $\Phi(z)$ has a cut 
on each disjointed segments $I_r$ ($r=1,\,\cdots,\,N$)  
along the real axis and satisfies 
\begin{eqnarray}
&&
\begin{array}{ll}
\Re \hspace{1mm}\Phi(u)\,=\,-1 & 
\hspace{6mm}\mbox{on}\hspace{2mm}\cup_{r=1}^NI_r\,, \\ 
\Im \hspace{1mm}\Phi(u)\,=\,0  &
\hspace{6mm}\mbox{on}\hspace{2mm}
\mathbb{R}\setminus \cup_{r=1}^NI_r\,. 
\end{array}
\label{condition 1 for Phi SU(N)}
\end{eqnarray}
Let $C_r$ be a contour which encircles the cut 
$I_r$ anticlockwise 
(Figure \ref{fig:Riemann sheet for SU(N)}).  
We further suppose that $\Phi(z)$ satisfies 
\begin{eqnarray}
\frac{R}{4\pi i}\oint_{C_r} dz\,\,\Phi(z)&=&1/N\,,
\label{condition 2 for Phi SU(N)} 
\\ 
\frac{R}{4\pi i}\oint_{C_r} dz\,\,z\Phi(z)
&=&a_r/N\,,
\label{condition 3 for Phi SU(N)}
\end{eqnarray}
for each $r$. 
Then the solution $\rho_{\star}^{SU(N)}$ 
is given by the imaginary part as follows. 
\begin{eqnarray}
\frac{d\rho_{\star}^{SU(N)}(u)}{du}=
\mp \frac{R}{2\pi} 
\Im \,\Phi(u\pm i0)\,.
\label{minimizer by Phi SU(N)} 
\end{eqnarray}

\subsection{$SU(N)$ gauge theory and the Ronkin function}

We can find the solution $\rho_{\star}^{SU(N)}$ 
by following the route similar to the $U(1)$ case. 
Consider the hyperelliptic curve (\ref{f_SU(N)=0}). 
This curve is a double cover of $\mathbb{C}^*$ 
with the branch points at two ends of each band $I_r$ 
(\ref{bands for SU(N) amoeba}). 
Roots $\beta_r$ of the monic polynomial $Q_N(x)$ in (\ref{f_SU(N)=0}) 
relate with $a_r$ by the conditions 
\begin{eqnarray}
\frac{1}{2\pi i}\oint_{C_r}z d\ln y=a_r\,,  
\label{a_r as period of SW differential}
\end{eqnarray}
where $C_r$ denotes the contour 
encircling $I_r$ anticlockwise on the Riemann sheet. 
The analytic function 
\begin{eqnarray}
\Phi(z)=
\frac{d}{dz}
\Bigl(-z+\frac{2}{NR}\ln y\Bigr)
\label{Phi SU(N)}
\end{eqnarray}
satisfies all the above requirements 
including  
(\ref{condition 1 for Phi SU(N)}), 
(\ref{condition 2 for Phi SU(N)}) 
and 
(\ref{condition 3 for Phi SU(N)}). 
Therefore we obtain  
\begin{eqnarray}
\frac{d\rho_{\star}^{SU(N)}(u)}{du}
=\mp \frac{1}{N\pi}\, 
\Im \,\frac{d\ln y}{dz}(u\pm i0)\,. 
\label{SU(N)drho}
\end{eqnarray}

The above solution can be expressed by using 
the Ronkin function of $f_{SU(N)}$. 
By comparing (\ref{SU(N)drho}) with (\ref{d_uN_fSU(N)}), 
we find  
\begin{eqnarray}
\rho_{\star}^{SU(N)}(u)=
\frac{1}{N}
\frac{\partial N_{f_{SU(N)}}(u,0)}{\partial u}\,. 
\label{dN_fSU(N)=rho_SU(N)}
\end{eqnarray}
This means 
\begin{eqnarray}
N_{f_{SU(N)}}(u,0)=
N \int_{-\infty}^u 
d\underline{u}\,\, 
\rho_{\star}^{SU(N)}(\underline{u})
+ \mbox{const.}
\label{N_fSU(N)=integrated rho_SU(N)}
\end{eqnarray}
Let $P_{\star}^{SU(N)}(u)$ be the limit shape 
of the Young diagrams obtained from $\rho_{\star}^{SU(N)}(u)$ 
by using the relation (\ref{f(u|lambda)}). 
We can write (\ref{N_fSU(N)=integrated rho_SU(N)}) as follows. 
\begin{eqnarray}
N_{f_{SU(N)}}(u,0)=
\frac{N}{2}
\Bigl(P_{\star}^{SU(N)}(u)+u\Bigr)
+ \mbox{const.}
\label{N_fSU(N)=P_SU(N)}
\end{eqnarray}

The variational problem formulated 
in subsection \ref{subsubsection:SU(N) variation problem} 
could be understood as a variant of that 
considered in \cite{Nekrasov-Okounkov},  
where the variational problem is addressed 
to the dual partition function 
for the purpose of proving 
that the thermodynamic limit 
realizes the Seiberg-Witten geometry 
of the prepotential theory. 
However,  
that requires the Legendre transformation 
to reproduce the thermodynamic limit 
and is likely to become a detour to find 
a connection with the amoeba and the Ronkin function,   
while our treatment is straightforward 
to see the relation.

\section{Tropical geometry and crystal}
\label{section:tropical geometry}

So far, 
we have seen that the Ronkin function of $f_{SU(N)}$ 
relates with the limit shape $P^{SU(N)}_{\star}$ 
as (\ref{N_fSU(N)=P_SU(N)}) 
under the identification (\ref{a_r as period of SW differential}). 
We expect that such a relation between the Ronkin functions 
and the gauge instanton countings 
is not merely a coincidence but persists further. 
To support this, we describe a certain degeneration 
of the amoebas and the Ronkin functions, 
and provide an interpretation 
from the viewpoint of statistical models.

The parameter $R$ is identified with 
the radius of the circle in the fifth dimension 
of the gauge theories. 
If one keeps $R\Lambda$ fixed, 
one can interpret $R$ 
as the inverse temperature of the statistical models, 
$R=1/T$, where $T$ denotes the temperature.  
This means that 
the large radius limit of the gauge theories 
corresponds to the low temperature limit of the statistical models. 
As the temperature approaches to zero, 
the statistical models get to freeze to the ground states. 
They are determined by the charges.  
Crystals are complements of the ground states in the octant  
and have an interpretation as gravitational quantum foams 
of the corresponding local Calabi-Yau threefolds \cite{MNNT}. 
In this section, we argue that the low temperature limits  
are degenerations of the amoebas 
known as tropical geometry \cite{Mikhalkin,RGST,Viro}.

The max-plus algebra or tropical semiring 
$(\mathbb{R},\oplus,\odot )$ is defined by 
\begin{eqnarray}
u \oplus v=\max(u,v)\,, 
\hspace{10mm}
u \odot v=u+v\,. 
\label{max-plus algebra}
\end{eqnarray}
The tropical semiring is idempotent, as follows from 
$u \oplus u =u$. If one uses $\oplus$ for addition 
and $\odot$ for multiplication, a tropical polynomial 
in two variables is defined as 
\begin{eqnarray}
``\,\,\sum_{i,j}a_{ij}u^iv^j\,\,"=
\max_{i,j}
\bigl(a_{ij}+iu+jv\bigr)\,.  
\label{trop polynomial}
\end{eqnarray}
The quotation mark in the above is used 
to distinguish the tropical operations from the standard ones. 
The tropical polynomials are piecewise linear functions 
and are responsible for some piecewise linear real geometry. 
Surprisingly, this tropical geometry can be obtained as 
a certain degeneration of the complex geometry in 
$(\mathbb{C}^*)^2$.

The idea \cite{Viro} 
comes from the so-called Maslov dequantization of 
real positive numbers. 
It is a family of semirings 
$(\mathbb{R},\oplus_R,\odot_R)$ parameterized by $R >0$. 
\begin{eqnarray}
u \oplus_R v 
=\frac{1}{R}\ln 
(e^{Ru}+e^{Rv})\,, 
\hspace{10mm}
u \odot_R v 
=u+v\,. 
\label{semirings_R}
\end{eqnarray}
For each finite $R$, 
the semiring $(\mathbb{R},\oplus_R,\odot_R)$ 
is isomorphic to the standard semiring of 
real positive number by the logarithmic map 
\begin{eqnarray}
\begin{array}{cccc}
\mbox{Log}~: 
&
(\mathbb{R}_{>0},+,\cdot) 
&
\longrightarrow 
&
(\mathbb{R},\oplus_R,\odot_R)
\\
&
x 
&
\longmapsto
& 
\frac{1}{R}\ln x\,.
\end{array}
\label{log map for R_+} 
\end{eqnarray} 
On the other hand,  
the semiring becomes the tropical semiring 
at the limit $R \rightarrow \infty$.  
\begin{eqnarray}
u \oplus_{\infty} v 
=u \oplus v\,, 
\hspace{10mm}
u \odot_{\infty} v 
=u\odot v\,. 
\label{semirings_infty}
\end{eqnarray} 
By using the terminology of deformation quantization, 
this means that  
the classical semiring of real positive number 
is a quantized version of the tropical semiring.  
In other words, 
the Maslov dequantization says that 
the tropical semiring is a classical counterpart 
of the standard semiring of real positive number.

Anticipated from the definition of amoeba, 
the above dequantization deformation has a counterpart 
in amoebas. 
For instance, 
consider a Laurent polynomial 
$f(x,y)=\sum_{i,j}b_{ij}x^iy^j$, 
where all the $b_{ij}$'s are real positive numbers 
and supposed to be $b_{ij}=e^{Ra_{ij}}$. 
As $R \rightarrow \infty$, 
the corresponding amoeba $\mathcal{A}_f$ 
tends to a piecewise linear curve 
that is described by the corner set of 
$``\,\sum_{i,j}a_{i,j}u^iv^j\,"$ 
\cite{Mikhalkin, Rullgard}.

\subsection{Tropical limit of $SU(N)$ amoeba}

As in the previous section, 
we take $a_r$ sufficiently separated from each other 
and arrange them in numerical order 
$a_1 < \cdots < a_N$. 
They determine the Laurent polynomial $f_{SU(N)}(x,y)$, 
given in (\ref{f_SU(N)}), by using the conditions 
(\ref{a_r as period of SW differential}). 
We first study degeneration of the corresponding amoeba 
as $R \rightarrow \infty$. 
The limit is taken with keeping $a_1, \cdots, a_N$ 
and $R\Lambda$ fixed. 
This implies that $\Lambda$ behaves as $O(1/R)$.  
The thermal fluctuations of the statistical model 
are suppressed at the low temperature, and 
each band $I_r$ of the density shrinks to the point $a_r$. 
This means that, when $R$ becomes very large, 
the parameters of the Laurent polynomial (\ref{f_SU(N)}) 
are $\beta_r \approx e^{Ra_r}$.  
We examine how the part of the amoeba 
within the strip $a_r<u<a_{r+1}$ 
behaves under the above limit. 
The parts extending to the infinities can be included, 
by considering the cases of $r=0,\,N$ 
putting  $a_0=-\infty$ and $a_{N+1}=+\infty$. 
It is seen that 
both $e^{Ra_s}x^{-1}$  ($1 \leq s \leq r$) 
and 
$e^{-Ra_s}x$  ($r+1 \leq s \leq N$) vanish 
as $R \rightarrow \infty$,   
if $x$ satisfies $a_r < \frac{1}{R}\ln |x| <a_{r+1}$. 
By noting this,  
we can describe effectively the zero locus 
which is mapped to within the strip as 
\begin{eqnarray}
y^{\pm 1}-
\frac{(-)^{N-r}}{(R\Lambda)^N}
e^{R\sum_{s=r+1}^Na_s}x^r=0\,, 
\end{eqnarray}
where $y^{\pm 1}$ are chosen so that $y$ 
for $|y|>1$, and $y^{-1}$ for $|y|<1$.  
This shows that the amoeba within the strip 
$a_r<u<a_{r+1}$ degenerates to two linear pieces 
$v=\pm(ru+\sum_{s=r+1}^Na_s)$ at the limit.

\begin{figure}[ht]
\begin{center}
\begin{minipage}{0.25\linewidth}
\begin{center}
\includegraphics[width=\linewidth]{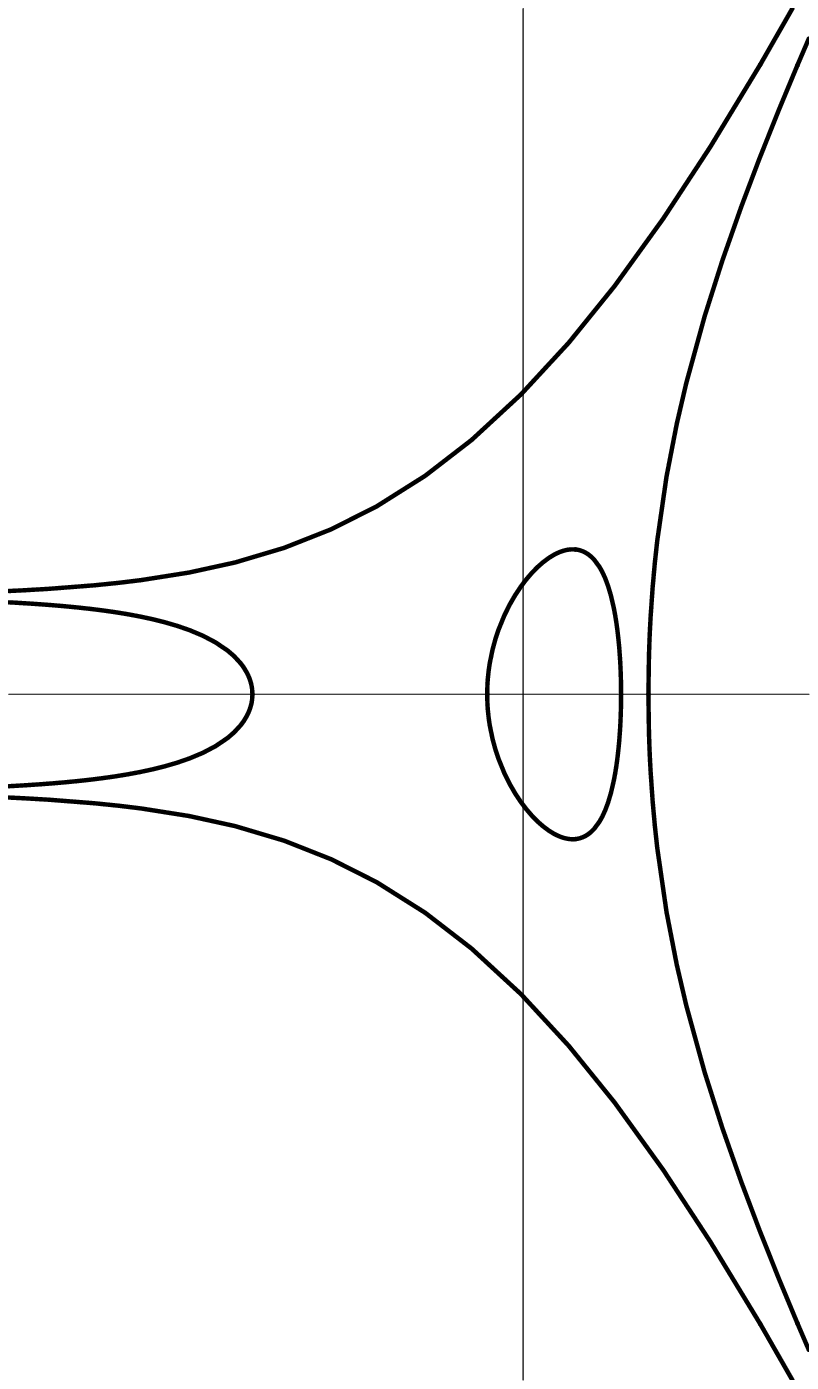}
\end{center}
\end{minipage}
\begin{minipage}{0.1\linewidth}
\begin{center}
{\LARGE $\Longrightarrow$}
\end{center}
\end{minipage}
\begin{minipage}{0.25\linewidth}
\begin{center}
\includegraphics[width=\linewidth]{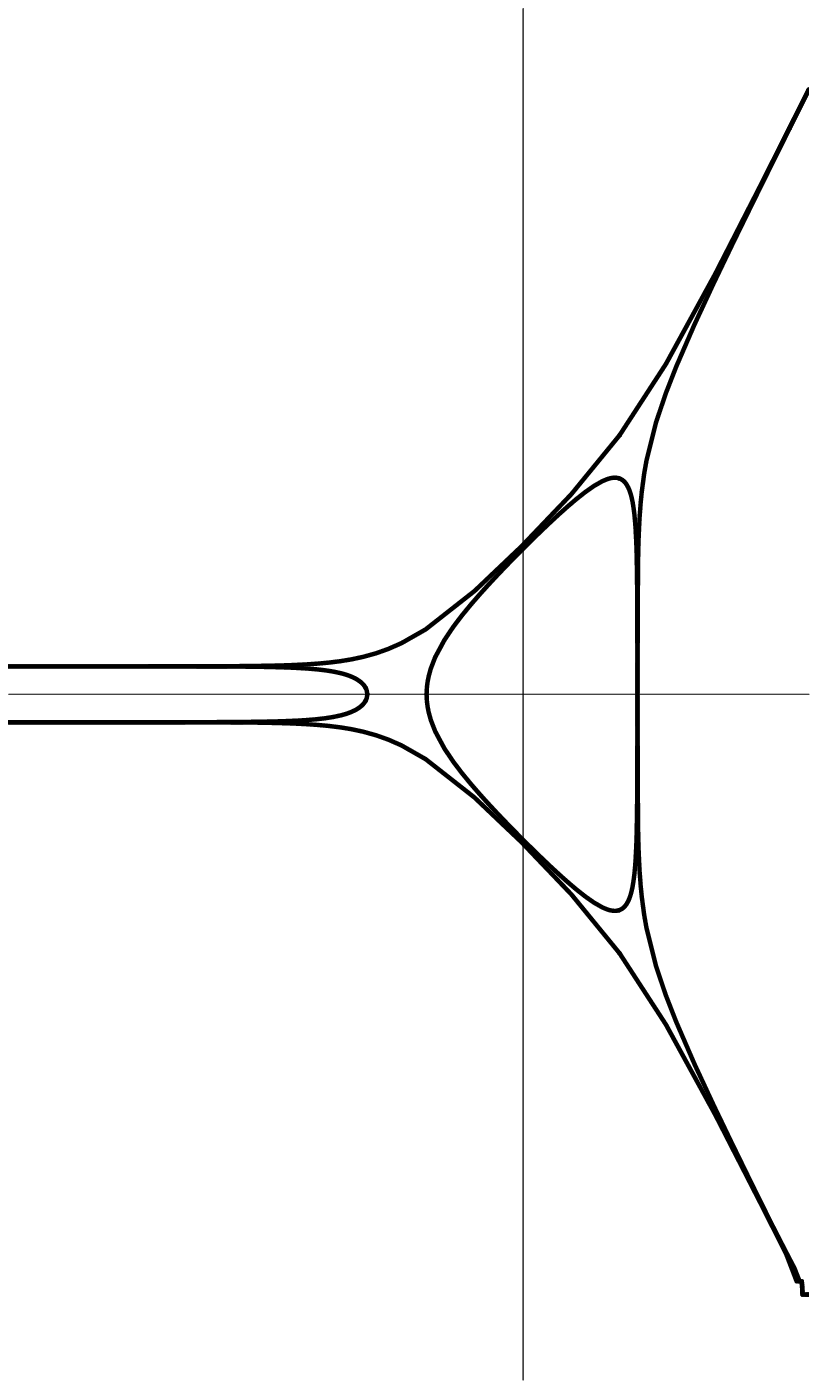}
\end{center}
\end{minipage} 
\begin{minipage}{0.1\linewidth}
\begin{center}
{\LARGE $\Longrightarrow$}
\end{center}
\end{minipage} 
\begin{minipage}{0.25\linewidth}
\begin{center}
\includegraphics[width=\linewidth]{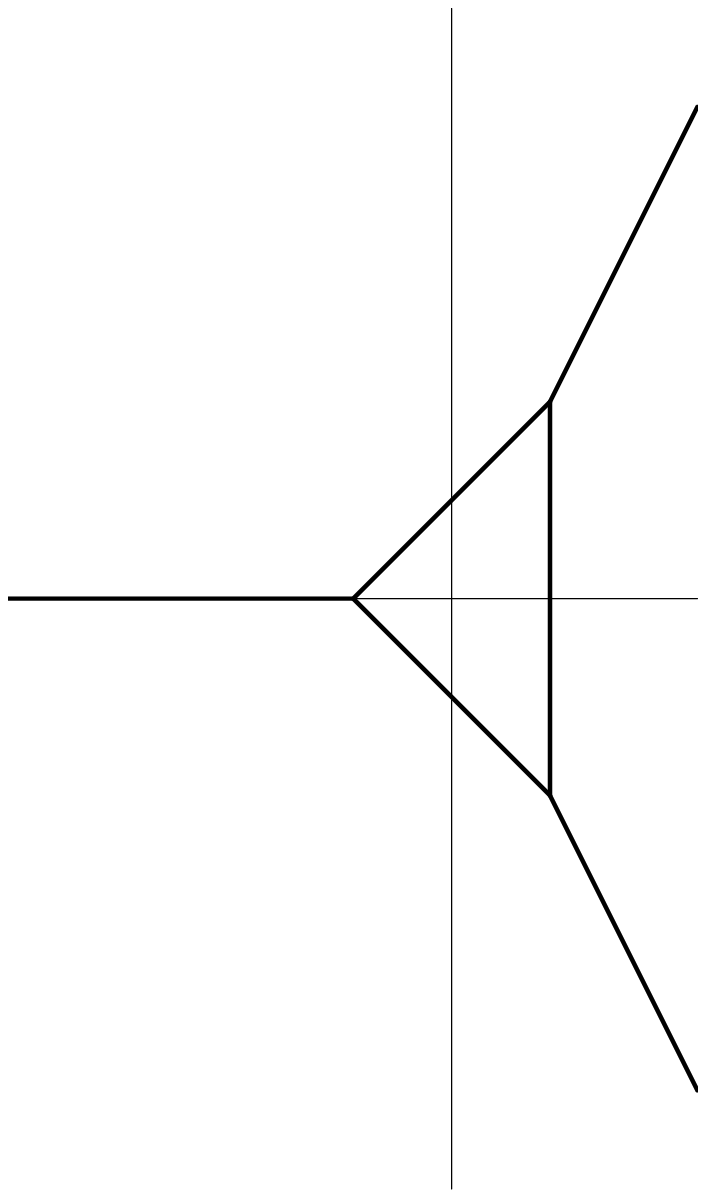}
\end{center}
\end{minipage}
\end{center}
\caption{\it Tropical degeneration of the the amoeba of $f_{SU(2)}$.}
\label{fig:tropical degeneration of amoeba}
\end{figure}

In addition to the above, 
taking account of the convexity of the connected components 
of the amoeba complement, 
the amoeba degenerates to a piecewise linear curve 
that is described by  
\begin{eqnarray}
v=\pm\, \max_{0 \leq r \leq N}
\Bigl(ru+\sum_{s=r+1}^Na_s \Bigr)  
\label{SU(N) tropical curve 1}
\end{eqnarray}
and
\begin{eqnarray}
\bigcup_{r=1}^N\,
\Bigl\{  
(a_r,v)\,\in \mathbb{R}^2\,\,; \hspace{3mm}
|v| \leq ra_r+\sum_{s=r+1}^Na_s 
\Bigr\}\,.
\label{SU(N) tropical curve 2} 
\end{eqnarray}
An example of the degeneration is depicted 
in Figure \ref{fig:tropical degeneration of amoeba}. 

We turn to consider the behavior of the Ronkin function 
at the tropical limit.  
Over the amoeba complement, 
the Ronkin function is 
the piecewise linear function $S_{f_{SU(N)}}$. 
At the tropical limit, 
where the amoeba degenerates to the above curve, 
these two functions coincide all over $\mathbb{R}^2$.  
Let us write the piecewise linear function thus obtained as 
\begin{eqnarray}
S_{\infty}(u,v)
= \lim_{R \rightarrow \infty}\,
N_{f_{SU(N)}}(u,v)\,. 
\label{s_infty definition}
\end{eqnarray}
We can reproduce this $S_{\infty}$ from 
the piecewise linear curve 
(\ref{SU(N) tropical curve 1}) and (\ref{SU(N) tropical curve 2}), 
except an addition of a constant to the function. 
This is because the piecewise linear curve 
is the corner set of $S_{\infty}$ 
and 
the gradient $\nabla S_{\infty}$ equals to  
the corresponding lattice point of the Newton polygon 
on each connected component divided by the curve. 
The constant may be found, for instance,   
by considering the Ronkin function over $E_{\alpha}$, 
where $\alpha$ is a vertex of the Newton polygon. 
In this manner we obtain 
\begin{eqnarray}
S_{\infty}(u,v)=
\max_{\alpha \in \Delta_{f_{SU(N)}} \cap\,\, \mathbb{Z}^2}
\Bigl(  
-d_{\alpha}+\langle (u,v),\alpha \rangle 
\Bigr)\,, 
\label{s_infty}
\end{eqnarray}
where 
\begin{eqnarray}
\begin{array}{lr}
d_{(i,0)}=\sum_{r=1}^ia_r 
& (1 \leq i \leq N)\,, 
\\ 
d_{(0,0)}=d_{(0,\pm 1)}=0\,. 
& 
\end{array}
\label{d_alpha}
\end{eqnarray}

\begin{figure}[ht]
\begin{center}
\includegraphics[width=0.8\linewidth]{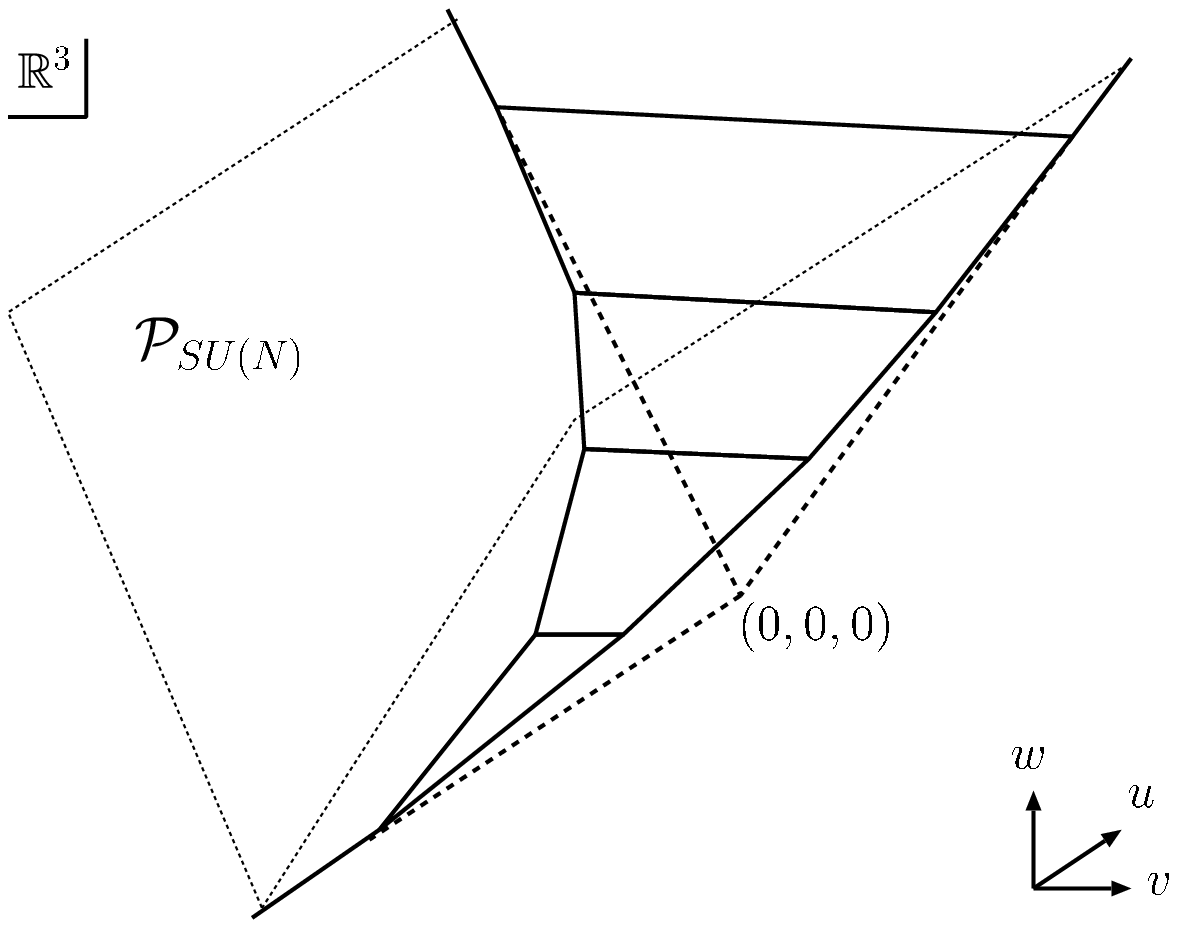}
\end{center}
\caption{\it The three-dimensional polyhedron $\mathcal{P}_{SU(N)}$.}
\label{fig:SU(N) polyhedra}
\end{figure}

The set of $(u,v,S_{\infty}(u,v))$ becomes 
a facet of a three-dimensional polyhedron. 
We associate a three-dimensional lattice element 
with each integer lattice point of the Newton polygon by 
\begin{eqnarray}
v_{\alpha}=(-\alpha, 1)\,\,\, 
\in \mathbb{Z}^3\,,
\label{v_alpha}
\end{eqnarray}
where 
$\alpha \in \Delta_{f_{SU(N)}} \cap \mathbb{Z}^2$. 
By using these $v_{\alpha}$, 
we can describe the polyhedron as 
follows. 
\begin{eqnarray}
\mathcal{P}_{SU(N)}=
\Bigl\{ 
(u,v,w) \in \mathbb{R}^3 
\hspace{1mm};\,\,
\langle (u,v,w), v_{\alpha} \rangle 
+d_{\alpha} \geq 0 
\hspace{4mm}
\forall \alpha  \in \Delta_{f_{SU(N)}} \cap \mathbb{Z}^2 
\Bigr\}\,, 
\label{P}
\end{eqnarray}
where $\langle \,,\,\rangle$ denotes the standard 
inner product on $\mathbb{R}^3$. 
By comparing the above with (\ref{s_infty}), 
we see that the facet of this $\mathcal{P}_{SU(N)}$ 
is actually given by $S_{\infty}$. 
See Figure \ref{fig:SU(N) polyhedra}. 

\subsection{Tropical geometry and quantum foam of local geometry}

The polyhedron $\mathcal{P}_{SU(N)}$ has 
an interpretation in the local geometry 
and accords with the quantum foam picture 
\cite{quantum foam},\cite{MNNT}.  
We introduce three-dimensional cones 
$\sigma_0,\sigma_1,\cdots,\sigma_{2N-1}$ by 
\begin{eqnarray}
\begin{array}{l}
\sigma_{2i}=
\mathbb{R}_{\geq 0}v_{(i,0)}
+\mathbb{R}_{\geq 0}v_{(i+1,0)}
+\mathbb{R}_{\geq 0}v_{(0,-1)}\,, \\
\sigma_{2i+1}=
\mathbb{R}_{\geq 0}v_{(i,0)}
+\mathbb{R}_{\geq 0}v_{(i+1,0)}
+\mathbb{R}_{\geq 0}v_{(0,1)}\,.  
\end{array}
\label{sigma}
\end{eqnarray}
All these cones and their faces constitute 
the fan that describes the local $SU(N)$ geometry $X$.  
Two-dimensional cones of the fan determine 
two cycles of the geometry as closed subvarieties 
that are invariant under the torus action.  
Cones that are generated by $v_{(k,0)}$ and $v_{(0,-1)}$ 
determine vanishing cycles in the fibred ALE space, 
where $1 \leq k \leq N-1$. 
Cone generated by $v_{(1,0)}$ and $v_{(0,0)}$ 
determines the base $\mathbb{P}^1$.

The polyhedron (\ref{P}), 
more precisely, that rescaled by $1/\hbar$,  
emerges naturally 
when quantizations of the local geometry $X$ 
equipped with a K$\ddot{\mbox{a}}$hler two form $\omega$ on it, 
are considered. 
For the consistent quantizations, 
this $\omega$ needs to be quantized. 
Topological $A$-model strings suggest 
the rule that $\omega$ is quantized in the unit of 
string coupling constant $g_{st}$ \cite{quantum foam}. 
\begin{eqnarray}
\frac{1}{g_{st}}[\omega] 
\in 
H^2(X,\mathbb{Z})\,. 
\label{quantized kahler form}
\end{eqnarray} 
This means that the K$\ddot{\mbox{a}}$hler volumes 
of two cycles become integral.
\begin{eqnarray}
T_b=
\frac{1}{g_{st}}\int_{V_b}\omega  
\hspace{2mm}
\in \mathbb{Z}_{\geq 0}\,, 
\hspace{8mm}
T_k=
\frac{1}{g_{st}}\int_{V_k}\omega
\hspace{2mm} 
\in \mathbb{Z}_{\geq 0}
\hspace{6mm}
(k=1,\cdots,N-1)\,, 
\label{kahler parameters T}
\end{eqnarray}
where 
$V_b$ and $V_k$  
denote respectively the base $\mathbb{P}^1$ and 
the vanishing cycles in the fibre that 
correspond to the cones 
generated by $v_{(k,0)}$ and $v_{(0,-1)}$.  
Relation between quantizations of the local geometry  
and the statistical models is found in \cite{MNNT}. 
In particular,  
the integral parameters 
(\ref{kahler parameters T}) 
are converted to 
\begin{eqnarray}
g_{st}=R\hbar\,,
\hspace{8mm}
T_b=-\frac{1}{g_{st}}\log (R\Lambda)^{2N}\,,
\hspace{8mm}
T_k=p_{k+1}-p_k\,,
\label{geometry and gauge theory parameters}
\end{eqnarray}
where $p_r$ are integers arranged in numerical order 
$p_1 \leq p_2 \leq \cdots \leq p_N$.  
We will associate these integers with partitions subsequently. 
We will also impose the $SU(N)$ condition on the integers  
and slightly restrict allowed values of the parameters 
so that $\sum_{k=1}^{N-1}kT_k \in N \mathbb{Z}_{\geq 0}$.

When the limit $R \rightarrow \infty$ 
is taken with keeping $R\Lambda$ fixed, 
it makes $T_b$ vanish under the identification 
(\ref{geometry and gauge theory parameters}). 
This implies that 
we can put $T_b=0$ from the beginning, 
for the present purpose. 
Let $\omega$ be such a quantized K$\ddot{\mbox{a}}$hler form.
The geometric quantization of $(X,\omega)$ requires, first of all, 
a holomorphic line bundle $L$ on $X$ which the first Chern class 
equals $c_1(L)=[\omega]/g_{st}$. 
We may take such a line bundle as  
$L=\mathcal{O}(D)$, where $D$ is a certain toric divisor.
Rays or edges of the fan determine closed subvarieties of 
codimension one that are invariant under the torus action. 
The divisor is a formal sum of these subvarieties with 
integral coefficients. 
\begin{eqnarray}
D=\sum_{\alpha}p_{\alpha}D_{\alpha}\,,
\label{D}
\end{eqnarray}
where $D_{\alpha}$ is the subvariety that corresponds to 
the ray generated by $v_{\alpha}$. 
The integral parameters (\ref{kahler parameters T}) 
relate with the above $p_{\alpha}$ 
by using $c_1(\mathcal{O}(D))=[\omega]/g_{st}$. 
The identification (\ref{geometry and gauge theory parameters}) 
allows us to express the coefficients $p_{\alpha}$ by the charges 
as follows. 
\begin{eqnarray}
\begin{array}{lc}
p_{(i,0)}=\sum_{r=1}^ip_r 
&
(1\leq i\leq N)\,,
\\
p_{(0,0)}=p_{(0,\pm 1)}=0 
& \,.
\end{array}
\label{p_alpha}
\end{eqnarray}

Physical states in the geometric quantization   
appear as the global sections of $\mathcal{O}(D)$. 
Thanks to that $X$ is a toric variety, 
these states are labeled 
by integer lattice points in a convex polyhedron 
determined by $D$ \cite{Fulton}. 
\begin{eqnarray}
\mathcal{P}(X,D)=
\Bigl\{ 
(u,v,w) \in \mathbb{R}^3 
\hspace{1mm};\,\,
\langle (u,v,w),v_{\alpha} \rangle 
+p_{\alpha} \geq 0 
\hspace{4mm}
\forall 
\alpha 
\Bigr\}\,. 
\label{P(X,D)}
\end{eqnarray}
Thus space of the physical states has a basis $\chi_m$, 
where $m$ runs over $\mathcal{P}(X,D) \cap \mathbb{Z}^3$. 
Each $\chi_m$ is interpreted as a quantum of $(X,\omega)$.   
The charges $p_r$ are scaled to $a_r$ by $p_r=a_r/\hbar$  
at the thermodynamic limit or the semiclassical limit. 
Comparing (\ref{p_alpha}) with (\ref{d_alpha}),  
we find $p_{\alpha}=d_{\alpha}/\hbar$. 
This means that two polyhedra $\mathcal{P}(X,D)$ and 
$\mathcal{P}_{SU(N)}$ are similar.   
They relate with each other by the similarity transformation. 
\begin{eqnarray}
\mathcal{P}(X,D)=\frac{1}{\hbar}\mathcal{P}_{SU(N)}\,. 
\label{P(X,D) sim P_SU(N)}
\end{eqnarray}

The number of the physical states is infinite since the cardinality 
of $\mathcal{P}(X,D) \cap \mathbb{Z}^3$ is infinite. 
In order to manipulate such an infinity, 
another polyhedron that originates 
from a local singular geometry is introduced \cite{MNNT}.  
\begin{eqnarray}
\mathcal{P}_{sing}=
\Bigl\{ 
(u,v,w) \in \mathbb{R}^3 
\hspace{1mm};\,\,
\langle (u,v,w),v_{\alpha} \rangle \geq 0 
\hspace{4mm}
\alpha=(0,\pm 1),(N,0),(0,0). 
\Bigr\}\,. 
\label{P_sing}
\end{eqnarray}
The underlying singular geometry 
is a $\mathbb{Z}_N$ orbifold of 
$\mathcal{O}\oplus \mathcal{O}(-2)\rightarrow \mathbb{P}^1$  
and is described by the fan consisting of three-dimensional cones 
$\tau_1, \tau_2$ and their faces. 
\begin{eqnarray}
\begin{array}{l}
\tau_{1}=
\mathbb{R}_{\geq 0}v_{(0,0)}
+\mathbb{R}_{\geq 0}v_{(N,0)}
+\mathbb{R}_{\geq 0}v_{(0,-1)}\,, \\
\tau_{2}=
\mathbb{R}_{\geq 0}v_{(0,0)}
+\mathbb{R}_{\geq 0}v_{(N,0)}
+\mathbb{R}_{\geq 0}v_{(0,1)}\,.  
\end{array}
\label{tau}
\end{eqnarray} 
\begin{figure}[ht]
\begin{center}
\includegraphics[width=0.8\linewidth]{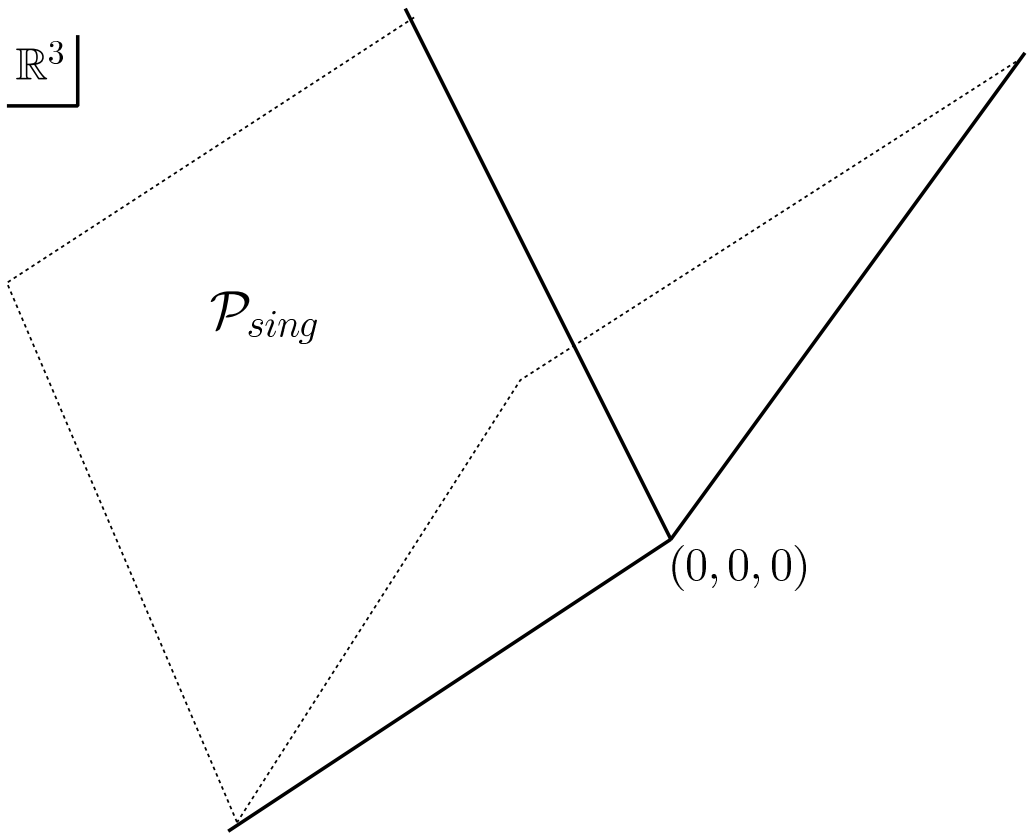}
\end{center}
\caption{\it The three-dimensional polyhedron $\mathcal{P}_{sing}$.}
\label{fig:P_sing}
\end{figure}

The above polyhedron includes $\mathcal{P}(X,D)$ 
as an unbounded subset.
In \cite{MNNT}, 
the number of the physical states is regularized and is prescribed 
to be the cardinality of $\mathcal{P}^c(X,D) \cap \mathbb{Z}^3$, 
where $\mathcal{P}^c(X,D)$ denotes the complement 
$\mathcal{P}_{sing}\setminus \mathcal{P}(X,D)$. 
Since it becomes bounded,  
this relative counting gives a finite answer. 
When the charges become very large, the cardinality is 
approximated by the volume measured by the volume element $dudvdw$. 
Therefore, the number of the physical states becomes 
$\approx \mbox{Vol}(\mathcal{P}^c(X,D))$.  
The volume can be computed as 
\begin{eqnarray}
\mbox{Vol}(\mathcal{P}^c(X,D))=
\frac{1}{6}\sum_{r>s}^N(p_r-p_s)^3
+\frac{N}{6}\sum_{r=1}^Np_r^3\,, 
\label{Vol P_c(D)}
\end{eqnarray}
which turns to equal $-D^3/3!$, where 
$D^3$ denotes the self-intersection number obtained 
by intersecting $D$ with itself three times.

The counterpart of the singular geometry 
is a Laurent polynomial consisting of monomials $1,x^N,y,y^{-1}$ 
and reflecting that the geometry is the $\mathbb{Z}_N$ orbifold of
$\mathcal{O}\oplus \mathcal{O}(-2)\rightarrow \mathbb{P}^1$. 
As such, we will take 
\begin{eqnarray}
f^{base}_{SU(N)}(x,y)=x^N-1-(R\Lambda)^{2N}
-(R\Lambda)^N(y+y^{-1})\,. 
\label{f_base_SU(N)}
\end{eqnarray}
The amoeba of $f^{base}_{SU(N)}$ has four tentacles 
which asymptote to the straight lines (\ref{SU(N) tentackles}) 
but there appears no hole.  
The tropical limit of the Ronkin function 
$N_{f^{base}_{SU(N)}}$ becomes 
\begin{eqnarray}
S^{base}_{\infty}(u,v) 
=
\max_{\alpha=(0,0),(N.0),(0,\pm 1)}
\Bigl(\langle (u,v),\alpha \rangle\Bigr) \,. 
\label{s_base_infty}
\end{eqnarray}
This piecewise linear function describes 
the facet of $\mathcal{P}_{sing}$. 
The counterpart of $\mathcal{P}^c(X,D)$ is 
the complement 
$\mathcal{P}^c_{SU(N)}=\mathcal{P}_{sing}
\setminus \mathcal{P}_{SU(N)}$. 
This is similar to $\mathcal{P}^c(X,D)$. 
\begin{eqnarray}
\mathcal{P}^c(X,D)=\frac{1}{\hbar}\mathcal{P}^c_{SU(N)}\,. 
\label{P_c(D) sim P_cSU(N)}
\end{eqnarray}
\begin{figure}[ht]
\begin{center}
\includegraphics[width=0.7\linewidth]{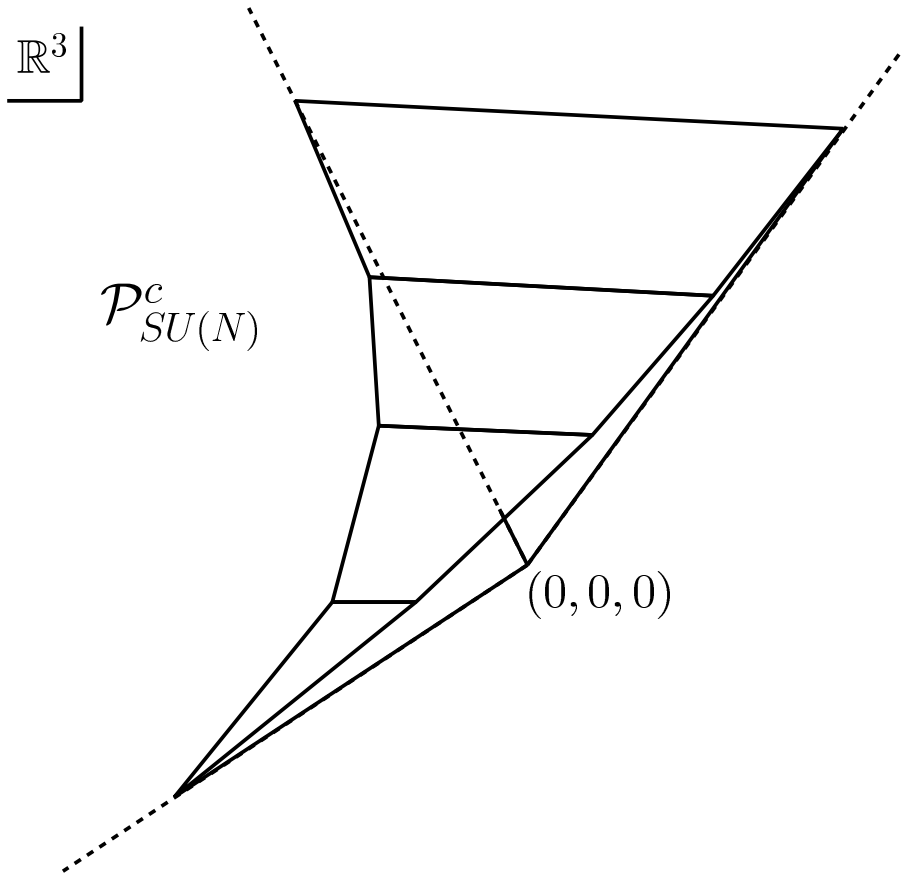}
\end{center}
\caption{\it The complement 
$\mathcal{P}^c_{SU(N)}=\mathcal{P}_{sing}
\setminus \mathcal{P}_{SU(N)}$.}
\label{fig:Pc_SU(N)}
\end{figure}

\subsection{Tropical geometry and crystal}

\begin{figure}[ht]
\begin{center}
\includegraphics[width=\linewidth]{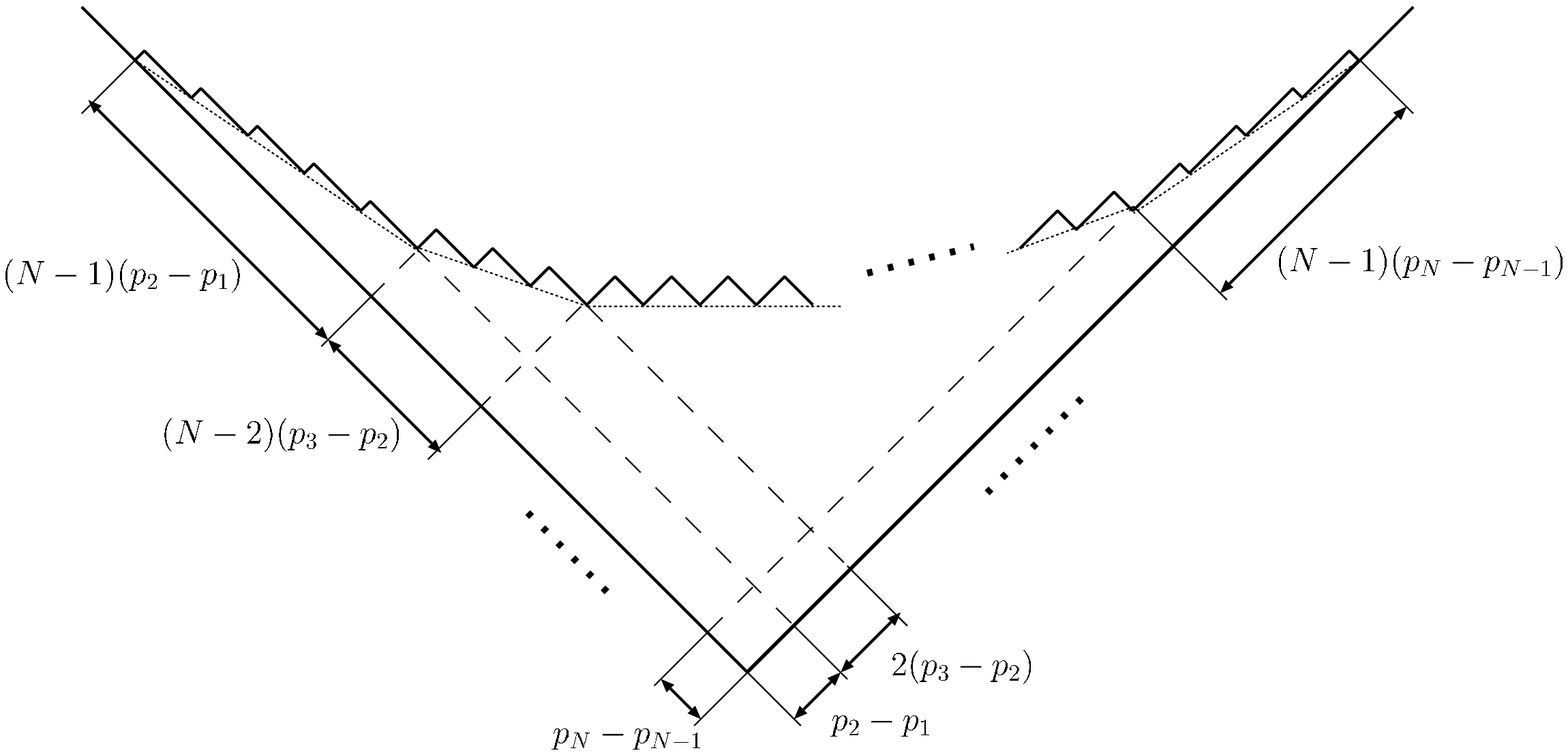}
\end{center}
\caption{\it $\mu_{N\mbox{\footnotesize{-core}}}$.}
\label{fig:N-core}
\end{figure}

By using the correspondence (\ref{1:N by Maya diagrams}),  
$N$ charged empty partitions describe a specific partition, 
which is called $N$ core. 
Let us consider $N$ charged empty partitions 
$\{(\emptyset,p_r)\}_{r=1}^N$ 
that charges are of order $\hbar^{-1}$ 
and satisfy the $SU(N)$ condition (\ref{SU(N) condition}).
The corresponding $N$ core is 
denoted by $\mu_{N\mbox{\footnotesize{-core}}}$. 
See Figure \ref{fig:N-core}. 
Asymptotics of $\mu_{N\mbox{\footnotesize{-core}}}$ 
at the limit $\hbar \rightarrow 0$ 
is obtained by scaling each charged empty partition according to 
(\ref{u(s|lambda) for r-th partition}) and (\ref{a_r}). 
In particular, the charges are scaled by $p_r=a_r/\hbar$. 
The scaled density of $\mu_{N\mbox{\footnotesize{-core}}}$ 
becomes 
\begin{eqnarray}
\rho_{N\mbox{\footnotesize{-core}}}(u)=
\rho(u|\mu_{N\mbox{\footnotesize{-core}}})=
\frac{1}{N}
\sum_{r=1}^N\theta(u-a_r)\,. 
\label{rho(u|N-core)}
\end{eqnarray}
Each charged empty partition contributes to the above 
as a step function. 
Comparing (\ref{rho(u|N-core)}) with (\ref{s_infty}), 
one finds that this $\rho_{N\mbox{\footnotesize{-core}}}$ 
is written by using the piecewise linear function $S_{\infty}$ as 
\begin{eqnarray}
\rho_{N\mbox{\footnotesize{-core}}}(u)=
\frac{1}{N}
\frac{\partial S_{\infty}(u,0)}{\partial u}
\,. 
\label{S_infty vs rho_N-core}
\end{eqnarray}

The above expression suggests a certain relation between 
the $N$ core and the tropical limit of the $SU(N)$ amoeba. 
To see this, 
we first compute the energy of $\mu_{N\mbox{\footnotesize{-core}}}$. 
The charges are arranged $p_1 \leq \cdots \leq p_N$. 
We evaluate the energy functional (\ref{energy functional 2}) 
at $\rho=\rho_{N\mbox{\footnotesize{-core}}}$. 
The function $\gamma(u;\Lambda;R)$, 
that is defined by the conditions 
(\ref{condition 1 for gamma}) and (\ref{condition 2 for gamma}), 
has the expression 
\begin{eqnarray}
\gamma(u;\Lambda;R)=
\frac{R}{6}u^3-\log (R\Lambda) u^2
-\frac{\pi^2}{3R}u
+\frac{2}{R^2}
\left(\zeta(3)-Li_3(e^{-Ru})\right)\,. 
\label{gamma(u)}
\end{eqnarray}
By using this, the energy of 
$\mu_{N\mbox{\footnotesize{-core}}}$ becomes 
\begin{eqnarray}
E[\rho_{N\mbox{\footnotesize{-core}}}] 
&=&
\frac{R}{6}
\left\{\sum_{r>s}^N(a_r-a_s)^3
        +N\sum_{r=1}^Na_r^3 \right\} 
-\log (R\Lambda)\sum_{r>s}^N(a_r-a_s)^2 
\nonumber \\
&&
-\frac{\pi^2}{6R}\sum_{r>s}^N(a_r-a_s)
+\frac{1}{R^2}\sum_{r>s}^N
    \Bigl(\zeta(3)-Li_3(e^{-R(a_r-a_s)})\Bigr)\,. 
\label{E(N core)}
\end{eqnarray}
This is the perturbative prepotential of 
five dimensional $\mathcal{N}=1$ supersymmetric $SU(N)$ Yang-Mills 
on $\mathbb{R}^4 \times S^{1}$. 
The first term  of (\ref{E(N core)}) dominates 
when the limit $R \rightarrow \infty$ is taken. 
Therefore, 
taking account of 
(\ref{Vol P_c(D)}) and (\ref{P_c(D) sim P_cSU(N)}), 
we obtain the following estimate at the low temperature limit. 
\begin{eqnarray}
E[\rho_{N\mbox{\footnotesize{-core}}}] 
\approx 
R \cdot  \mbox{Vol}\Bigl( \mathcal{P}^c_{SU(N)} \Bigr)\,. 
\label{E(N core) at T=0}
\end{eqnarray}

A set $\mathcal{M}(p_1,\cdots,p_N)$ denotes 
a set of plane partitions 
that main diagonal partitions are 
$\mu_{N\mbox{\footnotesize{-core}}}$. 
It follows from (\ref{asymptotic BW})
that the energy (\ref{E(N core)}) 
is the free energy of 
random plane partitions restricted within 
$\mathcal{M}(p_1,\cdots,p_N)$. 
\begin{eqnarray}
\sum_{\pi \in \mathcal{M}(p_r)} 
q^{|\pi|}Q^{|\pi(0)|}
=
\exp 
\Bigl\{-\frac{1}{\hbar^2} 
\Bigl(
E[\rho_{N\mbox{\footnotesize{-core}}}] 
+O(\hbar)
\Bigr) 
\Bigr\}\,. 
\end{eqnarray}

The ground state of the above restricted model is 
a plane partition that minimizes $|\pi|$ 
in $\mathcal{M}(p_1,\cdots,p_N)$. 
Such a plane partition is determined uniquely by the charges 
and is denoted by $\pi_{GPP}$. 
Elements of this $\pi_{GPP}$ are expressed 
by using the $N$ core as follows. 
\begin{eqnarray}
\pi_{GPP}(m)_i= 
\mu_{N\mbox{\footnotesize{-core}}\,\,\,i+|m|}\,. 
\label{pi_GPP}
\end{eqnarray} 
The number of the cubes of $\pi_{GPP}$ becomes 
\begin{eqnarray}
|\pi_{GPP}|&=&
\sum_{i=1}^{\infty}~^t\mu_{N\mbox{\footnotesize{-core}}\,\,\,i}^2
\nonumber \\
&=&
\frac{N}{6}\sum_{r>s}^N(p_r-p_s)^3
+\frac{N^2}{6}\sum_{r=1}^Np_r^3 
+O(p^2)\,. 
\label{|pi_GPP|}
\end{eqnarray}   
At the limit $\hbar \rightarrow 0$, 
by taking account of 
(\ref{Vol P_c(D)}) and (\ref{P_c(D) sim P_cSU(N)}), 
we obtain 
\begin{eqnarray}
\lim_{\hbar \rightarrow 0}\,\, \hbar^3\,|\pi_{GPP}|
=
N \cdot  \mbox{Vol}\Bigl( \mathcal{P}^c_{SU(N)} \Bigr)\,. 
\label{|pi_GPP| at thermo limit}
\end{eqnarray}

The plane partition $\pi_{GPP}$ becomes  
the ground state of the original random plane partitions. 
At the low temperature limit the statistical model 
freezes to $\pi_{GPP}$. 
There is a description \cite{MNNT} 
that the complement $\mathcal{P}^c_{SU(N)}$ is identified 
with the plane partition $\pi_{GPP}$ at the thermodynamic limit. 
Consider the linear transformation $A$ in $\mathbb{R}^3$. 
\begin{eqnarray}
A=
\left(\begin{array}{ccc}
           0 & -1  & 1 \\
           0 &  1  & 1 \\
          -N &  0  & 1 
\end{array}\right)\,. 
\label{A}
\end{eqnarray} 
The image of the polyhedron $\mathcal{P}_{sing}$ is the positive 
octant $\mathrm{O}=\{(u,v,w) \in \mathbb{R}^3;\,\,u,v,w \geq 0\}$. 
Thus the image of the complement $\mathcal{P}^c_{SU(N)}$ is 
\begin{eqnarray}
A(\mathcal{P}^c_{SU(N)})
=\mathrm{O} \setminus A(\mathcal{P}_{SU(N)})\,. 
\end{eqnarray}

This $A(\mathcal{P}^c_{SU(N)})$ can be identified with 
the shape of the plane partition $\pi_{GPP}$.  
We assemble a plane partition $\pi$ 
in the positive octant $\mathrm{O}$,  
by a rule slightly different from the standard one.  
We substitute unit cubes by rectangular solids of the size 
$\sqrt{2} \times \sqrt{2} \times 1$,
and number $\pi/4$-rotated squares 
in the first quadrant of the $(u,v)$-plane 
as in Figure \ref{fig:(i,j)-squares in quadrant}. 
We stack $\pi_{ij}$ pieces of the rectangular solid 
vertically on each $(i,j)$-square 
of the first quadrant. 
As the result, the plane partition $\pi_{GPP}$ 
is assembled in the positive octant $\mathrm{O}$. 
It is straightforward to see that 
this $\pi_{GPP}$ scales to $A(\mathcal{P}^c_{SU(N)})$ 
at the limit  $\hbar \rightarrow 0$. 
\begin{figure}[ht]
\begin{center}
\includegraphics[width=0.6\linewidth]{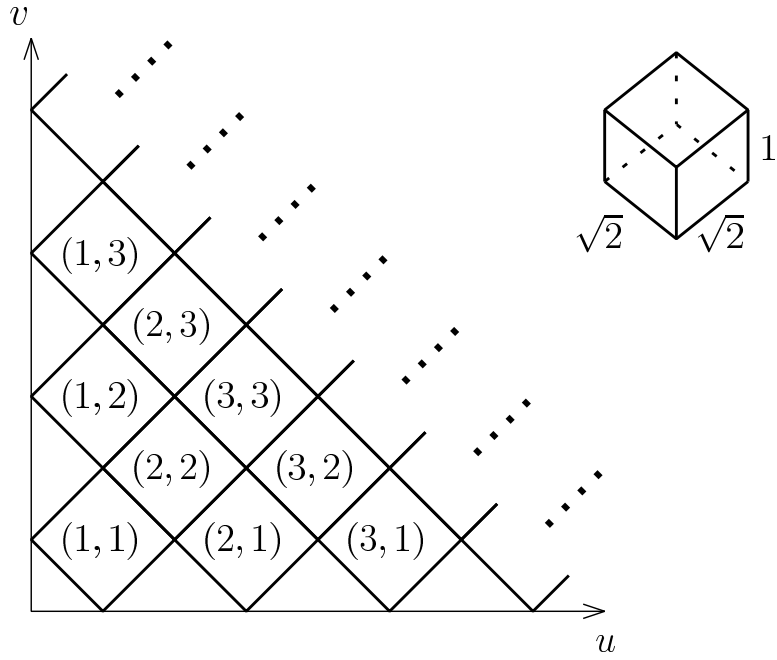}
\end{center}
\caption{\it $\pi/4$-rotated squares 
in the first quadrant of the $(u,v)$-plane.}
\label{fig:(i,j)-squares in quadrant}
\end{figure}

\subsection*{Acknowledgements}
We thank to Y. Noma and T. Tamakoshi for fruitful discussion.   
T. N. benefited from conversation with S. Fujii, Y. Hashimoto, 
M. Jimbo, S. Minabe and D. Yamada. 
T. N. is supported in part by 
Grant-in-Aid for Scientific Research 15540273.

\newpage


\begin{thebibliography}{99}


\bibitem{superstrings}
M.~B.~Green, J.~H.~Schwarz and E.~Witten,
\textit{``Superstring Theory,''}
Cambridge University Press, 1987.\\
J.~Polchinski,
\textit{``String Theory,''}
Cambridge University Press, 1998.



\bibitem{Wess-Bagger}
J.~Wess and J.~Bagger,
\textit{``Supersymmetry and Supergravity,''}
Princeton University Press.



\bibitem{Geometric engineering}
A.~Klemm, W.~Lerche, P.~Mayr, C.~Vafa and N.~P.~Warner,
\textit{``Self-Dual Strings and N=2 Supersymmetric Field Theory,''}
Nucl.\ Phys. \textbf{B477} (1996) 746, 
\texttt{hep-th/9604034}.\\*
S.~Katz, A.~Klemm and C.~Vafa,
\textit{``Geometric Engineering of Quantum Field Theories,''}
Nucl.\ Phys.  \textbf{B497} (1997) 173, 
\texttt{hep-th/9609239}.




\bibitem{AdS/CFT1}
J.~Maldacena,
\textit{``The Large $N$ Limit of Superconformal Field Theories
and Supergravity,''}
Adv. Theor. Math. Phys.\ \textbf{2}  (1998) 231,
\texttt{hep-th/9711200}.




\bibitem{AdS/CFT2}
S.~S.~Gubser, I.~R.~Klebanov and A.~M.~Polyakov,
\textit{``Gauge Theory Correlators from Non-Critical String Theory,''}
Phys. Lett. \textbf{B428}  (1998) 105,
\texttt{hep-th/9802109}. \\ 
E.~Witten,
\textit{``Anti De Sitter Space and Holography,''}
Adv. Theor. Math. Phys. \textbf{2}   (1998)  253,
\texttt{hep-th/9802150}.




\bibitem{Seiberg-Witten}
N.~Seiberg and E.~Witten,
\textit{``Electric-Magnetic Duality, Monopole Condensation, 
and Confinement in N=2 Supersymmetric Yang-Mills Theory,''}
Nucl.\ Phys.\ \textbf{B426} (1994) 19, 
\texttt{hep-th/9407087}; 
Erratum,  ibid.\ \textbf{B430} (1994) 485; 
\textit{``Monopoles, Duality and Chiral Symmetry Breaking 
in N=2 Supersymmetric QCD,''}
ibid.  \textbf{B431} (1994) 484, 
\texttt{hep-th/9408099}.





\bibitem{Nekrasov}
N.~A.~Nekrasov,
\textit{``Seiberg-Witten Prepotential from Instanton Counting,''}
Adv.\ Theor.\ Math.\ Phys.\  \textbf{7}  (2004) 831, 
\texttt{hep-th/0206161}.


\bibitem{Nekrasov-Okounkov}
N.~Nekrasov and A.~Okounkov,
\textit{``Seiberg-Witten Theory and Random Partitions,''}
\texttt{hep-th/0306238}.




\bibitem{Nakajima-Yoshioka}
H.~Nakajima and  K.~Yoshioka,
\textit{``Instanton Counting on Blowup. I,''}
\texttt{math.AG/0306198}; 
\textit{``Instanton Counting on Blowup. II,
~K-theoretic partition functions,''} 
\texttt{math.AG/05055553}. 






\bibitem{topological string}
E.~Witten,
\textit{``Topological Sigma Models,''}
Commun.\ Math.\ Phys.\ {\bf 118} (1988) 411;
\textit{``Mirror Manifolds and Topological Field Theory,''}
In Yau, S.T.(ed.): Mirror symmetry I, 121-160,
\texttt{hep-th/9112056}





\bibitem{topological vertex 1}
M.~Aganagic, A.~Klemm, M.~Marino and C.~Vafa, 
\textit{``The Topological Vertex,''}
Commun. Math. Phys. \textbf{254} (2005) 425,
\texttt{hep-th/0305132}.




\bibitem{topological vertex 2}
A.~Iqbal, 
\textit{``All Genus Topological String Amplitudes 
and 5-brane Webs as Feynman Diagrams,''}
\texttt{hep-th/0207114}.




\bibitem{Iqbal}
A.~Iqbal and A.~K.~Kashani-Poor,  
\textit{``Instanton Counting and Chern-Simons Theory,''} 
Adv. Theor. Math. Phys. \textbf{7} (2004) 457, 
\texttt{hep-th/0212279}; 
\textit{``SU(N) Geometries and Topological String Amplitudes,''} 
\texttt{hep-th/0306032}.




\bibitem{Eguchi}
T.~Eguchi and H.~Kanno,
\textit{``Topological Strings and Nekrasov's Formulas,''}
JHEP {\bf 0312} (2003) 006, 
\texttt{hep-th/0310235}. 



\bibitem{Crystal}
A.~Okounkov, N.~Reshetikhin and C.~Vafa,  
\textit{``Quantum Calabi-Yau and Classical Crystals,''} 
\texttt{hep-th/0309208}. 



\bibitem{Hawking}
S.~W.~Hawking,
\textit{``Space-time Foam,''}
Nucl.\ Phys.\ \textbf{B144} (1978) 349.



\bibitem{kahler gravity}
M.~Bershadsky and V.~Sadov, 
\textit{``Theory of K$\ddot{\mbox{a}}$hler Gravity,''}
Int. J. Mod.  Phys. \textbf{A11} (1996) 4689, 
\texttt{hep-th/9410011}



\bibitem{quantum foam}
A.~Iqbal, N.~Nekrasov, A.~Okounkov and C.~Vafa,
\textit{``Quantum Foam and Topological Strings,''}
\texttt{hep-th/0312022}.



\bibitem{Macdonald}
I.~G.~Macdonald,
\textit{``Symmetric Functions and Hall Polynomials,''}
Second edition, Clarendon Press, 1995.





\bibitem{MNTT1}
T.~Maeda, T.~Nakatsu, K.~Takasaki and T.~Tamakoshi, 
\textit{``Five-Dimensional Supersymmetric Yang-Mills Theories 
and Random Plane Partitions,''}
JHEP \textbf{0503} (2005) 056, 
\texttt{hep-th/0412327}.





\bibitem{Miwa-Jimbo}
M.~Jimbo and T.Miwa,
\textit{``Solitons and Infinite Dimensional Lie Algebras,''}
Publ. RIMS, Kyoto Univ., \textbf{19} (1983) 943.





\bibitem{Okounkov-Reshetikhin} 
A.~Okounkov and N.~Reshetikhin,  
\textit{``Correlation Function of Schur Process with Application 
to Local Geometry of a Random 3-Dimensional Young Diagram,''} 
J. Amer. Math. Soc. \textbf{16} (2003) no.3 581, 
\texttt{math.CO/0107056}. 





\bibitem{SU(N) curve}
A.~Klemm, W.~Lerche, S.~Theisen and S.~Yankielowicz,
\textit{``Simple Singularities and N=2 Supersymmetric Yang-Mills Theory,''}
Phys.\ Lett.\ \textbf{B344} (1995) 169,
\texttt{hep-th/9411048}.\\*
P.~Argyres and A.~Faraggi,
\textit{``The Vacuum Structure and Spectrum of N=2 Supersymmetric 
$SU(N)$ Gauge Theory,''}
Phys.\ Rev.\ Lett.\ \textbf{74} (1995) 3931,
\texttt{hep-th/9411057}.





\bibitem{Itzykson}
C.~Itzykson and J.~M.~Drouffe,
\textit{``Statistical Field Theory,''}
Cambridge University Press, 1989.



\bibitem{GKZ}
I.~Gelfand, M.~Kapranov and A.~Zelevinsky,
\textit{``Discriminants, resultants and Multidimensional Determinants,''}
Birkh\"auser, Boston, 1994.




\bibitem{Passare-Rullgard}
M.~Forsberg, M.~Passare and A.~Tsikh,
\textit{``Laurent Determinants and Arrangements of
Hyperplane Amoebas,''}
Advances in Math.\ \textbf{151} (2000) 45.\\*
M.~Passare and H.~Rullg\r{a}rd,
\textit{``Amoebas, Monge-Amp\`ere Measures,
and Triangulations of the Newton Polytope,''}
Duke. Math. J. \textbf{121} (2004) no.3 481.





\bibitem{Hori-Vafa}
K.~Hori and C.~Vafa,
\textit{``Mirror Symmetry,''}
\texttt{hep-th/0002222}.



\bibitem{Mirror}
K.~Hori, S.~Katz, A.~Klemm, R.~Pandharipande,
R.~P.~Thomas, C.~Vafa, R.~Vakil and E.~Zaslow,
\textit{``Mirror symmetry,''}
vol.~1 of
\textit{Clay Mathematics Monographs}.
American Mathematical Society, Providence, RI, 2003.




\bibitem{Toric Mirror}
V.~V.~Batyrev,
\textit{``Dual Polyhedra and Mirror Symmetry for Calabi-Yau Hypersurfaces
in Toric Varieties,''}
J.\ Alg.\ Geom.\ \textbf{3} (1994) 493.\\*
V.~V.~Batyrev and L.~A.~Borisov,
\textit{``On Calabi-Yau Complete Intersections in Toric Varieties,''}
\texttt{alg-geom/9412017};
\textit{``Mirror Duality and String-Theoretic Hodge Numbers,''}
\texttt{alg-geom/9509009}.



\bibitem{Kasteleyn}
P.~Kasteleyn, 
\textit{``Graph theory and crystal physics,''}
in 
\textit{Graph theory and theoretical physics,} 
Academic Press, London, 1967.  



\bibitem{Kenyon}
R.~Kenyon, 
\textit{``An Introduction to the Dimer Model,''} 
\texttt{math.CO/0310326}. 



\bibitem{Kenyon-Okounkov-Sheffield} 
R.~Kenyon, A.~Okounkov and S.~Sheffield, 
\textit{``Dimers and Amoebae,''} 
\texttt{math-ph/0311005}. 



\bibitem{Hanany-Kennaway}
A.~Hanany and K.~D.~Kennaway, 
\textit{``Dimer Models and Toric Diagrams,''} 
\texttt{hep-th/0503149}.



\bibitem{MNNT}
T.~Maeda, T.~Nakatsu, Y.~Noma and T.~Tamakoshi,
\textit{``Gravitational Quantum Foam and 
Supersymmetric Gauge Theories,''}
Nucl. Phys. \textbf{B735} (2006) 96, 
\texttt{hep-th/0505083}.



\bibitem{Mikhalkin}
G.~Mikhalkin,
\textit{``Amoebas of Algebraic Varieties and Tropical Geometry,''}
In: Different Faces of Geometry,
\texttt{math.AG/0403015}.



\bibitem{RGST}
J.~Richter-Gebert, B.~Sturmfels and T.~Theobald,
\textit{``First Steps in Tropical Geometry,"}
\texttt{math.AG/0306366}.



\bibitem{Viro} 
O.~Ya.~Viro, 
\textit{``Dequantization of Real Algebraic Geometry on a 
Logarithmic Paper,"}
Proceedings of the European Congress of Mathematicians 
(2000). 



\bibitem{Kenyon-Okounkov}
R.~Kenyon and A.~Okounkov, 
\textit{``Limit Shapes and the Complex Burgers Equation,''} 
\texttt{math-ph/0507007}. 



\bibitem{MNTT2}
T.~Maeda, T.~Nakatsu, K.~Takasaki and T.~Tamakoshi,
\textit{``Free Fermion and Seiberg-Witten Differential 
in Random Plane Partitions,''}
Nucl.  Phys.  \textbf{B715} (2005) 275,
\texttt{hep-th/0412329}. 




\bibitem{instanton countings on ALE 1}
R.~Fucito, J.~F.~Morales and R.~Poghossian, 
\textit{``Multi Instanton Calculus on ALE spaces,"}
Nucl.  Phys.  \textbf{B703} (2004) 518,
\texttt{hep-th/0406243}. 



\bibitem{quiver gauge theory}
S.~Matsuura and K.~Ohta, 
\textit{``Localization on D-brane and Gauge Theory/Matrix Model,"}
\texttt{hep-th/0504176}. 



\bibitem{instanton countings on ALE 2}
S.~Fujii and S.~Minabe, 
\textit{``A Combinatorial Study on Quiver Varieties,"}
\texttt{math.AG/0510455}. 



\bibitem{Rullgard}
H.~Rullg\r{a}rd, 
\textit{``Polynomial Amoebas and Convexity,"}
Preprint, 
Stockholm University, 2001. 



\bibitem{Fulton}
W.~Fulton,
\textit{``Introduction to Toric Varieties,''}
Princeton University Press, 1993.\\ 
T.~Oda,
\textit{``Convex Bodies and Algebraic Geometry,''}
Springer-Verlag, 1988.






\end{thebibliography}
\end{document}